\documentclass[fleqn,usenatbib,useAMS]{mnras}

\usepackage{graphicx}	
\graphicspath{{./}{figures/}}
\usepackage{amsmath}	
\usepackage{multicol}        
\usepackage{bm}		
\usepackage{pdflscape}	
\usepackage{tikz}
\usepackage{slashbox}

\newcommand\psr{PSR J1023$+$0038}
\newcommand{\flux}{\,erg\,cm$^{-2}$\,s$^{-1}$}

\newcommand{\lum}{\,erg\,s$^{-1}$}

\newcommand{\cm}{\,cm$^{-2}$}

\usepackage{nicematrix,booktabs,enumitem}
\setlist{leftmargin=5.5mm}
\usepackage[T1]{fontenc}
\usepackage{ae,aecompl}
\usepackage{subcaption}

\usepackage{newtxtext,newtxmath}

\usepackage{lineno}

\title{Two Possible Optical--X-Ray Anti-Correlations of PSR J1023$+$0038}

\author[K. Y. Au et al.]{Ka-Yui Au,$^{1}$\thanks{E-mail: kyau@phys.ncku.edu.tw, lilirayhk@phys.ncku.edu.tw}
Kwan-Lok Li,$^{1}$\footnotemark[1]
Albert K. H. Kong,$^{2}$
Jumpei Takata,$^{3}$
Chung-Yue Hui,$^{4}$
Lupin C. C. Lin$^{1}$
\\
$^{1}$Department of Physics, National Cheng Kung University, Tainan City 70101, Taiwan\\
$^{2}$Institute of Astronomy, National Tsing Hua University, Hsinchu 30013, Taiwan\\
$^{3}$Department of Astrophysics, School of Physics, Huazhong University of Science and Technology, Wuhan 430074, People’s Republic of China\\
$^{4}$Department of Astronomy and Space Science, Chungnam National University, Daejeon 34134, Republic of Korea
}

\pubyear{2025}

\begin{document}
\label{firstpage}
\pagerange{\pageref{firstpage}--\pageref{lastpage}}
\maketitle

\begin{abstract}
X-ray emission is generally believed to be one of the major heating sources for the optical modulation in redback pulsar binaries as we have seen similar phenomena in many low mass X-ray binaries (LMXBs). While, e.g.,  MeV/GeV $\gamma$-rays from the neutron stars are also possible heating sources, X-ray observations are currently much more sensitive, and therefore, joint optical--X-ray data are observationally unique on the irradiation mechanism investigation. Using 18 X-ray/B-band simultaneous \textit{XMM-Newton} observations (717 ks in total) of the redback system PSR J1023+0038 taken during the LMXB state, we find a general trend that the amplitude of the B-band orbital modulation was lower when the observed X-ray flux was higher. Depending on the analysis method adopted, the statistical significance of the anti-correlation can be from 1.7$\sigma$ to 3.1$\sigma$. We also extended the analysis to the GeV $\gamma$-ray band using the \textit{Fermi}-LAT data, but the result is insignificant to claim any relations. Moreover, another X-ray/optical correlation regarding the low modes of the system was found in some of the \textit{XMM-Newton} observations, and the astrophysical reason behind is currently unclear yet. These anti-correlations likely suggest that the irradiation is generally stronger when the X-ray flux is in a fainter state, indicating that there is a more dominant irradiation source than the X-ray emission. 
\end{abstract}




\begingroup
\let\clearpage\relax
\endgroup
\newpage

\section{Introduction}

Neutron stars with spin periods in the order of milliseconds are millisecond pulsars (MSPs), representing an important stage for the evolution of neutron star binary. It is widely believed that there is a companion star which transfers the mass and angular momentum to spin up the neutron star by accretion. This process is commonly known as the recycling scenario \citep{alpar}. Two subclasses of pulsar binaries with short orbital periods ($\leq$ 1 day) and very low-mass companions, named black widow (companion mass: $\leq$ 0.1 $M_{\sun}$) and redback (companion mass: 0.1--0.4 $M_{\sun}$; \citealt{roberts2011new,chen}), could be formed during the recycling process. These spider pulsar systems' companions keep being ablated because of the $\gamma$-ray radiation and/or the winds from the pulsars. Eventually, the companion would be ``evaporated'' leaving an isolated MSP \citep{van}. On the other hand, \citet{2019MNRAS.490..889P} suggests that the efficiency of the ablation is too low to fully evaporate the companion and lead a black window or a redback system to be an isolated MSP. Furthermore, spider pulsars formation could be difficult during the recycling process, as some extra processes (e.g., cyclic mass transformation due to X-ray emissions and sustained magnetic braking; \citealt{2014ApJ...786L...7B,2015MNRAS.449.4184B,2020MNRAS.495.3656G,2021MNRAS.500.1592G}) could be required. Recently, \citet{2025A&A...693A.314M} further shows that the formation of these binary MSPs highly depends on the interaction between accretion and pulsar wind irradiation in the system.

These black widow/redback systems maybe an important link between low-mass X-ray binaries (LMXBs) and isolated MSPs. In the pulsar recycling process, a radio pulsar can be spun up during the LMXB state, given that the accretion rate is sufficiently high. In 2010s, the transition between the LMXB state and the radio pulsar state was observed in at least three redbacks systems, PSR J1227$-$4853 \citep{roy}, PSR J1023+0038 \citep{archi,pat,sta}, and M28I \citep{pap}. This provides direct evidence to support the MSP formation through the recycling scenario \citep{alpar}.

The target in this paper, \psr\ (J1023 hereafter) discovered by \citet{bond}, was identified as a 1.69-ms redback MSP with an orbital period of 4.75 hr and a $\sim$0.2$M_{\sun}$ companion star by radio observations \citep{Woudt,sta}. J1023 is one of the three transitional MSPs (tMSPs) that shows transitions from/to the radio pulsar state to/from the LMXB state in the last 25 years. An accretion disk feature was shown in this system before 2002 but then disappeared when the radio pulsations were observed \citep{wang}. Until mid-June 2013, the radio pulsations of J1023 disappeared as J1023 entered the LMXB state again with the accretion disk reappeared \citep{wang,sta}. During the LMXB state, the $\gamma$-ray, X-ray, and optical emission enhanced by an order of magnitude \citep{kong,pat,sta,takata2014}, and the state of the radio pulsar is still under debate. In the intrabinary shock (IBS) scenario, due to the newly formed accretion disk, the stronger interaction in IBS (which is located far from the light cylinder) increases the emission and blocks radio pulses from the active pulsar \citep{takata2014,li2014}. Alternatively, in the propeller model, shocks (near the light cylinder) made by the disk in-flows and propelling magnetosphere of a quenched pulsar would accelerate electrons to emit the enhanced emission \citep{propeller}. Moreover, a hybrid model has also been proposed, and the pulsar could be swinging between on and off during the LMXB state in the model \citep{ibsprop}. In 2017, the optical pulsations of J1023 was found, providing a strong evidence to show that the pulsar in J1023 is still active at least sometimes \citep{2017NatAs...1..854A}. Recently, mode-switching models have shown that the disk's innermost region can provide an environment for the enhanced emissions. The existence of a shock (near the light cylinder) formed by the disk inner-flow and the pulsar winds control the switching phenomenon \citep{2019ApJ...884..144V,2019ApJ...882..104P, 2023A&A...677A..30B}. Moreover, the pulsar would ejects most of the accretion masses before they reach the pulsar's surface. Those ejected masses shroud the system and probably block the radio pulsation to cause the radio pulsation non-detection \citep{2023A&A...677A..30B}.

In the past decade, J1023 has been observed in multi-wavelength by different telescopes. Observed features include flares in optical and X-ray band \citep{obs10101}, optical orbital modulation \citep{thor,homer,archi,obs10101}, anti-correlation in X-ray and radio variabilities \citep{xranti}, and the X-ray high/low mode emission \citep{obs10101,2019ApJ...882..104P, 2022MNRAS.512.5269L,2023A&A...677A..30B}. Notice that the optical orbital modulation is mainly caused by the pulsar heating on the tidally-locked companion \citep{2018MNRAS.477.1120K} and this phenomenon has been observed in many spider pulsar systems. Moreover, the significant anti-correlations in the X-ray and radio variabilities of J1023 were found in both the \textit{Chandra}/Very Large Array (VLA) and one \textit{XMM-Newton}/VLA datasets \citep{xranti}. In the datasets, there are always corresponding radio brightening during the X-ray low-mode phases. Recently, a promising tMSP candidate, 3FGL J1544.6--1125, also shows a possible anti-correlation between the X-ray and radio emissions in one of the two epochs from the simultaneous  \textit{Chandra} and VLA observations \citep{2025MNRAS.536...99G}, which behaves similarly to J1023.

The X-ray and radio anti-correlation of J1023 \citep{xranti} has clearly revealed that simultaneous multi-wavelength observations are required for a deeper investigation of the sub-luminous state. In this paper, we reanalyzed the \textit{XMM-Newton} and the \textit{Fermi}-LAT observations of J1023, with which we find an anti-correlation between the X-ray emission and the irradiation on the companion. With this result and the aforementioned hybrid shock models, we deduce that the irradiations on the companion from the pulsar winds could be affected by the disk inner-flow. In addition, another optical/X-ray anti-correlation regarding the high/low modes of the tMSP was found, although the physical reason behind is currently unclear yet. Our result is similar to the X-ray and radio variability behaviour in \citet{xranti}, however, the low-mode anti-correlation presented in this work is not observed in every \textit{XMM-Newton} observation. In the following sections, we will describe the details of the analyses followed by a discussion section for the possible implications from these observational results.

\section{Data Reduction}
We downloaded the \textit{XMM-Newton} observations from the astronomical data archive of the High Energy Astrophysics Science Archive Research Center (HEASARC)\footnote{\url{https://heasarc.gsfc.nasa.gov/cgi-bin/W3Browse/w3browse.pl}}. A total of 21 archival observations that were taken during the LMXB state were found in the database. However, three of them were not used in this work because (i) J1023 was not fully in the field of view of the Optical Monitor (OM) in Obs ID 0784700201, and (ii) the effective exposure times of Obs IDs 0864010201 and 0784700301 are too short\footnote{The exposure time of Obs ID 0864010201 is less than 6ks, which is significantly shorter than the orbital period of J1023. For Obs ID 0784700301, the EPIC data is only available in the first 650s.}. As a result, only 18 datasets were used and analysed (Table \ref{tab:epicom_exp}). In the 18 observations, all the OM images were taken in the fast mode (time resolution: 0.5s) with the B filter of \textit{XMM-Newton}\footnote{\url{https://xmm-tools.cosmos.esa.int/external/xmm_user_support/documentation/uhb/omfilters.html}}. For the European Photon Imaging Camera (EPIC) data, all the pn and MOS 1/2 X-ray observations were taken in the timing mode and the small window mode (i.e., \texttt{PrimePartialW2}), respectively, except for Obs ID 0864010101, in which the full window mode was used for the MOS 1/2 data. In this study, we focused on the OM and EPIC data.

Science Analysis Software (SAS; version 21.0.0) and HEASoft (version 6.29) were used to reduce and analyse the observations. Following the instruction of the on-line manual\footnote{\url{https://www.cosmos.esa.int/web/xmm-newton/sas-thread-startup}}, we set up a proper SAS environment and downloaded the current calibration files (CCF; downloaded on 2025 August 26) on our local computer. The extractions of the scientific products will be described in the following sections.

\subsection{X-Ray Spectra}
The SAS task \texttt{xmmextractor} was employed to reduce the EPIC data with an energy range of 0.3--10~keV. All the parameters of the task were kept to the default values, except otherwise stated. In the X-ray spectral extraction, the data taken during the time intervals with high external flaring background levels were removed. High background intervals were identified by single event (i.e., \texttt{PATTERN==0}) high-energy light curves (10--12~keV; 10s binned, to match the X-ray variability time-scale of J1023) using the criteria of count rates higher than 0.35 and 0.4 counts/s for MOS 1/2 and pn, respectively (which are default values in \texttt{xmmextractor}). In \texttt{xmmextractor}, optimized source and background extraction regions were generated by \texttt{eregionanalyse}. For the MOS 1/2 observations, all the source regions are circular centred at the target. Most of them have a radius of 40\arcsec, and the exceptions are 38\arcsec, 50\arcsec, and 51\arcsec\ in MOS 2 Obs. Q and MOS 1/2 Obs. R, respectively. In most of the MOS 1/2 observations, the background regions are circular regions with radius 57\arcsec\ in a source-free field. The three exceptions are MOS 2 Obs. Q and MOS 1/2 Obs. R, which have target-centred annulus background regions with inner/outer radii of 60\arcsec/81\arcsec, 60\arcsec/93\arcsec, and 60\arcsec/94\arcsec, respectively. For all the timing-mode pn data, the source regions are in the RAW-X range between 29 and 45, and the background regions are in the RAW-X range between 1 and 16. Table \ref{tab:epicom_exp} shows the remaining exposures after the background filtering. The X-ray spectra were produced by \texttt{especget}, and the corresponding ancillary response files (ARFs) and response matrix files (RMFs) were generated accordingly. Finally, the MOS 1/2 and pn spectra were binned to at least 20 counts per bin using the HEASoft tool \texttt{grppha}.

\subsection{EPIC and OM Light Curves}
The MOS 1/2 and pn event files reprocessed by \texttt{xmmextractor} were used to extract light curve in 0.3--10~keV. The exposure corrections of the light curves were performed by the task \texttt{epiclccorr}. Since MOS 1/2 and pn did not start/stop observing at the exact same times, we trimmed the data so that the three EPIC instruments were all observing throughout the time period. The OM B-band light curves were extracted by \texttt{omfchain} with an initial time bin size of 0.5s. This fine bin size ensures good time alignments between the OM and EPIC light curves, since OM and EPIC started/ended the observations at different times as well. For each observation set, the four light curves were re-binned to 10s with the same epoch using \texttt{lcurve}. In addition, the three background subtracted EPIC light curves were summed up to maximize the signal-to-noise ratio. These resultant light curves (MOS 1 + MOS 2 + pn) are in the unit of  count rate, and the uncertainties were computed using standard error propagation. The combined X-ray and OM light curves of all the 18 observations were shown in appendix (Figures \ref{app_a} and \ref{app_b}).

\begin{table*}
\caption{Start times and exposure times of the EPIC (after the flaring background filtering) and OM data}
\begin{tabular}{cccccccc}

\hline
&Obs. ID & Calendar date & Start time (t$_s$) & MOS 1 & MOS 2 & pn & OM\\
&&(UTC)&(MJD)& (ks) & (ks) & (ks) & (ks)\\
\hline
A	&	0720030101	& 2013 Nov 10 &	56606.65	&	111.8	&	109.1	&	98.7	&	138.1\\
B	&	0742610101	& 2014 Jun 10 &	56818.14	&	104.8	&	105.0	&	99.9	&	131.1\\
C	&	0748390101	& 2014 Nov 21 &	56982.76	&	31.3	&	30.3	&	28.7	&	35.7\\
D	&	0748390501	& 2014 Nov 23 &	56984.76	&	31.0	&	30.5	&	28.0	&	36.2\\
E	&	0748390601	& 2014 Nov 28 &	56989.88	&	20.0	&	19.9	&	16.7	&	22.0\\
F	&	0748390701	& 2014 Dec 17 &	57008.64	&	33.1	&	33.0	&	30.8	&	35.8\\
G	&	0770581001	& 2015 Nov 11 &	57337.78	&	28.0	&	27.8	&	25.4	&	32.4\\
H	&	0770581101	& 2015 Nov 13 &	57339.13	&	22.0	&	21.8	&	20.4	&	24.0\\
I	&	0783330301	& 2015 Dec 09 &	57365.03	&	25.4	&	25.3	&	23.2	&	27.7\\
J	&	0794580801	& 2017 May 23 &	57896.89	&	21.3	&	21.1	&	19.7	&	26.0\\
K	&	0794580901	& 2017 May 24 &	57897.70	&	22.2	&	22.3	&	20.8	&	24.5\\
L	&	0803620201	& 2017 May 08 &	57881.90	&	22.7	&	22.7	&	21.1	&	25.3\\
M	&	0803620301	& 2017 May 10 &	57883.93	&	26.5	&	26.5	&	25.0	&	29.0\\
N	&	0803620401	& 2017 May 16 &	57889.91	&	20.9	&	20.5	&	18.3	&	24.0\\
O	&	0803620501	& 2017 Jun 13 &	57917.57	&	21.9	&	21.7	&	20.4	&	24.0\\
P	&	0823750301	& 2018 Dec 11 &	58463.84	&	26.9	&	26.8	&	24.2	&	30.0\\
Q	&	0823750401	& 2018 Dec 15 &	58467.87	&	31.6	&	31.3	&	29.7	&	34.0\\
R	&	0864010101	& 2021 Jun 03 &	59368.83	&	57.8	&	52.4	&	58.5	&	64.5\\
\hline
\end{tabular}
\label{tab:epicom_exp}
\end{table*}

\section{Data Analysis and Results}
\subsection{An Anti-Correlation Between the Optical MA and X-Ray (and $\gamma$-Ray Possibly) Emission}
It is believed that the X-ray emission and the optical orbital modulation are closely related (e.g., \citealt{gen,j1653,xor}). If X-ray heating is important to the companion irradiation, we naturally expect that the optical orbital modulation amplitude (MA) is getting higher as the X-ray emission of the same system enhances. To study the X-ray emission and the optical modulation relation by \textit{XMM-Newton}, we estimate the degree of the pulsar heating effect in J1023 in different epochs by measuring the optical orbital MA using the OM data, counting the mode occurrence rate in each EPIC observation, and representing the average X-ray strength of the EPIC observations by several ways (i.e., mean/median count rates, and X-ray spectral modeling). The following subsections explain in detail how these parameters are determined and the result of comparing the MA and the average X-ray strength.


\subsubsection{Optical Modulation Amplitude} \label{sec:o}
\begin{figure*}
\centering
\includegraphics[width=0.9\textwidth]{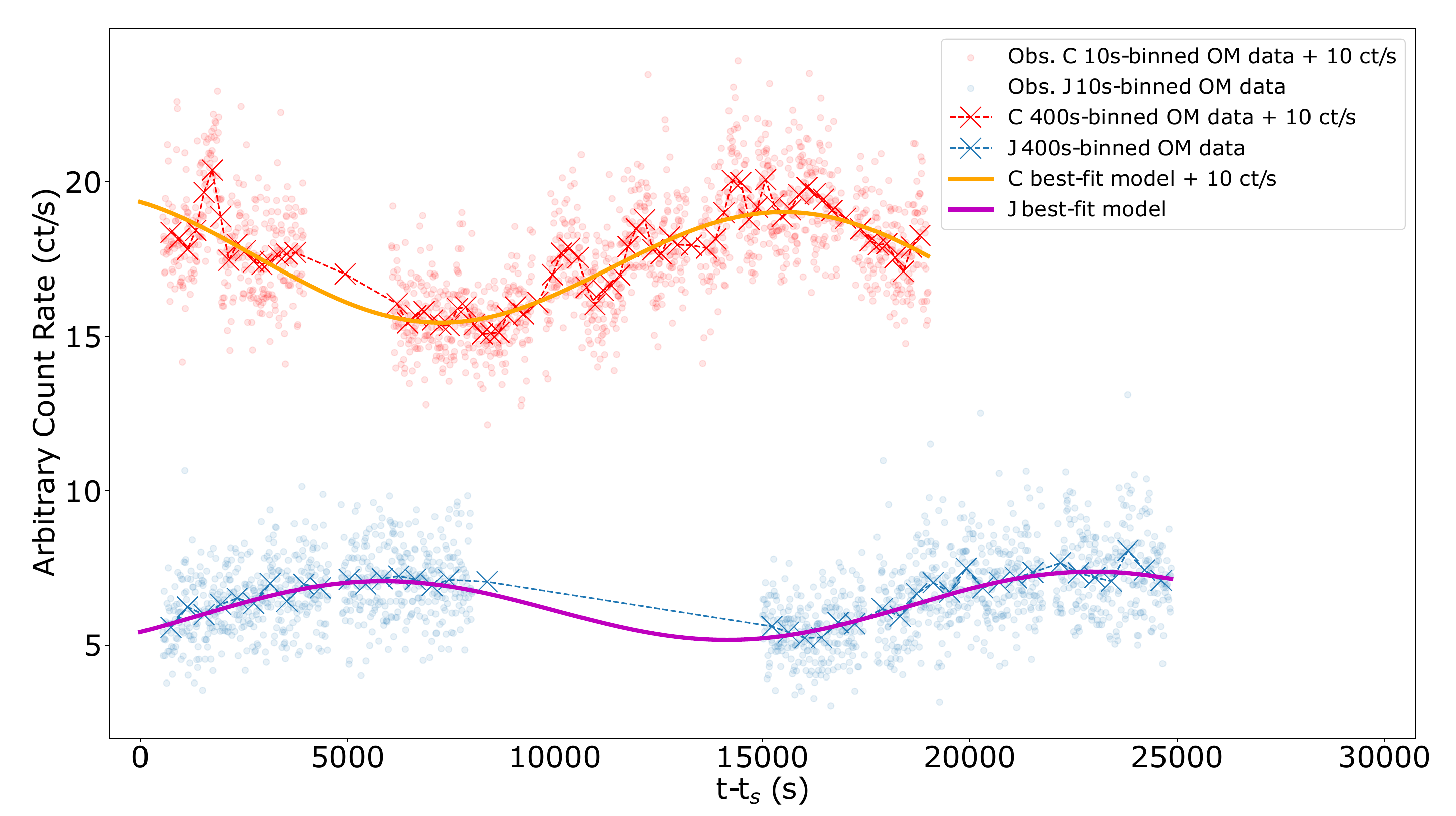}
\caption{The B-band light curves of Obs. C and J. The translucent dots and the cross/dashed lines refer to the 10s-binned and 400s-binned OM data, respectively (red/top for Obs. C and blue/bottom for Obs. J). The orange and magenta solid curves are the best-fit models of Obs. C and J, respectively. All the data relevant to Obs. C are shifted upwards by 10 ct/s to clarify the two data sets.}
\label{fig:optical_example}
\end{figure*}

Many redback pulsars show orbital modulation in the optical bands due to pulsar heating, and J1023 is one of the well-known examples \citep{archi,obs10101,2018MNRAS.477.1120K}. To measure the MA, we decide to fit a sinusoidal function without any physical model to the B-band OM light curve with $a*$sin$((2\pi/p)*(t-t_0))+m(t-t_0)+c$, where $a$ is the MA in count rate (i.e., ct/s), $p$ is the period of 0.1980963155 d \citep{otherobs}, $t$ is the time since the start of an observation, $m$ is the slope of a linear function to describe the long-term trend (referring to the non-flat optical modulation; \citealt{2018ApJ...858L..12P}), and $c$ is a constant count rate. Since the OM light curves contain flares, which significantly affect the fitting result of MA, we removed the flaring episodes by visual inspection before the fitting. The fitting results look good after removing the flares. The MA fitting results of all 18 observations are shown in Table \ref{tab:o_fitresult}. The optical MAs are varying between 1.0 ct/s and 2.5 ct/s. Figure \ref{fig:optical_example} shows the OM data and the best-fit models of Obs. C and J (with one of the lowest and highest MAs, respectively) as an example to indicate the MA differences among the datasets. The MAs can vary significantly in a few days. For example, the Obs. C, D, and E were observed during 2014 Nov 21--28, and their MAs were 1.95 ct/s, 1.50 ct/s, and 2.5 ct/s, respectively. These results imply that the heating sources varied on a time-scale as short as a few days. Moreover, we find that there are 4 observations (Obs. C, O, P, and Q) around MA$\sim$2.0 ct/s, seemingly forming their own group. We thought that they were all observed around the same epochs, but they were not (Obs. C, O,  and P/Q were observed in 2014 Nov., 2017 Jun., and 2018 Dec., respectively).

\begin{table*}
\begin{center}

\caption{The fitting parameters of the OM light curves}
 
    \begin{tabular}{cccccccc} 

 \hline
      &	Optical	&	t$_0$	& Slope ($m$) 				& Slope ($m$) 				& c \\
      & MA (Abs. values)				&			& (with flare filtering)	& (without flare filtering)	&	\\
      & (ct/s) 			& (s) 		& ($10^{-6}$) 				& ($10^{-6}$)				& (ct/s)	\\
      \hline
      A & 1.42$\pm$0.02 & 4065$\pm$38     & 3.7$\pm$0.5     & $-$9.4$\pm$0.3  & 7.33$\pm$0.03	\\
      B & 1.39$\pm$0.02 & $-$5805$\pm$34  & $-$6.0$\pm$0.3  & $-$4.6$\pm$0.3  & 6.74$\pm$0.03	\\
      C & 1.95$\pm$0.05 & $-$5743$\pm$64  & $-$38.8$\pm$6.9 & 144$\pm$2       & 7.89$\pm$0.11	\\
      D & 1.50$\pm$0.03 & 2157$\pm$57     & 6.1$\pm$2.5     & 6.1$\pm$2.5     & 7.54$\pm$0.04	\\
      E & 2.49$\pm$0.06 & $-$6030$\pm$112 & $-$469$\pm$20   & $-$219$\pm$7    & 14.78$\pm$0.33	\\
      F & 1.61$\pm$0.03 & 355$\pm$52      & $-$1.8$\pm$2.3  & $-$1.8$\pm$2.3  & 7.27$\pm$0.04	\\
      G & 1.53$\pm$0.06 & $-$1042$\pm$80  & 16.3$\pm$4.3    & 39.8$\pm$4.2    & 6.51$\pm$0.09	\\
      H & 1.08$\pm$0.06 & 2616$\pm$111    & $-$0.9$\pm$7.0  & 15.7$\pm$6.7    & 6.79$\pm$0.06	\\
      I & 1.51$\pm$0.03 & $-$498$\pm$58   & $-$70.1$\pm$3.1 & $-$70.1$\pm$3.1 & 8.76$\pm$0.05	\\
      J & 1.03$\pm$0.05 & 1433$\pm$153    & 18.1$\pm$3.2    & $-$3.6$\pm$3.1  & 5.97$\pm$0.07	\\
      K & 1.08$\pm$0.04 & $-$554$\pm$101  & $-$33.2$\pm$5.4 & 225$\pm$4       & 6.82$\pm$0.07	\\
      L & 1.15$\pm$0.09 & $-$4983$\pm$522 & 59.4$\pm$8.2    & 215$\pm$4       & 6.36$\pm$0.18	\\
      M & 1.51$\pm$0.03 & 502$\pm$72      & $-$7.0$\pm$3.6  & $-$33.8$\pm$2.8 & 7.18$\pm$0.05	\\
      N & 1.66$\pm$0.04 & 5486$\pm$74     & 24.8$\pm$5.1    & 24.8$\pm$5.1    & 6.36$\pm$0.04	\\
      O & 1.97$\pm$0.04 & $-$4403$\pm$107 & 42.7$\pm$13     & 582$\pm$5       & 7.18$\pm$0.16	\\
      P & 1.96$\pm$0.03 & 1961$\pm$46     & $-$15.0$\pm$3.1 & 22.6$\pm$2.8    & 7.07$\pm$0.04	\\
      Q & 1.93$\pm$0.04 & $-$3088$\pm$50  & $-$79.9$\pm$4.0 & $-$100$\pm$3    & 10.73$\pm$0.10	\\
      R & 1.27$\pm$0.02 & $-$4836$\pm$50  & $-$19.9$\pm$1.3 & $-$38.2$\pm$1.0 & 7.21$\pm$0.06	\\
      \hline
    \end{tabular}
\label{tab:o_fitresult} 
  \end{center}
\end{table*}

\subsubsection{X-Ray Low/High Modes Occurrence Rate}\label{sec:xlhm}
Because of the obvious flux difference between the low/high modes X-ray emission \citep{obs10101}, the modes occurrence rate could affect the overall X-ray flux in each observing window and thus affect the heating on the companion. Therefore, we calculated the non-low-mode (i.e., high mode + flare) occurrence rate ($\frac{high\,mode\,+\,flare\,mode\,duration}{total\,duration}$) in each observation and determine whether it (anti-)correlates with the optical MA.

\begin{figure}
\includegraphics[width=0.5\textwidth]{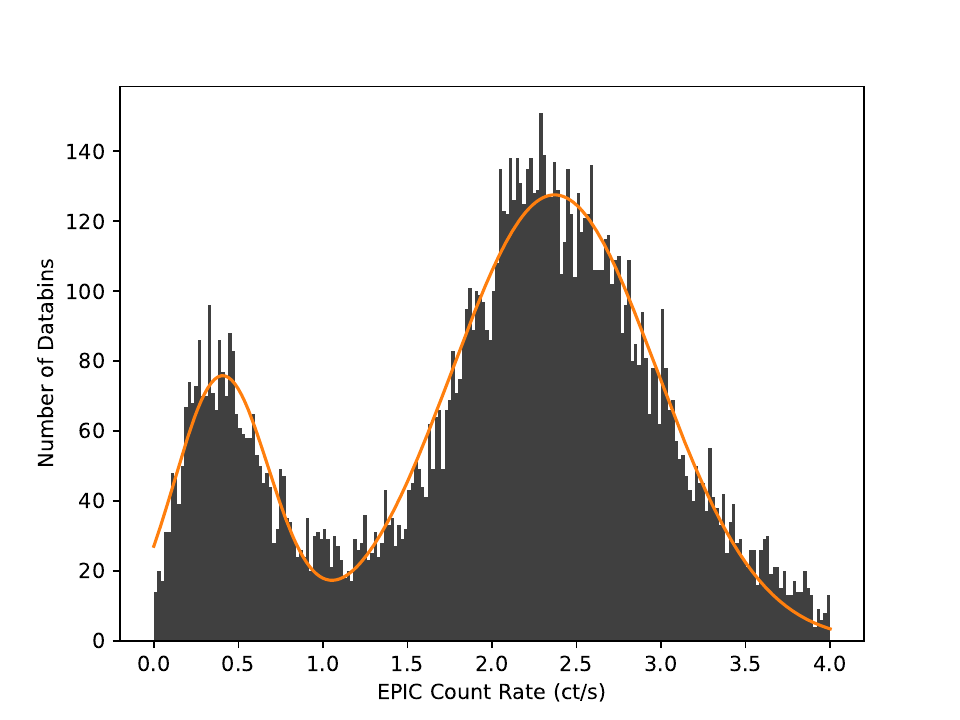}
\vspace{-0.5cm}
\caption{The X-ray flux bimodal distribution of J1023 (Obs. A) with the best-fit double Gaussian model.
\label{fig:doublegauss}}
\end{figure}

For each of the EPIC combined light curves, the bimodal flux distribution was fitted with a double Gaussian model (Figure \ref{fig:doublegauss}). Based on the best-fit mean and sigma of the low-mode Gaussian model (i.e., $\mu_{low}$ and $\sigma_{low}$), the low-mode time intervals were identified by the following steps.

\begin{itemize}
\item{All the EPIC data bins below the 3 low-mode sigma level (i.e., $\mu_{low}+3\sigma_{low}$) are initially marked as low-mode.}
\item{A data bin below $4\sigma_{low}$ is marked as low mode if one of its adjacent points is in low mode and the count rate difference between the adjacent point and the data bin is less 1.5 counts/s.}
\item{A data bin between two low-mode data points is marked as low mode if the count rate differences between the adjacent points and the data bin are both less 2.0 counts/s.}
\item{After 10 iterations of the above two steps, all low-mode marks with two adjacent non-low-mode data points are unmarked if the count rate differences between the adjacent points and the data bins are both less 2.0 counts/s.}
\item{Based on the above result, the corresponding OM data bins will be marked accordingly.}
\end{itemize}

The second and third rules are mainly used to recover some marginal low-mode cases, which only have minor effects on the results. To elaborate the threshold values of 1.5 and 2.0 counts/s (named $z_1$ and $z_2$, respectively, see appendix \ref{appC}), most of the adjacent low-mode point pairs have count rate differences less than 1.5 counts/s. Also, a point between two (non-)low-mode data likely belongs to the same class, if their differences are less than 2.0 counts/s. We tried a few other threshold sets with steps of 0.5 counts/s, but the result did not get significantly better (see Figure \ref{fig:lowmode_classification} for four examples). Judged visually, the accuracy of the combination of 1.5 and 2.0 counts/s is slightly better than the others.

\begin{figure}
\includegraphics[width=0.5\textwidth]{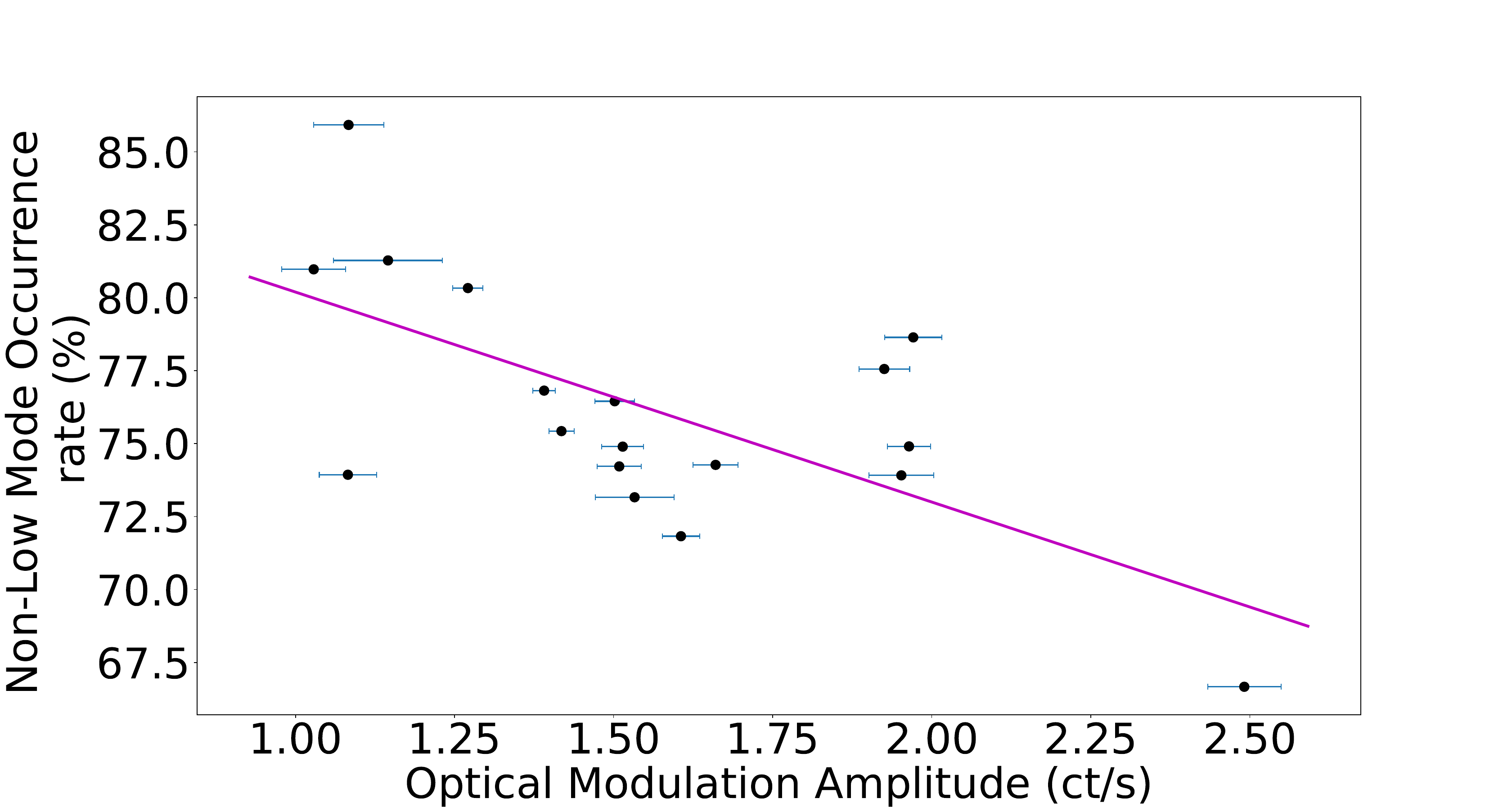}

\caption{The non-low-mode occurrence rate vs. optical MA. The solid line shows a possible anti-correlation between the non-low-mode occurrence rate and the optical MA.
\label{fig:mode_occur}}
\end{figure}

The non-low-mode occurrence rates are shown in Table \ref{tab:lowmode_count}. By comparing the rates with the optical MAs, a likely anti-correlation feature is shown in Figure \ref{fig:mode_occur}. We calculate the anti-correlation significance through the Pearson coefficient ($r$) and Spearman’s rank correlation coefficient ($\rho$) of the dataset. Using the coefficient, we can compute the t-score and p-value to find the significance levels ($\sigma_r$ and $\sigma_\rho$, respectively). The significances of the anti-correlation are $\sigma_{r}\approx$2.9$\sigma$ with $r\approx-$0.65 and $\sigma_{\rho}\approx$2.2$\sigma$ with $\rho\approx-$0.51 (shown in Table \ref{tab:quan_signi}). The anti-correlation is most prominent when MA is less than 1.8 ct/s with Obs. K as a possible outlier (MA = 1.08 ct/s; non-low-mode occurrence rate = 74.3\%). The highest non-low-mode occurrence rate can reach about 86\% and the lowest can go down to around 67\%. The anti-correlation is not clearly shown around MA = 2.0 ct/s, since there are four observations do not follow the trend with higher occurrence rates (Figure \ref{fig:mode_occur}).

\begin{figure*}
\centering
\includegraphics[width=0.9\textwidth]{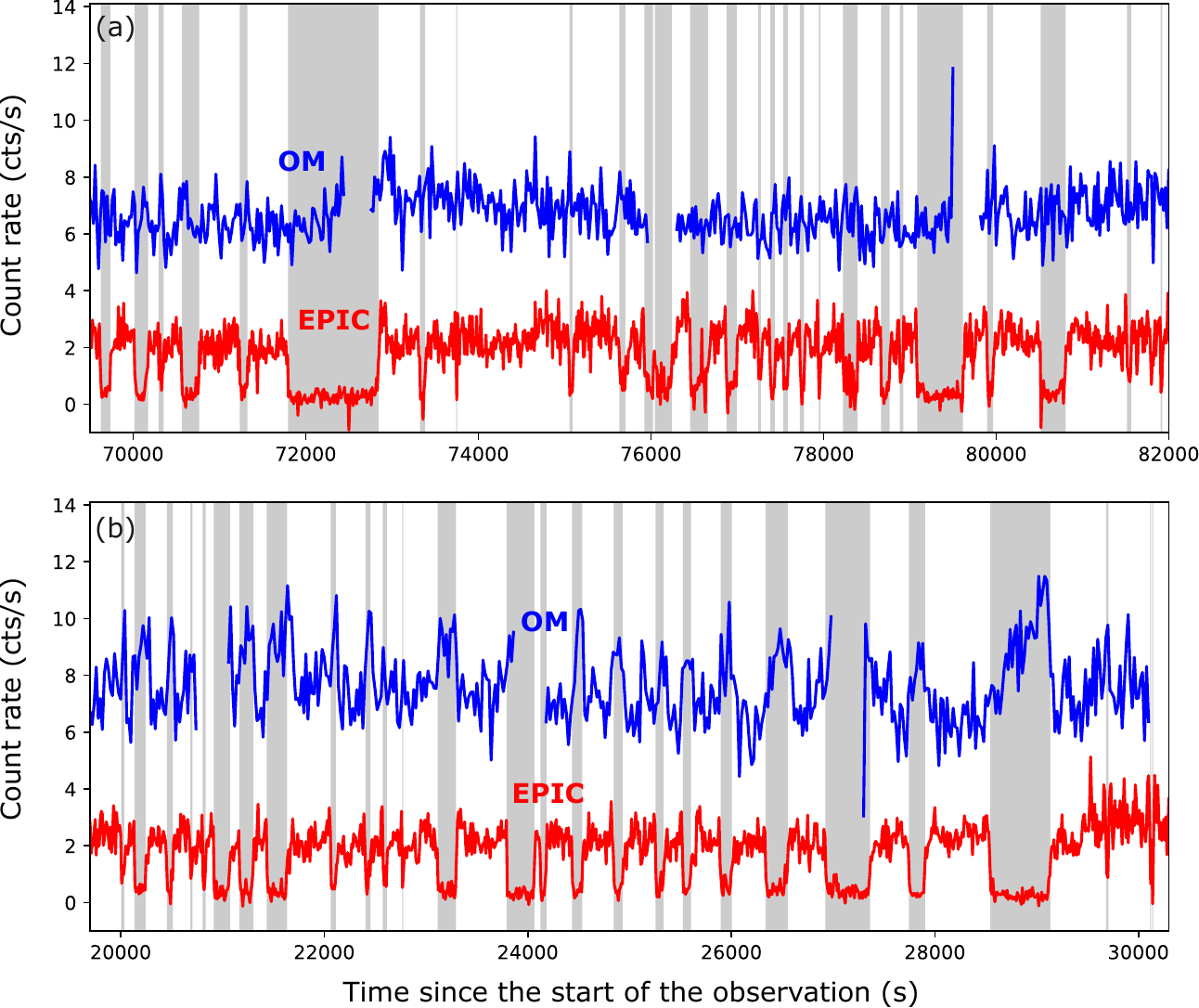}
\caption{The EPIC (red) and detrended OM (blue) light curves of J1023 that show the non-detection (upper panel; part of Obs. B) and detection (lower panel; part of Obs. D) of the low-mode anti-correlation. The OM light curves were re-binned with a 20s bin size for better visualization. The gray shadows indicate the identified X-ray low mode of J1023.
\label{fig:low_mode_example}}
\end{figure*}

\begin{table*}
  \caption{The X-ray best-fit double Gaussian parameters and the mode classification for the \textit{XMM-Newton} observations}
    \begin{tabular}{c c c c c c c c} 
      \hline
      & Low-mode best-fit & Low-mode best-fit & High-mode best-fit & High-mode best-fit &Total duration & Non-low-mode & Non-low-mode\\
      & mean $\mu_{low}$ (ct/s) & sigma $\sigma_{low}$ (ct/s) & mean $\mu_{high}$ (ct/s) & sigma $\sigma_{high}$ (ct/s) &(ks) & duration (ks) & occurrence rate (\%)\\   
      \hline
      A & 0.40 & 0.28 & 2.37 & 0.61 & 128.2 & 96.7 & 75.4 \\
      B & 0.37 & 0.25 & 2.37 & 0.56 & 115.6 & 88.8 & 76.8 \\
      C & 0.32 & 0.17 & 2.38 & 0.57 & 32.2 & 23.8 & 73.7 \\
      D & 0.36 & 0.22 & 2.21 & 0.50 & 32.7 & 25.0 & 76.3 \\
      E & 0.31 & 0.18 & 2.30 & 0.49 & 16.8 & 11.2 & 66.8 \\
      F & 0.37 & 0.22 & 2.41 & 0.53 & 32.3 & 23.2 & 71.9 \\
      G & 0.38 & 0.20 & 2.31 & 0.60 & 27.2 & 19.9 & 73.3 \\
      H & 0.30 & 0.19 & 2.62 & 0.56 & 20.6 & 17.7 & 85.9 \\
      I & 0.39 & 0.23 & 2.29 & 0.51 & 24.3 & 18.2 & 75.0 \\
      J & 0.36 & 0.19 & 2.45 & 0.54 & 22.6 & 18.3 & 81.1 \\
      K & 0.34 & 0.20 & 2.41 & 0.55 & 21.1 & 15.6 & 74.3 \\
      L & 0.38 & 0.18 & 2.30 & 1.01 & 21.9 & 17.8 & 81.5 \\
      M & 0.36 & 0.20 & 2.35 & 0.52 & 25.6 & 19.0 & 74.4 \\
      N & 0.35 & 0.21 & 2.36 & 0.49 & 20.6 & 15.3 & 74.5 \\
      O & 0.39 & 0.18 & 2.12 & 0.55 & 20.6 & 16.2 & 78.7 \\
      P & 0.38 & 0.22 & 2.26 & 0.51 & 26.3 & 19.7 & 74.7 \\
      Q & 0.39 & 0.22 & 2.35 & 0.54 & 30.3 & 23.5 & 77.6 \\
      R & 0.38 & 0.21 & 2.27 & 0.48 & 54.4 & 43.7 & 80.3 \\
      \hline
    \end{tabular}
\label{tab:lowmode_count} 
\end{table*}

\begin{table*}

	\caption{Summary of the correlation significances between different quantities and the optical MA in $\sigma_r$ and $\sigma_\rho$}

		\begin{tabular}{c|c c c }
		 	\multicolumn{4}{c}{In $\sigma_r$}\\

      		\hline
			\backslashbox{Quantities}{Conditions}& No classification & (1) & (2) \\ \hline
			Mode occurrence rate & 2.9$\sigma$ & 3.0$\sigma$ & 2.4$\sigma$\\
			Mean & 2.0$\sigma$ & 2.6$\sigma$  & 2.1$\sigma$\\
			Median & 3.1$\sigma$ & 2.7$\sigma$  & 2.4$\sigma$\\
			Energy flux & 1.2$\sigma$ & 1.3$\sigma$  & 1.7$\sigma$\\
			$\Gamma$ & 1.4$\sigma$ & 2.4$\sigma$  & 2.2$\sigma$\\
			$N_H$ & 1.6$\sigma$ & 1.7$\sigma$  & 1.1$\sigma$\\
			
			\hline
			\multicolumn{4}{c}{}\\
			
		 	\multicolumn{4}{c}{In $\sigma_\rho$}\\
      
      		\hline
			\backslashbox{Quantities}{Conditions}& No classification & (1) & (2) \\ \hline
			Mode occurrence rate & 2.2$\sigma$ & 2.4$\sigma$ & 2.3$\sigma$\\
			Mean & 1.8$\sigma$ & 2.2$\sigma$  & 1.7$\sigma$\\
			Median & 3.0$\sigma$ & 2.5$\sigma$  & 2.0$\sigma$\\
			Energy flux & 0.8$\sigma$ & 1.0$\sigma$  & 1.1$\sigma$\\
			$\Gamma$ & 1.5$\sigma$ & 2.2$\sigma$  & 1.3$\sigma$\\
			$N_H$ & 1.4$\sigma$ & 1.4$\sigma$  & 0.6$\sigma$\\

			\hline

		\end{tabular}
\label{tab:quan_signi}

\end{table*}

\subsubsection{X-Ray Mean/Median} \label{sec:X}
Inspired by the result in MA/non-low-mode occurrence rate anti-correlation, we extracted the mean/median value from the EPIC X-ray light curves as the representative X-ray flux for each observation. Due to imperfect background subtraction, there are some negative count rates in the light curves that were all considered as zero in the analysis. It is well known that X-ray flares exist in J1023 \citep{obs10101}. However, unlike the optical MA measurement with optical flare filtering, we decided not to filter the X-ray flares in the analysis. The main reason is that the X-ray flares should also contribute to the heating on the companion surface. As a result, the X-ray heating power will be underestimated if only ``quiescent'' X-ray flux is considered. On the other hand, we also noticed that intensive X-ray flare episodes can occasionally appear in short periods of time (see appendix \ref{appB}). These occasional events can raise the average X-ray count rate level and affect our results heavily. Therefore, we present the results using both the mean and median count rates rather than just showing one of them: the median values are affected less by these occasional flares; however, the medians may underestimate the X-ray low-mode contribution (i.e., the medians are generally larger than the means). It avoids bias whether X-ray flares are retained in our analysis. We also applied the standard error propagation to get the error of the mean value for each observation. All the results are shown in Table \ref{tab:x_stat}.

\begin{table}
\begin{center}

\caption{The mean/median count rates of the EPIC data}
 
    \begin{tabular}{ccc} 

 \hline
      &		X-ray mean	&	X-ray median\\
      & 	count rate	&	count rate\\
      & 		(ct/s)	&	(ct/s)\\
      \hline
      A &  2.113$\pm$0.003 & 2.21\\
      B &  2.115$\pm$0.003 & 2.22\\
      C &  2.317$\pm$0.006 & 2.22\\
      D &  1.867$\pm$0.006 & 2.05\\
      E &  1.677$\pm$0.007 & 2.00\\
      F &  1.901$\pm$0.006 & 2.17\\
      G &  2.058$\pm$0.006 & 2.11\\
      H &  2.727$\pm$0.008 & 2.60\\
      I &  1.899$\pm$0.006 & 2.11\\
      J &  2.527$\pm$0.008 & 2.39\\
      K &  2.405$\pm$0.007 & 2.27\\
      L &  3.201$\pm$0.008 & 2.49\\
      M &  1.995$\pm$0.006 & 2.17\\
      N &  1.911$\pm$0.007 & 2.16\\
      O &  2.701$\pm$0.007 & 2.14\\
      P &  2.039$\pm$0.006 & 2.08\\
      Q &  2.255$\pm$0.006 & 2.24\\
      R &  2.170$\pm$0.004 & 2.18\\
      \hline
    \end{tabular}
\label{tab:x_stat} 
  \end{center}
\end{table}

\begin{figure}
\centering
	\begin{subfigure}{.5\textwidth}
    \centering
    	\includegraphics[width=.95\linewidth]{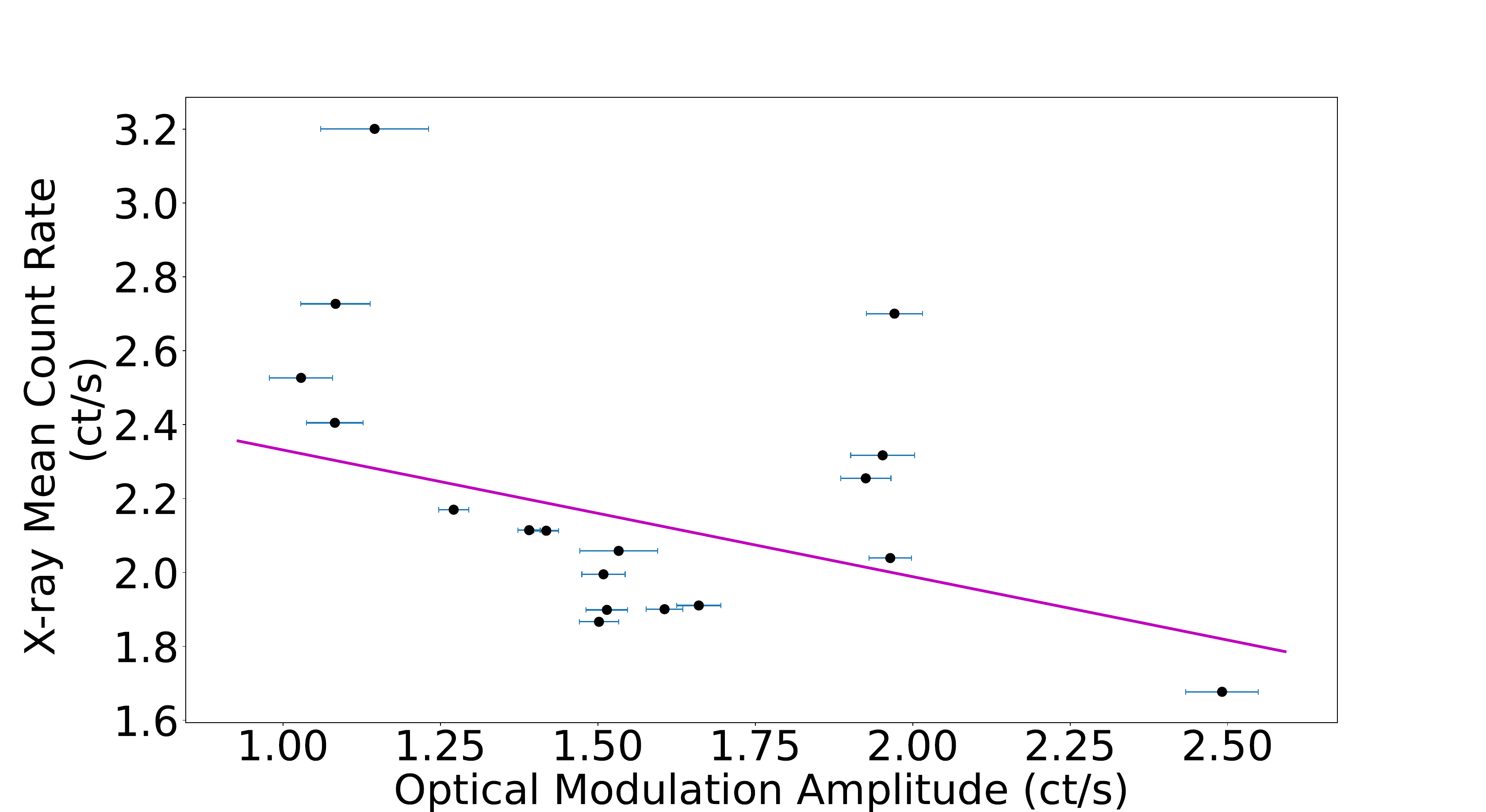}
\end{subfigure}
	\begin{subfigure}{.5\textwidth}
    \centering
   		 \includegraphics[width=.95\linewidth]{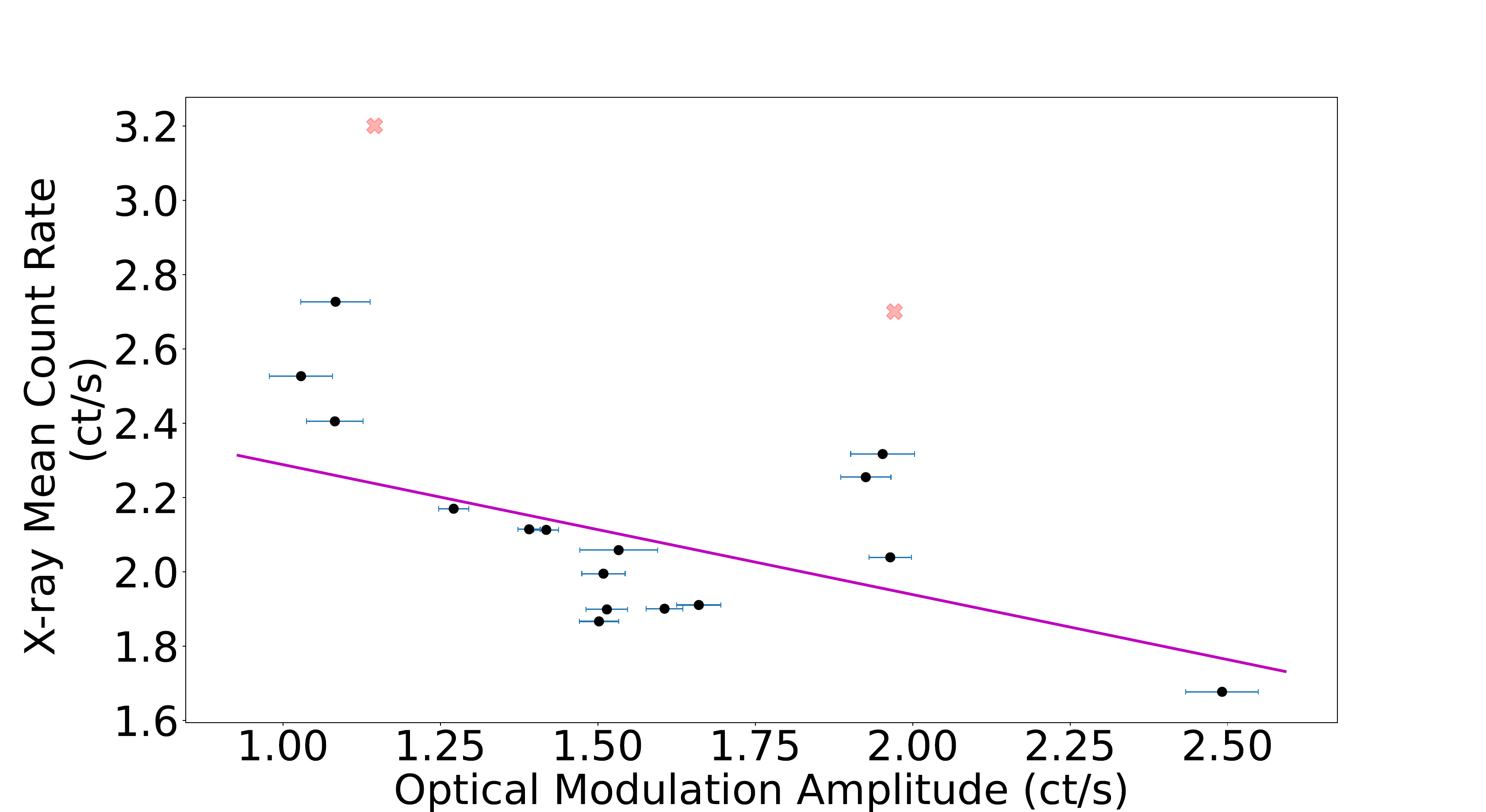}
\end{subfigure}       
	\begin{subfigure}{.5\textwidth}
    \centering
   		 \includegraphics[width=.95\linewidth]{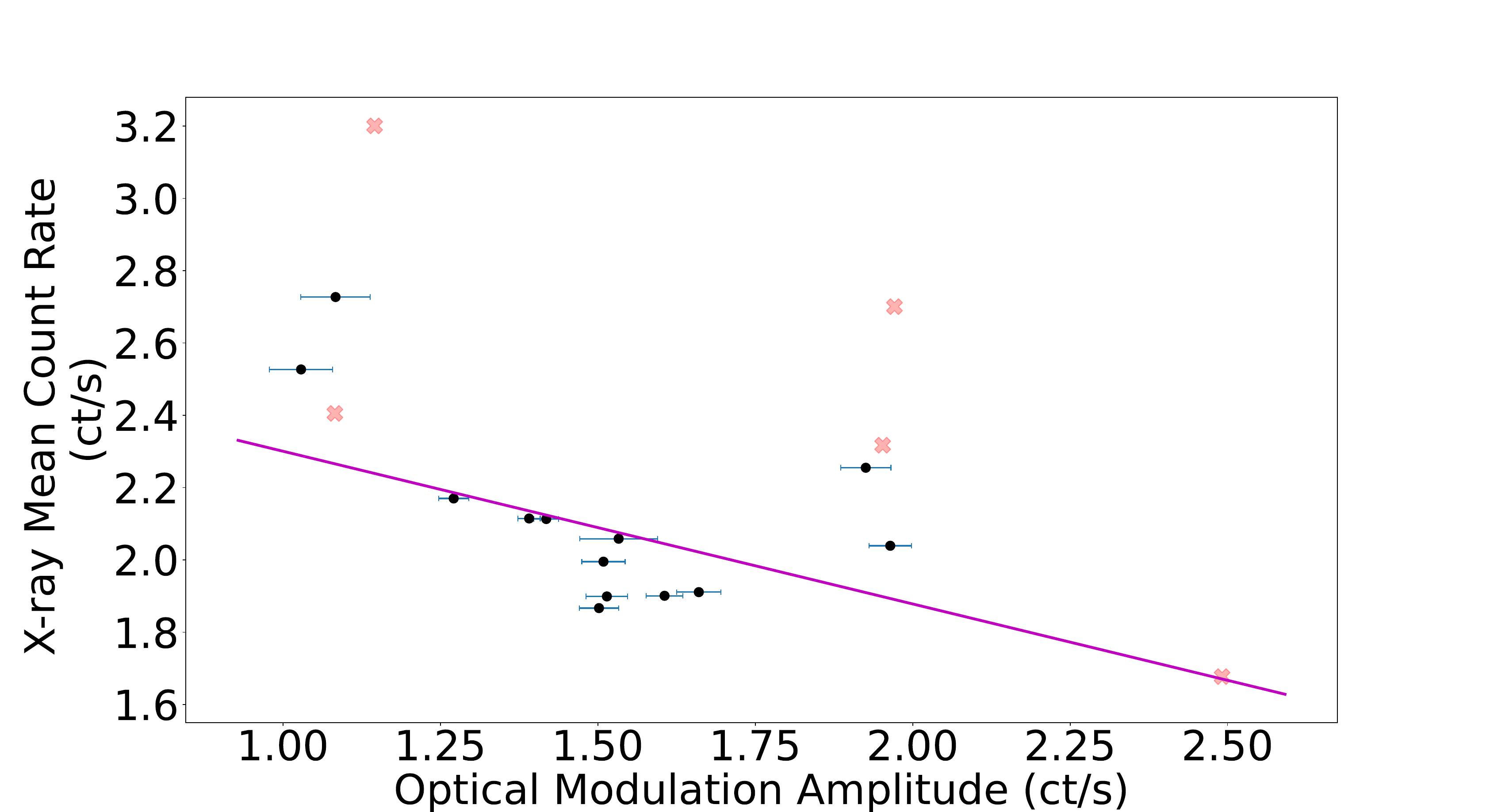}
\end{subfigure}    
\caption{The X-ray mean count rate (top: no filtration, middle: filtered by condition (1), bottom: filtered by condition (2)) vs. optical MA plot. The mean error bars are not visible because the errors are too small (Table \ref{tab:epicom_exp}). The solid lines are used to visualize the anti-correlation by fitting a linear function, and the transparent red crosses refer to the observations that are filtered by the corresponding conditions. The anti-correlation significances by using mean count rate are between 1.7--2.6$\sigma$ (see Sections \ref{sec:X}).
\label{fig:mean}}
\end{figure}

\begin{figure}
\centering
	\begin{subfigure}{.5\textwidth}
    \centering
    	\includegraphics[width=.95\linewidth]{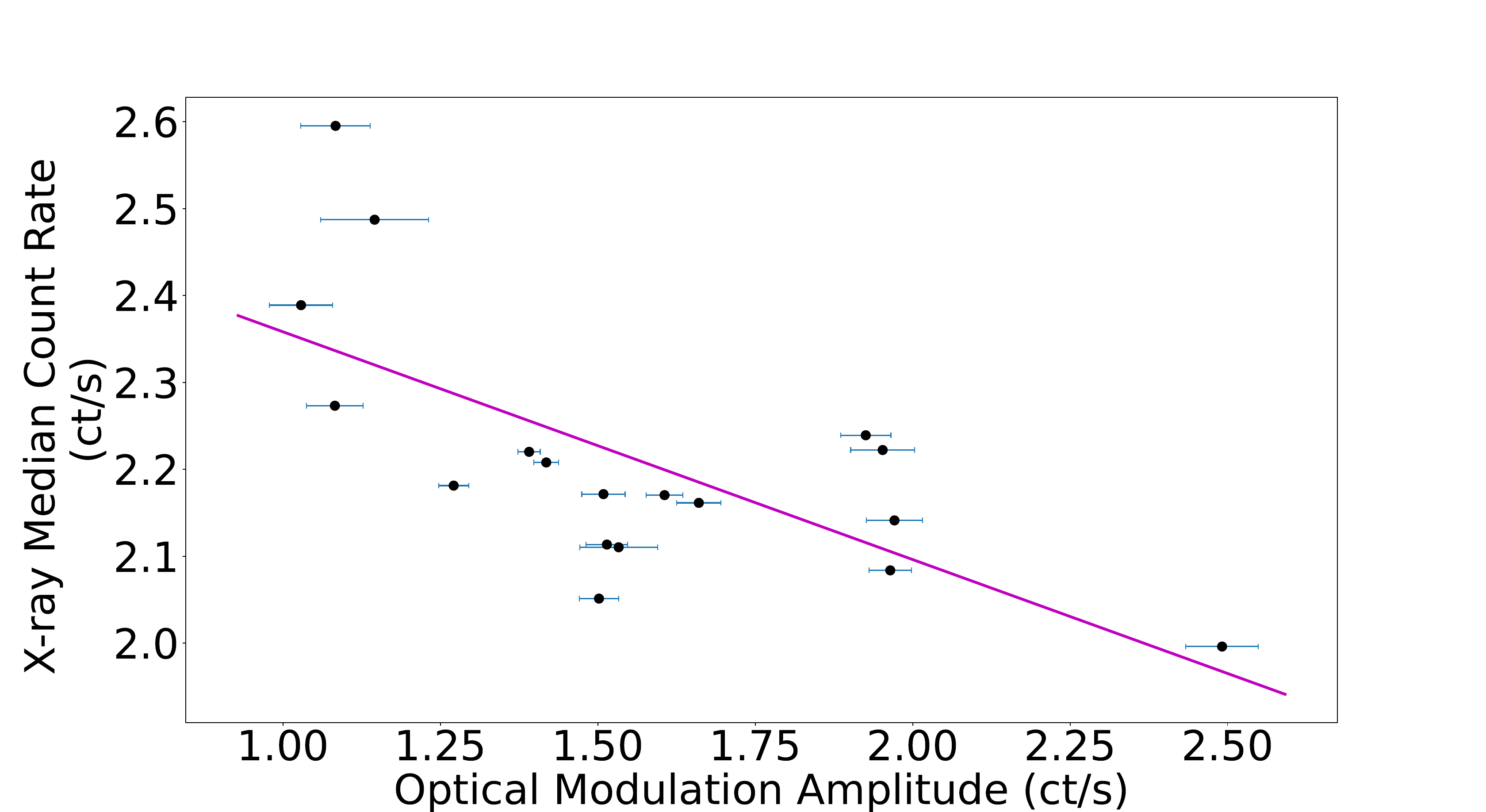}
\end{subfigure}
	\begin{subfigure}{.5\textwidth}
    \centering
   		 \includegraphics[width=.95\linewidth]{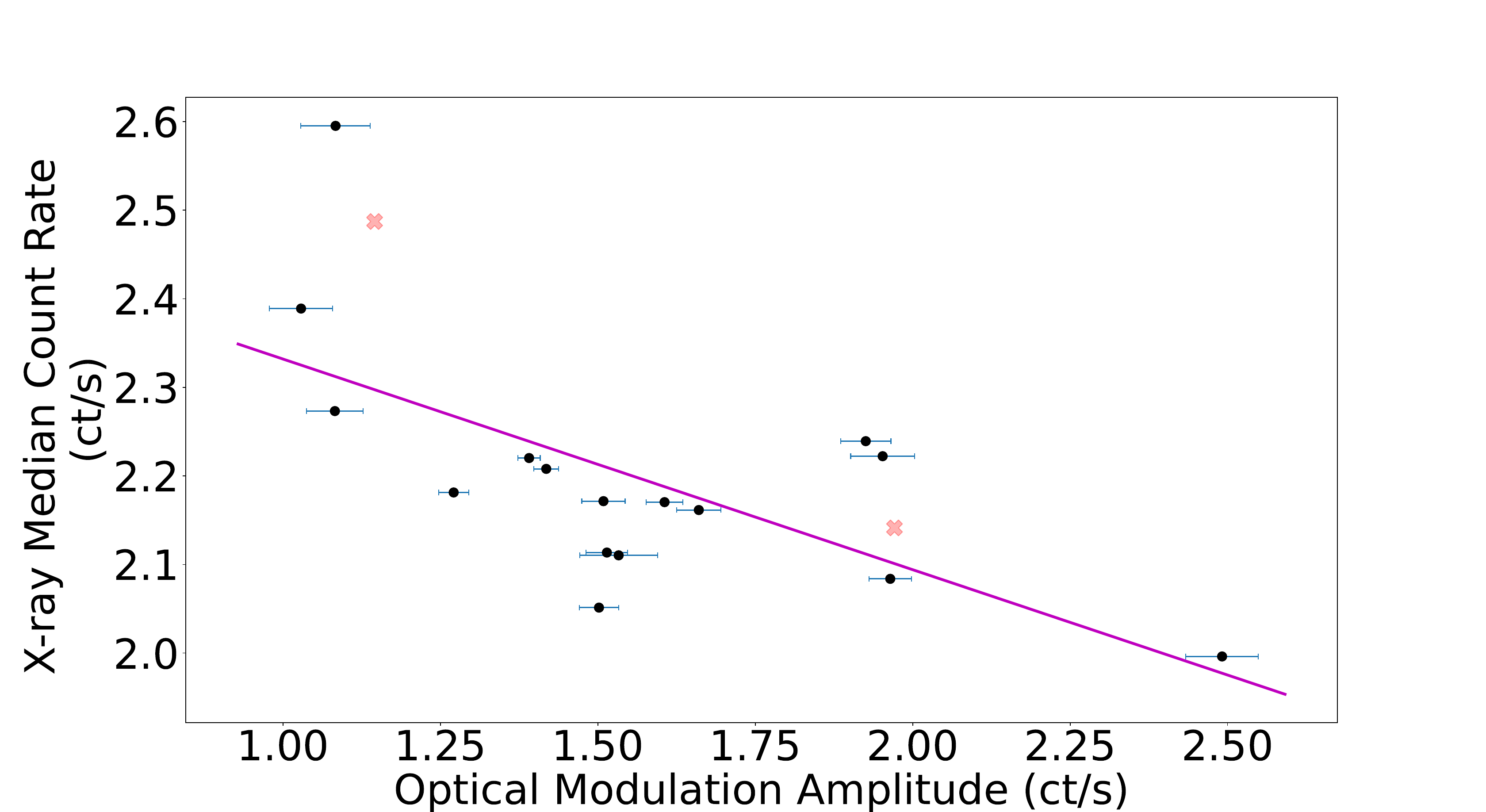}
\end{subfigure}       
	\begin{subfigure}{.5\textwidth}
    \centering
   		 \includegraphics[width=.95\linewidth]{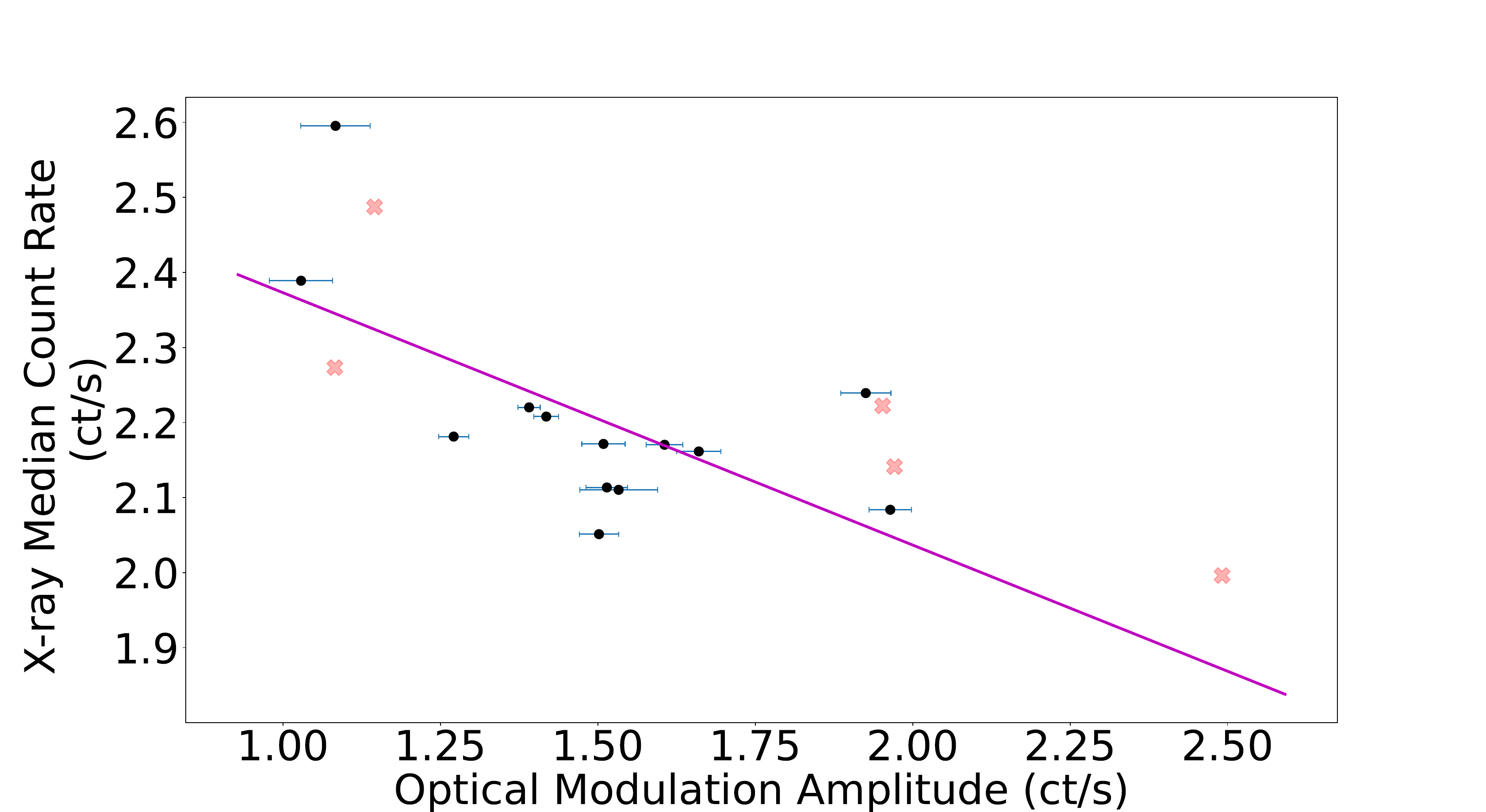}
\end{subfigure}       
\caption{The X-ray median count rate vs. optical MA plot, which is similar to the Fig. \ref{fig:mean}. The anti-correlation significances by using median count rate are between 2.0--3.1$\sigma$ (see Sections \ref{sec:X}).
\label{fig:median}}
\end{figure}

Combining the X-ray count rate representatives with MA, a possible anti-correlation feature is also shown in both datasets (Figures \ref{fig:mean} and \ref{fig:median}). The significances of the anti-correlation for the mean count rates are $\sigma_r\approx$2.0$\sigma$ with $r\approx-$0.47 and $\sigma_\rho\approx$1.8$\sigma$ with $\rho\approx-$0.44. The anti-correlation using the median count rates have significances $\sigma_r\approx$3.1$\sigma$ with $r\approx-$0.68 and $\sigma_\rho\approx$3.0$\sigma$ with $\rho\approx-$0.66, which are much more significant compared with the mean count rate version. These results are also shown in Table \ref{tab:quan_signi}. The anti-correlation can also be slightly seen in the version with mean count rates when the MA is less than 1.8 ct/s. However, at the same place around MA=2.0 ct/s, the same four observations also have higher mean X-ray count rates, making the anti-correlation unclear. After looking into the X-ray light curves of these data (Obs. C, O, P, and Q), the X-rays behave normally compared with the other EPIC data. Additionally, the four datasets were not taken in the same epoch (C: 2014 Nov; O: 2017 Jun; P/Q: 2018 Dec). Therefore, the reason why they do not follow the anti-correlation trend is unknown at this stage. For the version with median count rates, the trend seems to be consistent with all the observations. The differences are probably caused by the X-ray flares as mentioned. To minimise the effect, we tried to remove those observations that are possibly affected by the X-ray flares heavily by several classification methods that will be discussed in the following section, \S \ref{sec:xmmwc}.

\subsubsection{Classifications of the \textit{XMM-Newton} Observations}\label{sec:xmmwc}

Given the possibility that the outliers in top panel of Figures \ref{fig:mean} and \ref{fig:median} could be caused by the strong X-ray/B-band variabilities in the corresponding observations, we tried two classification schemes and see if (some of) the outliers can be removed. The two schemes are based on:

\begin{itemize}
 \item (1) The difference between the mean and median count rates: In Figures \ref{fig:mean} and \ref{fig:median}, we have shown that the mean and median X-ray count rates can result in different MA--X-ray count rate plots. This is perhaps because of the X-ray flare episodes as we have mentioned earlier in the section. Therefore, we employed the difference between the two count rates as an indicator to determine whether an observation is heavily affected by X-ray flaring. 

 \item (2) The difference in the optical light curve slope ($m$) with/without flare filtering: The MA measurement is crucial in the analysis, and optical flares are one of the major contaminations. Although the optical flares can be well removed by visual inspection, selection bias could be caused. To quantify the effect of the visual inspection on the MA measurement, we re-fit the OM light curve without filtering the flares visually. The difference between the new $m$ (without flare filtering) and the old $m$ (with flare filtering) can indicate the influence of the visual inspection process. If the difference is huge, then the visual inspection is a big factor for the MA measurement in this observation and the data should be ignored accordingly. We also tried to classify the observations by just considering the new $m$. Since the results are not significantly altered, we decided to not present it here for simplicity.
 
\end{itemize}

\begin{figure}
\centering
	\begin{subfigure}{.5\textwidth}
    \centering
    	\includegraphics[width=.95\linewidth]{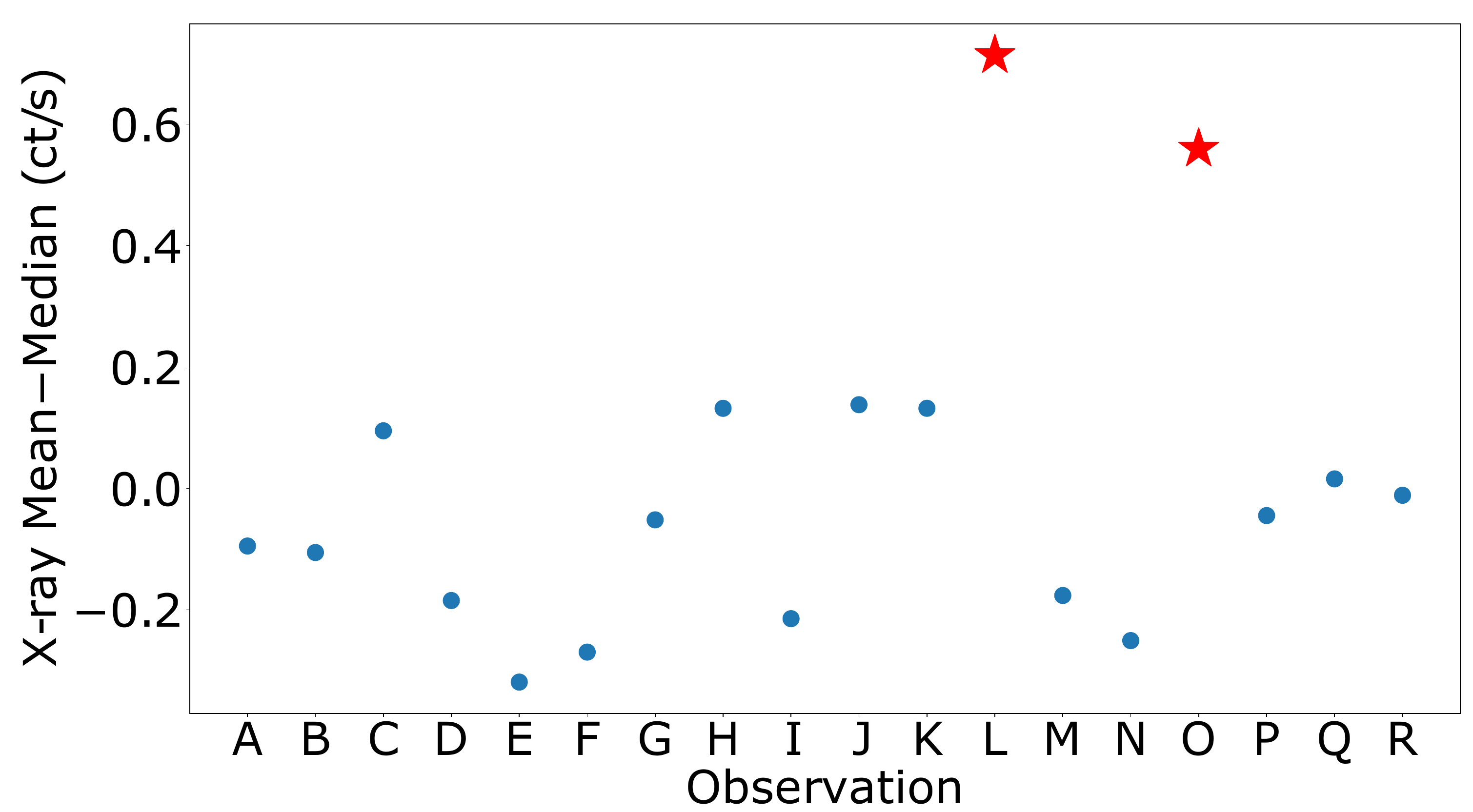}
\end{subfigure}
	\begin{subfigure}{.5\textwidth}
    \centering
   		 \includegraphics[width=.95\linewidth]{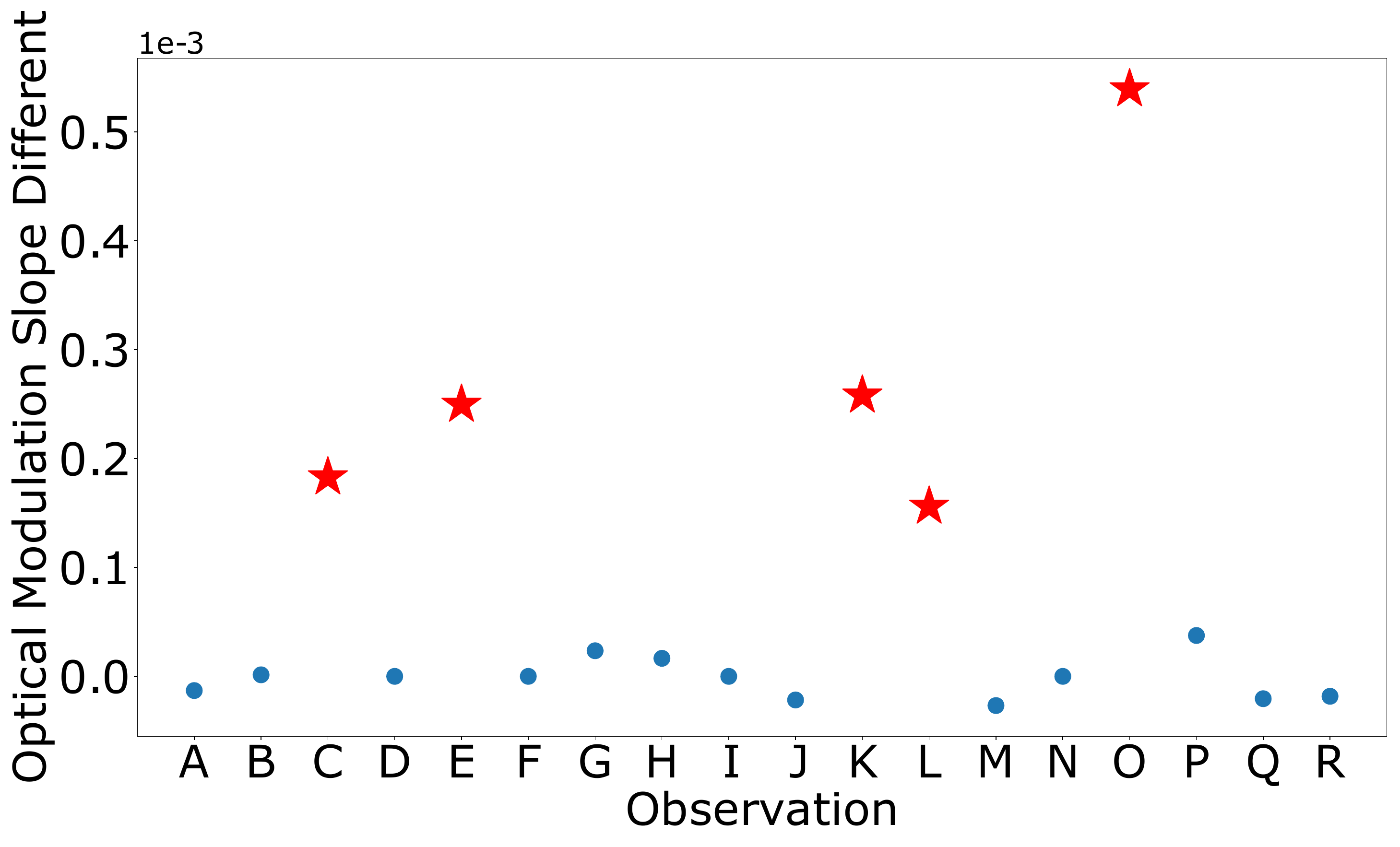}
\end{subfigure}       
\caption{The distributions of the parameters used in the conditions 1 (top) and 2 (bottom). The x-axis indicates different XMM-Newton observations (see Table \ref{tab:epicom_exp}). Red stars are the observations filtered by the corresponding condition.
\label{condi_dis}}
\end{figure}

For the two classification schemes, we used k-mean clustering by the Python package \texttt{kmeans1d} (version 0.3.1; \citealt{1dkmean}) to judge which observations should be filtered. The conditions (1) and (2) filter the observations (L and O) and (C, E, K, L, and O), respectively. The filtration results are also shown in Figure \ref{condi_dis}.

The X-ray mean/median count rates against optical MA plots after the filtration are shown in Figures \ref{fig:mean} and \ref{fig:median}. The anti-correlation significances in $\sigma_r$ using the mean X-ray count rates after the filtration by the conditions (1) and (2) increase to 2.6$\sigma$ with $r\approx-$0.62 and 2.1$\sigma$ with $r\approx-$0.59, respectively. For the median version, the anti-correlation significances in $\sigma_r$ decrease to 2.7$\sigma$ with $r\approx-$0.65 and 2.4$\sigma$ with $r\approx-$0.64, respectively. 
For the Spearman's test ($\sigma_\rho$), the significances in the mean version increase to 2.2$\sigma$ with $\rho\approx-$0.54 and decrease to 1.7$\sigma$ with $\rho\approx-$0.49 by the conditions, respectively. The median version significances decrease to 2.5$\sigma$ with $\rho\approx-$0.60 and 2.0$\sigma$ with $\rho\approx-$0.56, respectively. The results (including the filtration results in mode occurrence rate) are all summarised in Table \ref{tab:quan_signi}.

\subsubsection{X-Ray Spectral Parameters of the \textit{XMM-Newton} Observations}\label{sec:xsp}
\begin{figure}
\includegraphics[width=0.5\textwidth]{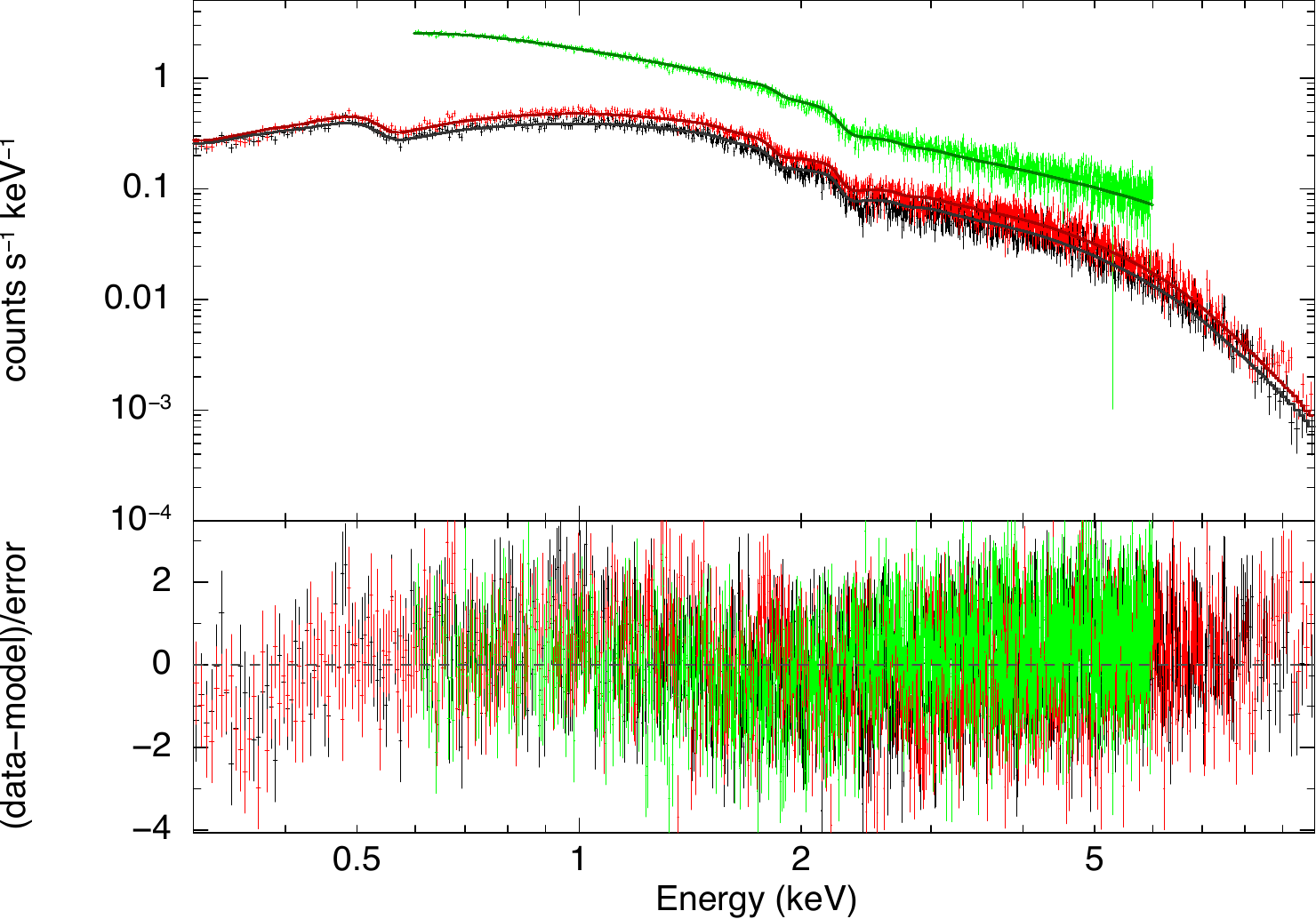}

\caption{The X-ray spectrum of Obs. A in an energy band of 0.3--10 keV. Solid lines and crosses refer to the best-fit results and data points. The pn, MOS 1, and MOS 2 data correspond to green, red, and black colors. The fitting parameters to the absorbed power-law and the statistical result of all the observations can be referred to Table \ref{tab:spec_para}.
\label{fig:sp_ex}}
\end{figure}

Besides using the X-ray mean/median count rates, we also considered the spectral parameters by fitting the 18 \textit{XMM-Newton} X-ray spectra with an absorbed power-law model using \texttt{XSPEC}\footnote{\url{https://heasarc.gsfc.nasa.gov/xanadu/xspec}} (version 12.12.0; \citealt{xspec}). Figure \ref{fig:sp_ex} as an example shows the best-fit absorbed power-law model with the EPIC data in the energy range of 0.3--10 keV (the energy range 0.6--6.0 keV was used for the pn data because there are instrumental artifacts below 0.6 keV and the data become very noisy above 6.0 keV, due to the fast timing mode). For each observation, we fit the MOS 1, MOS 2, and pn spectra simultaneously. Cross-calibrations between the detectors were also considered with $C_1$/$C_2$ as the cross calibration factors of MOS 2 and pn with respect to MOS 1. All the best-fit parameters with 90\% uncertainties are shown in Table \ref{tab:spec_para}.

\begin{figure}
\centering
	\begin{subfigure}{.5\textwidth}
    \centering
    	\includegraphics[width=.95\linewidth]{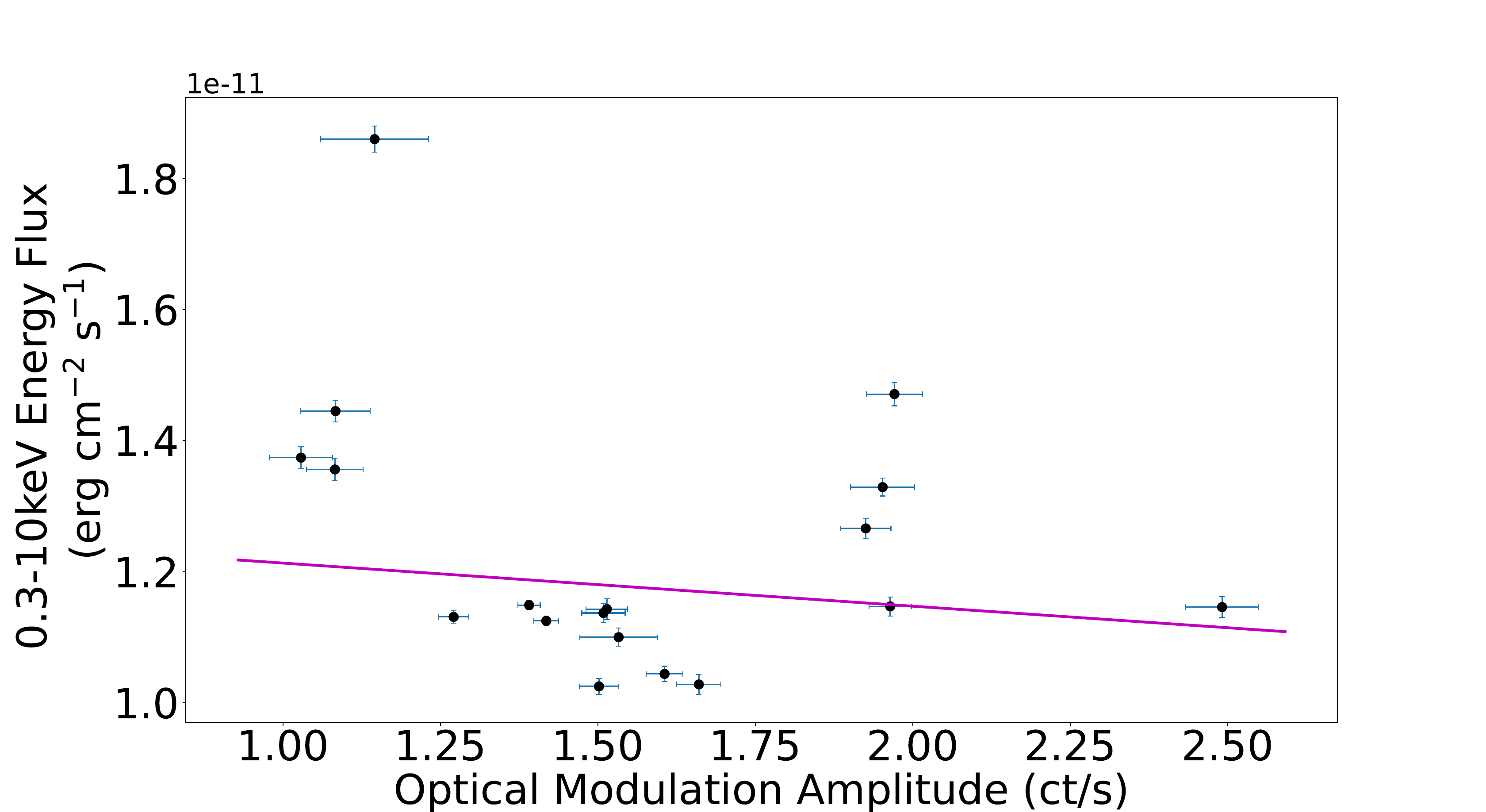}
\end{subfigure}
	\begin{subfigure}{.5\textwidth}
    \centering
   		 \includegraphics[width=.95\linewidth]{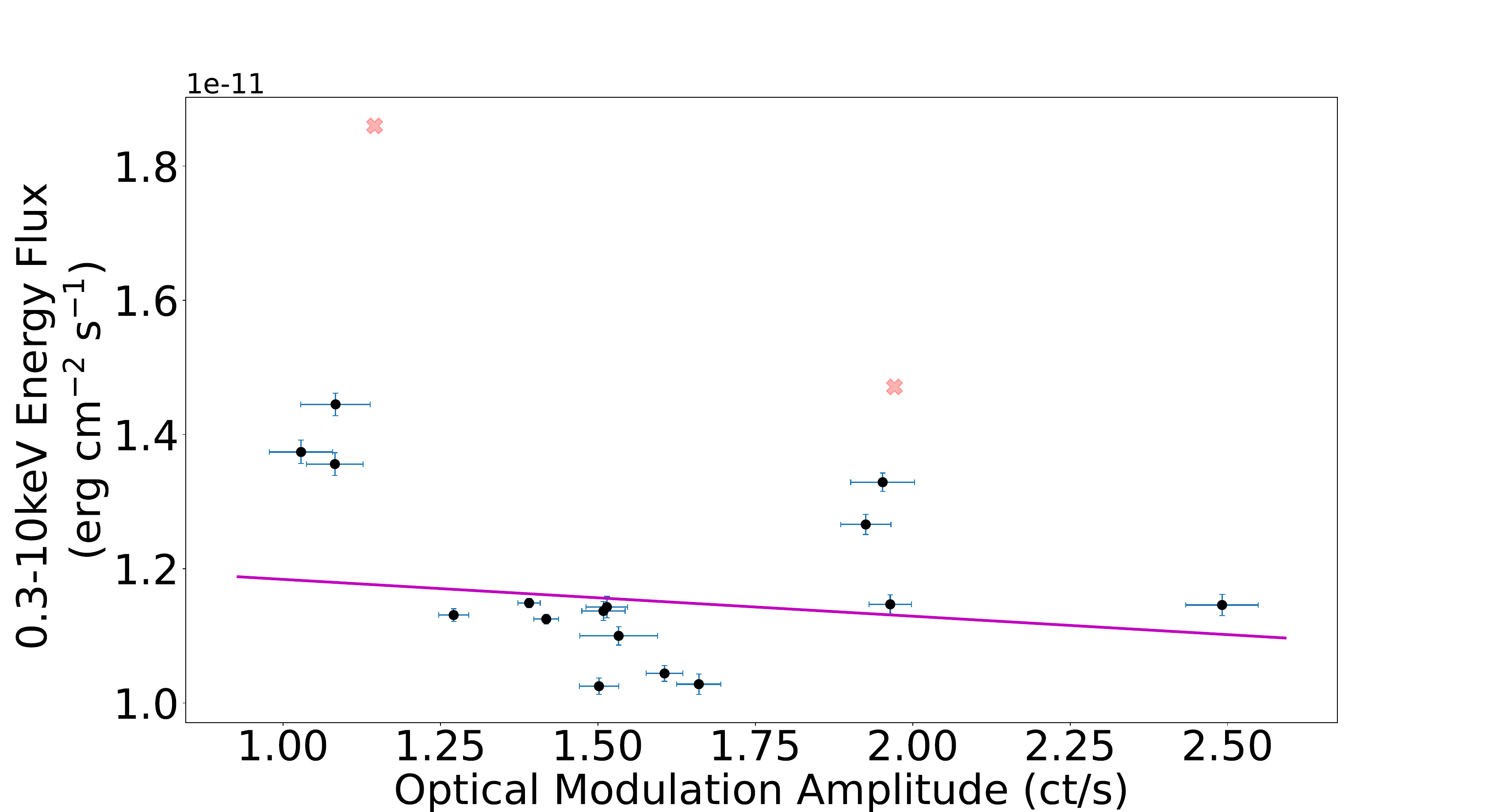}
\end{subfigure}  
\begin{subfigure}{.5\textwidth}
    \centering
    	\includegraphics[width=.95\linewidth]{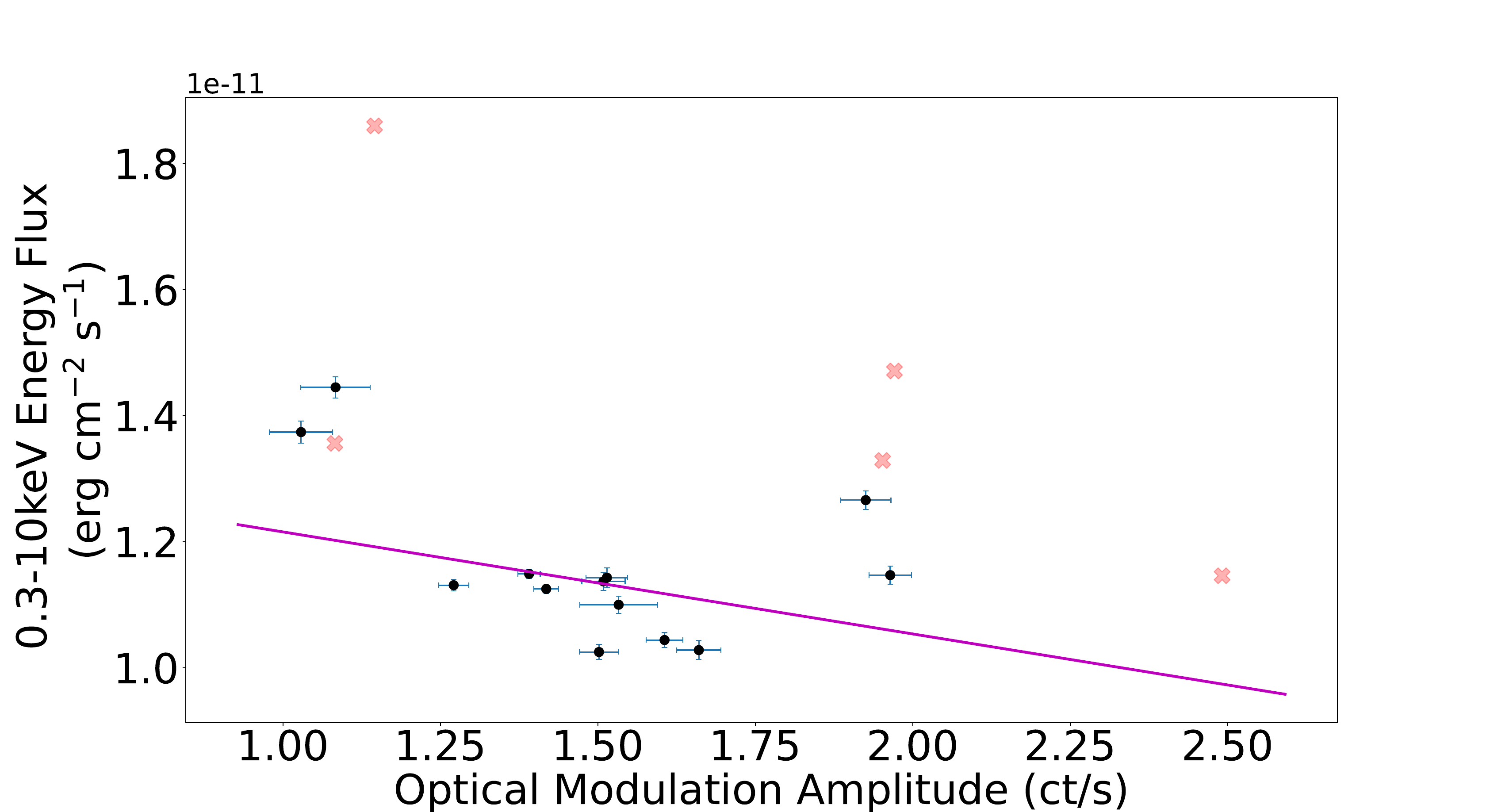}
\end{subfigure}     
\caption{The X-ray 0.3–10keV energy flux vs. optical MA plot, which is similar to Figure \ref{fig:mean}. The anti-correlation significances by using energy flux are between 0.8--1.7$\sigma$ (see Sections \ref{sec:xsp}).
\label{fig:cflux}}
\end{figure}

\begin{figure}
\centering
	\begin{subfigure}{.5\textwidth}
    \centering
    	\includegraphics[width=.95\linewidth]{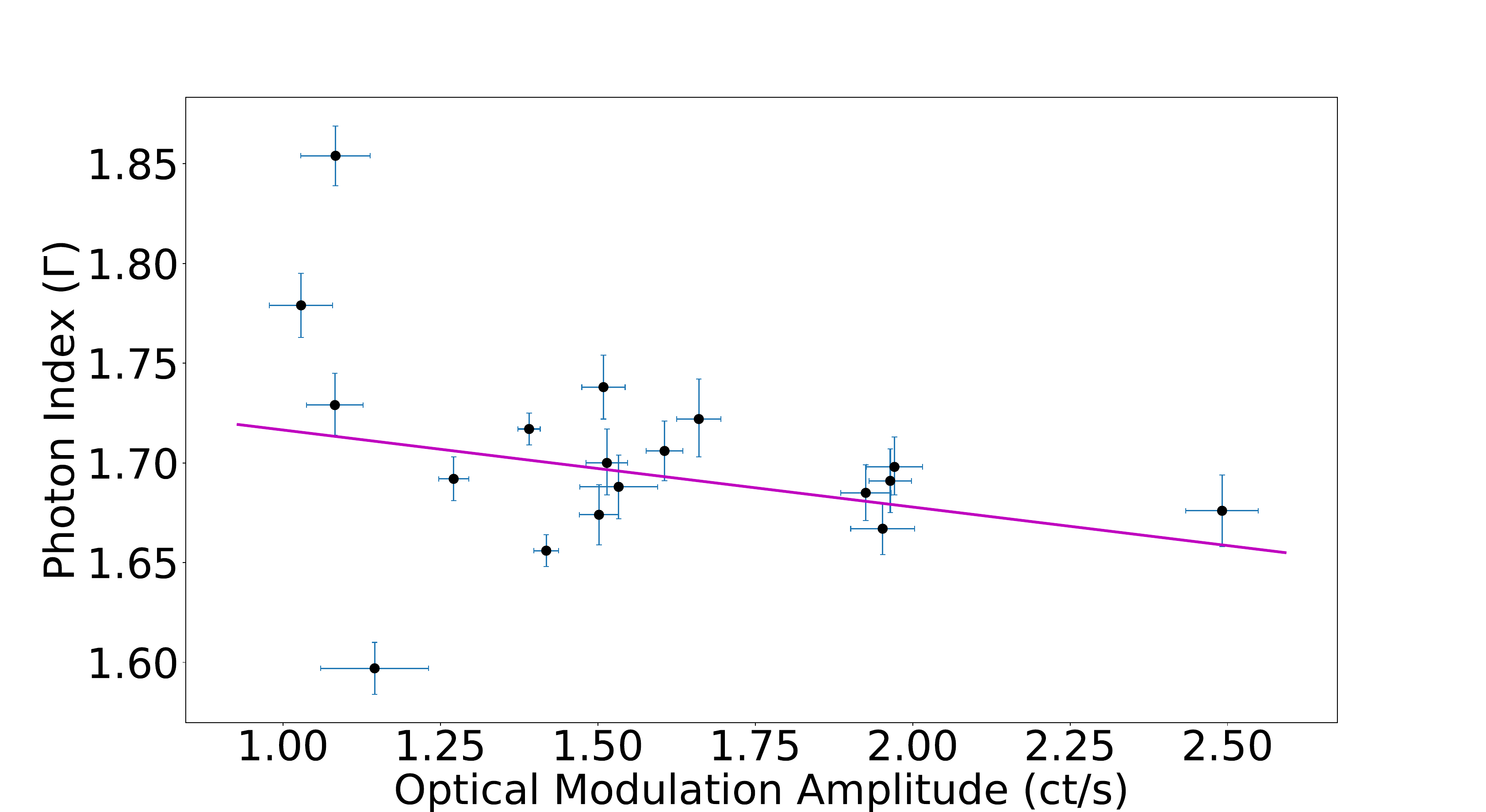}
\end{subfigure}
	\begin{subfigure}{.5\textwidth}
    \centering
   		 \includegraphics[width=.95\linewidth]{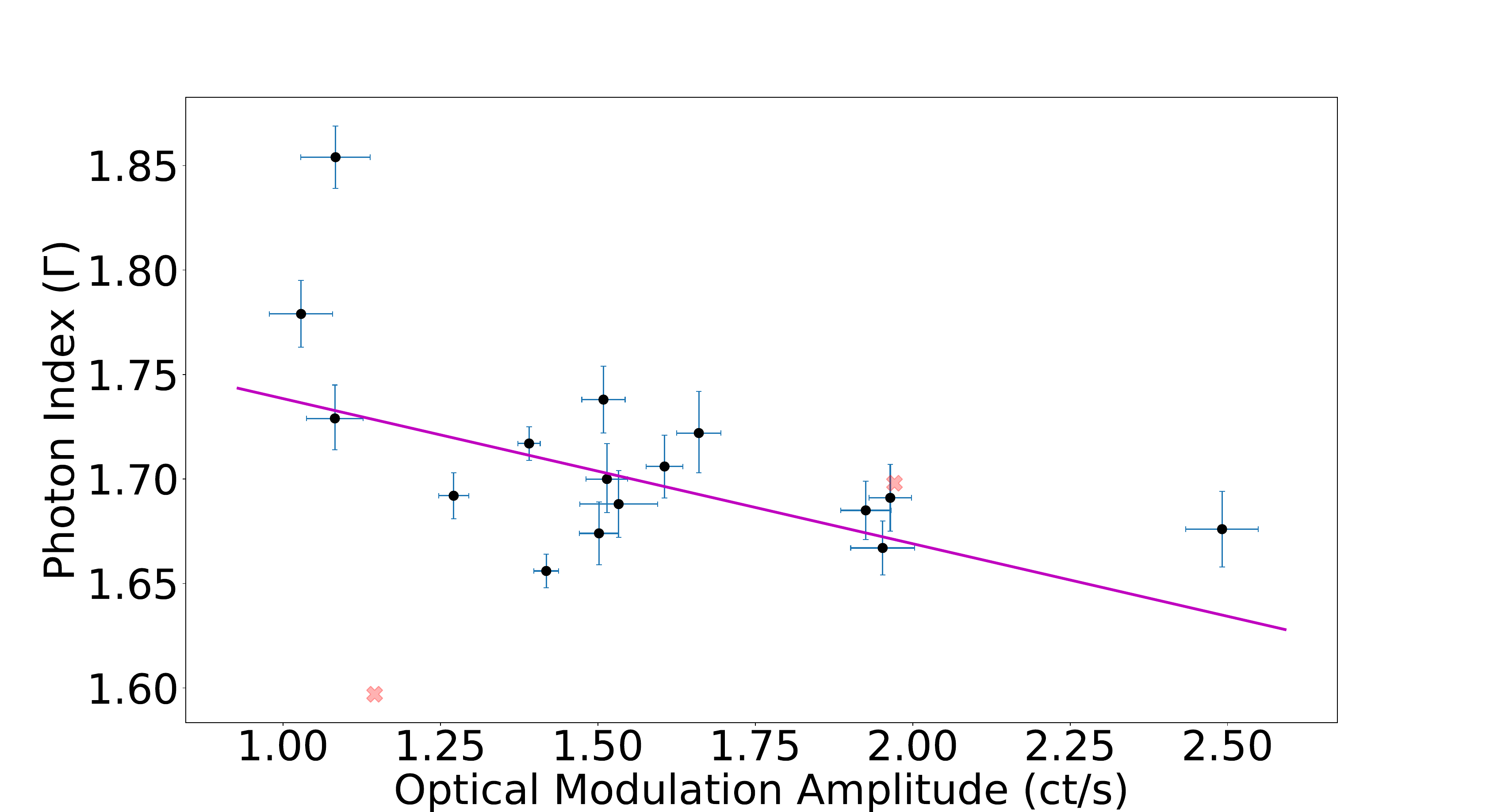}
\end{subfigure}  
\begin{subfigure}{.5\textwidth}
    \centering
    	\includegraphics[width=.95\linewidth]{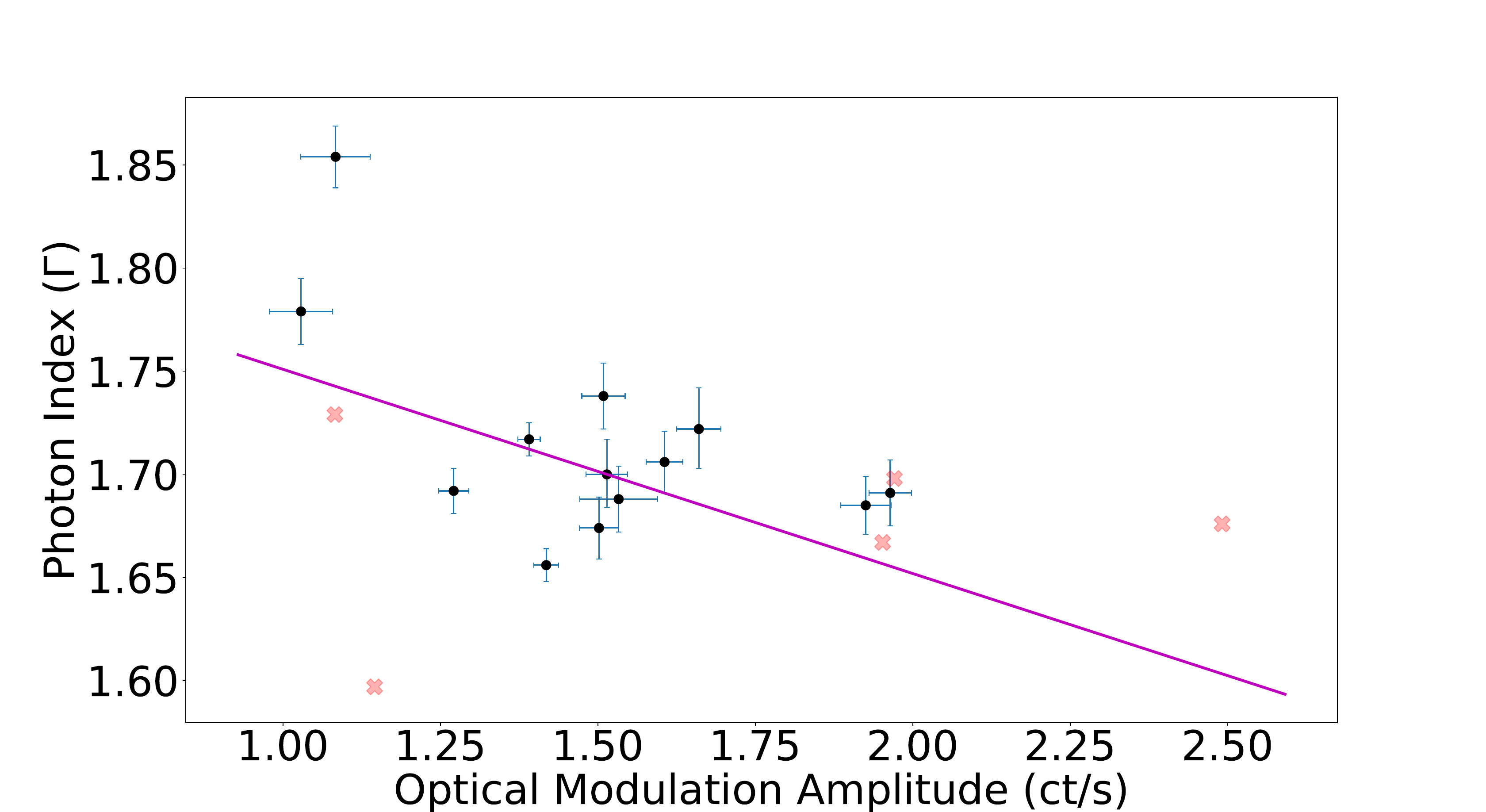}
\end{subfigure}     
\caption{The X-ray photon index $\Gamma$ vs. optical MA plot, which is similar to Figure \ref{fig:mean}. The anti-correlation significances by using $\Gamma$ are between 1.3--2.4$\sigma$ (see Sections \ref{sec:xsp}).
\label{fig:pi}}
\end{figure}

\begin{figure}
\centering
	\begin{subfigure}{.5\textwidth}
    \centering
    	\includegraphics[width=.95\linewidth]{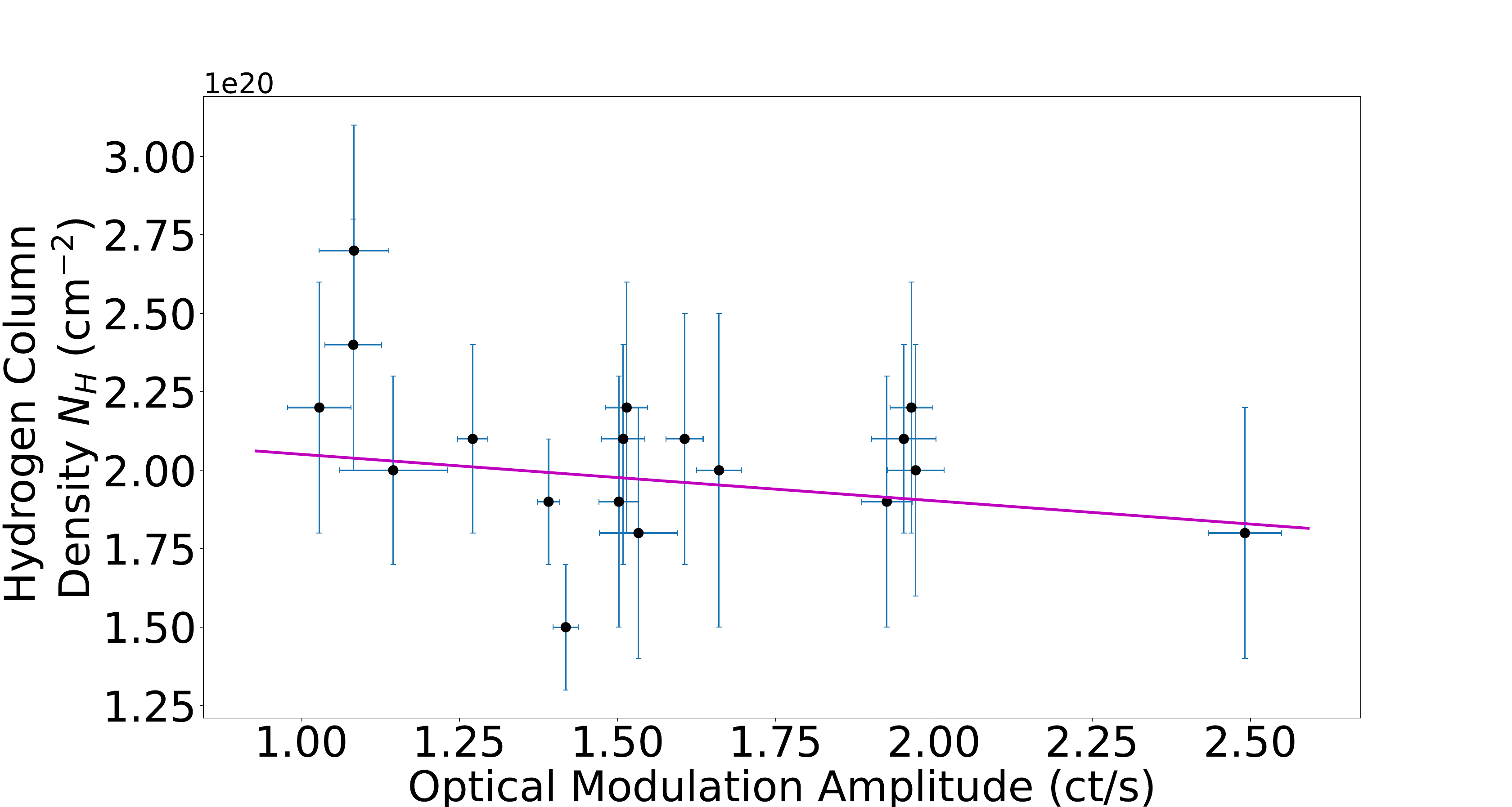}
\end{subfigure}
	\begin{subfigure}{.5\textwidth}
    \centering
   		 \includegraphics[width=.95\linewidth]{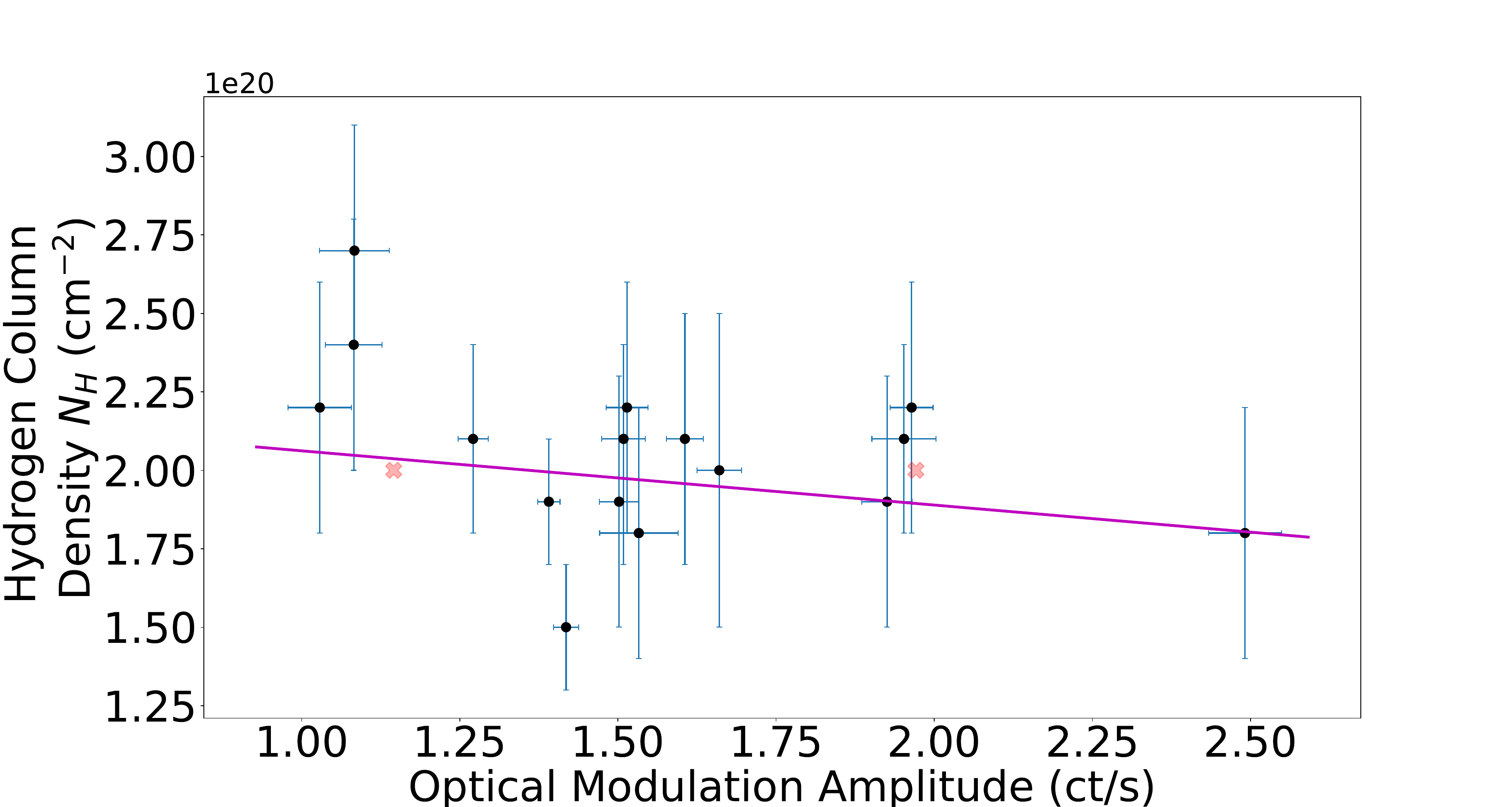}
\end{subfigure}  
\begin{subfigure}{.5\textwidth}
    \centering
    	\includegraphics[width=.95\linewidth]{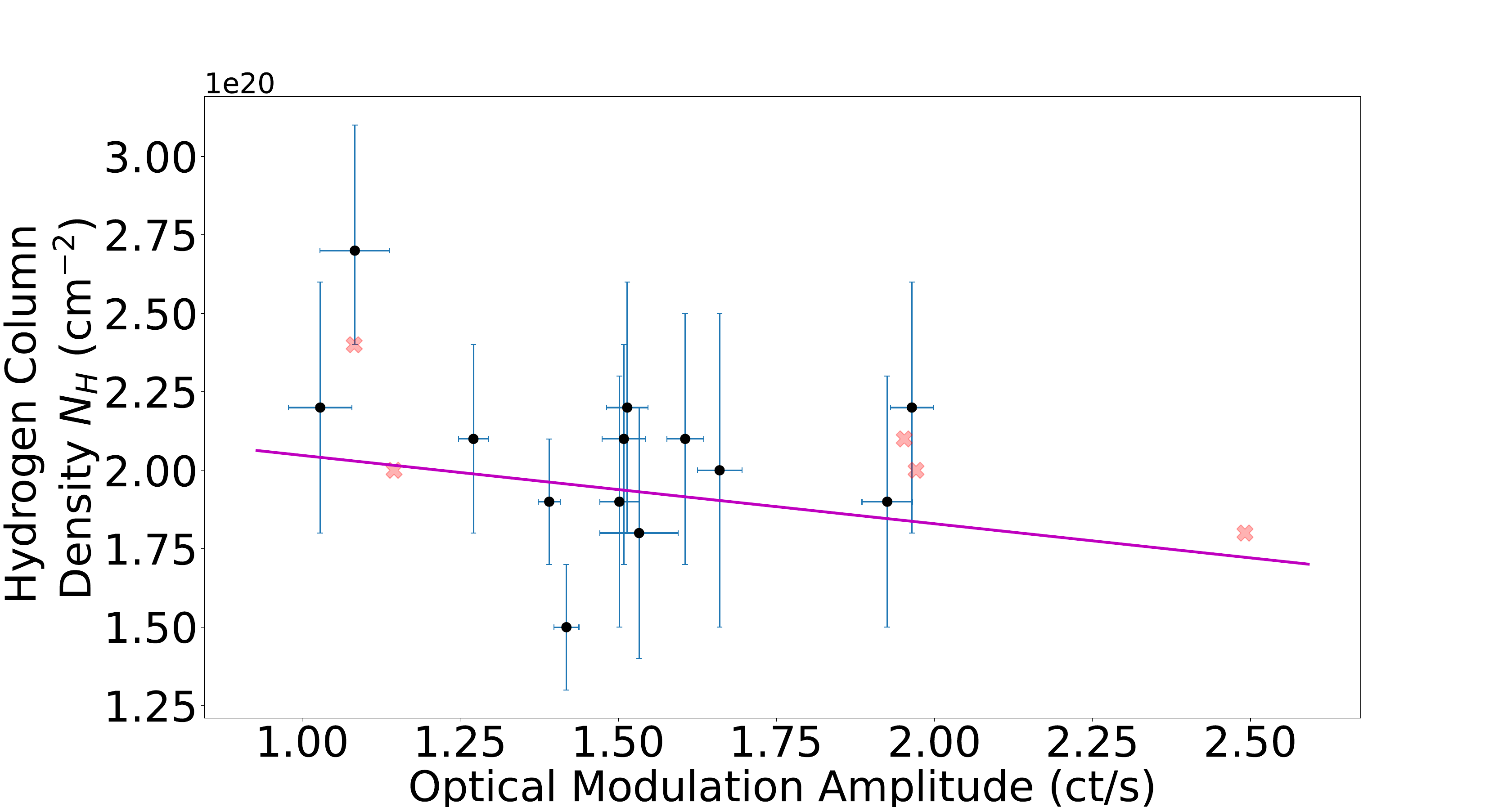}
\end{subfigure}     
\caption{The hydrogen column density $N_H$ vs. optical MA plot, which is similar to Figure \ref{fig:mean}. The anti-correlation significances by using $N_H$ are between 0.6--1.7$\sigma$ (see Sections \ref{sec:xsp}).
\label{fig:nh}}
\end{figure}

Figures \ref{fig:cflux}, \ref{fig:pi} and \ref{fig:nh} show the spectral parameters (energy flux, photon index $\Gamma$, and hydrogen column density $N_H$) against optical MA. The corresponding correlation significances with/without the filtrations mentioned in the previous section are all shown in Table \ref{tab:quan_signi}. All the spectral parameters do not show a significant anti-correlation with the optical MA (0.8--1.7$\sigma$ for energy flux and 0.6--1.7$\sigma$ for $N_H$), although an anti-correlation between the photo index and the MA is marginally seen (with 1.3--2.4$\sigma$ significance). The distribution of the MA--energy flux are similar to that of the mean value versions presented in Section \ref{sec:xmmwc}. However, there is an overall decline in the anti-correlation significances, perhaps indicating that the absorbed power-law model adopted is oversimplified for J1023 (assuming that the anti-correlation is true). Given the reasonably well spectral fitting results, we did not further test other more complex spectral models for improving the anti-correlation detection.

\begin{table*}

  \begin{center}
  
  \caption{The \textit{XMM-Newton} X-ray spectral parameters for J1023}

  	\hspace{-1cm}   
    \begin{tabular}{c c c c c c c c} 
      
      \hline
      &$C_1$ & $C_2$ & Energy flux & Photon index & Hydrogen column & $\chi^{2}_{rd}$ & d.o.f\\
      &&& ($F_{0.3-10keV}$) & ($\Gamma$) & density ($N_H$)&&\\
      &&& 10$^{-11}$\flux\ &  & 10$^{20}$\cm\ &&\\
      \hline
A	& 1.054$\pm$0.008			& 1.063$\pm$0.007 &	1.125	$\pm$	0.007	&	1.66	$\pm$	0.01	&	1.5	$\pm$	0.2				& 1.12 & 3248\\
B	& 1.029$\pm$0.008 			& 1.026$\pm$0.007 &	1.149	$\pm$	0.007	&	1.72	$\pm$	0.01	&	1.9	$\pm$	0.2				& 1.12 & 3244\\
C	& 1.010$\pm$0.013 			& 0.990$\pm$0.011 &	1.326	$\pm$	0.014	&	1.67	$\pm$	0.01	&	2.1	$\pm$	0.3				& 1.00 & 2382\\
D	& 1.033$\pm$0.015 			& 0.991$\pm$0.013 &	1.025	$\pm$	0.012	&	1.67	$\pm$	0.02	&	1.9	$\pm$	0.4				& 1.03 & 2158\\
E	& 1.030$\pm$0.018 			& 0.834$\pm$0.013 &	1.146	$\pm$	0.016	&	1.68	$\pm$	0.02	&	1.8	$\pm$ 	0.4				& 1.01 & 1787\\
F	& 1.021$^{+0.015}_{-0.014}$ & 1.044$\pm$0.013 & 1.044	$\pm$	0.012	&	1.71	$\pm$	0.02	&	2.1	$\pm$	0.4				& 1.05 & 2336\\
G	& 1.040$\pm$0.016 			& 1.038$\pm$0.014 &	1.100	$\pm$	0.014	&	1.69	$\pm$	0.02	&	1.8	$\pm$	0.4				& 1.09 & 2181\\
H	& 0.994$\pm$0.014			& 1.000$\pm$0.012 &	1.445	$\pm$	0.017	&	1.85	$\pm$	0.02	&	2.7	$^{+0.4}_{-0.3}$		& 1.09 & 2097\\
I	& 1.027$\pm$0.017 			& 0.949$\pm$0.014 &	1.143	$\pm$	0.016	&	1.70	$\pm$	0.02	&	2.2	$\pm$	0.4				& 1.02 & 1972\\
J	& 1.003$\pm$0.016 			& 1.011$^{+0.014}_{-0.013}$ &	1.374$^{+0.018}_{-0.017}$&	1.78	$\pm$	0.02	&	2.2	$\pm$	0.4	& 1.01 & 2022\\
K	& 1.025$\pm$0.016 			& 1.016$^{+0.014}_{-0.013}$ &	1.356	$\pm$	0.017	&	1.73	$\pm$	0.02	&	2.4	$\pm$	0.4	& 0.97 & 2068\\
L	& 1.037$\pm$0.014 			& 1.048$\pm$0.012 &	1.860	$\pm$	0.020	&	1.60	$\pm$	0.01	&	2.0	$\pm$	0.3				& 0.98 & 2455\\
M	& 1.038$\pm$0.016			& 0.975$\pm$0.013 &	1.137	$\pm$	0.014	&	1.74	$\pm$	0.02	&	2.1	$^{+0.3}_{-0.4}$		& 1.01 & 2077\\
N	& 1.010$^{+0.019}_{-0.018}$ & 1.022$^{+0.017}_{-0.016}$ &	1.028	$\pm$	0.015	&	1.72	$\pm$	0.02	&	2.0	$\pm$	0.5	& 1.01 & 1830\\
O	& 1.040$^{+0.016}_{-0.015}$ & 1.053$^{+0.014}_{-0.013}$ &	1.471	$\pm$	0.018	&	1.70	$^{+0.02}_{-0.01}$&	2.0	$\pm$	0.4	& 0.95 & 2145\\
P	& 1.019$\pm$0.016 			& 1.019$\pm$0.014 &	1.147	$^{+0.015}_{-0.014}$		&	1.69	$\pm$	0.02	&	2.2	$\pm$	0.4	& 0.99 & 2111\\
Q	& 1.022$^{+0.015}_{-0.014}$ & 1.023$\pm$0.013 &	1.266	$\pm$	0.015	&	1.69	$\pm$	0.01	&	1.9	$\pm$	0.4				& 1.00 & 2311\\
R	& 0.987$\pm$0.011 			& 1.094$\pm$0.010 &	1.131	$\pm$	0.009	&	1.69	$\pm$	0.01	&	2.1	$\pm$	0.3				& 1.07 & 2813\\

      \hline
    \end{tabular}
 \label{tab:spec_para} 
  \end{center}
\end{table*}

\subsubsection{\textit{Fermi}-LAT}\label{sec:fermi}
Following the results from the \textit{XMM-Newton} data, we tried to perform a similar analysis between the $\gamma$-ray and the MA using the \textit{Fermi}-LAT data. After downloading the LAT event files and spacecraft data from the \textit{Fermi} Science Support Center (FSSC)\footnote{\url{https://fermi.gsfc.nasa.gov/ssc}}, we extracted the $\gamma$-ray fluxes with an energy range 0.1--300 GeV corresponding to the 18 \textit{XMM-Newton} observing windows using Fermitools (version v11r5p3; \citealt{fermitools}) developed by the \textit{Fermi}-LAT science team.

\begin{figure}
\centering
	\begin{subfigure}{.5\textwidth}
    \centering
    	\includegraphics[width=.95\linewidth]{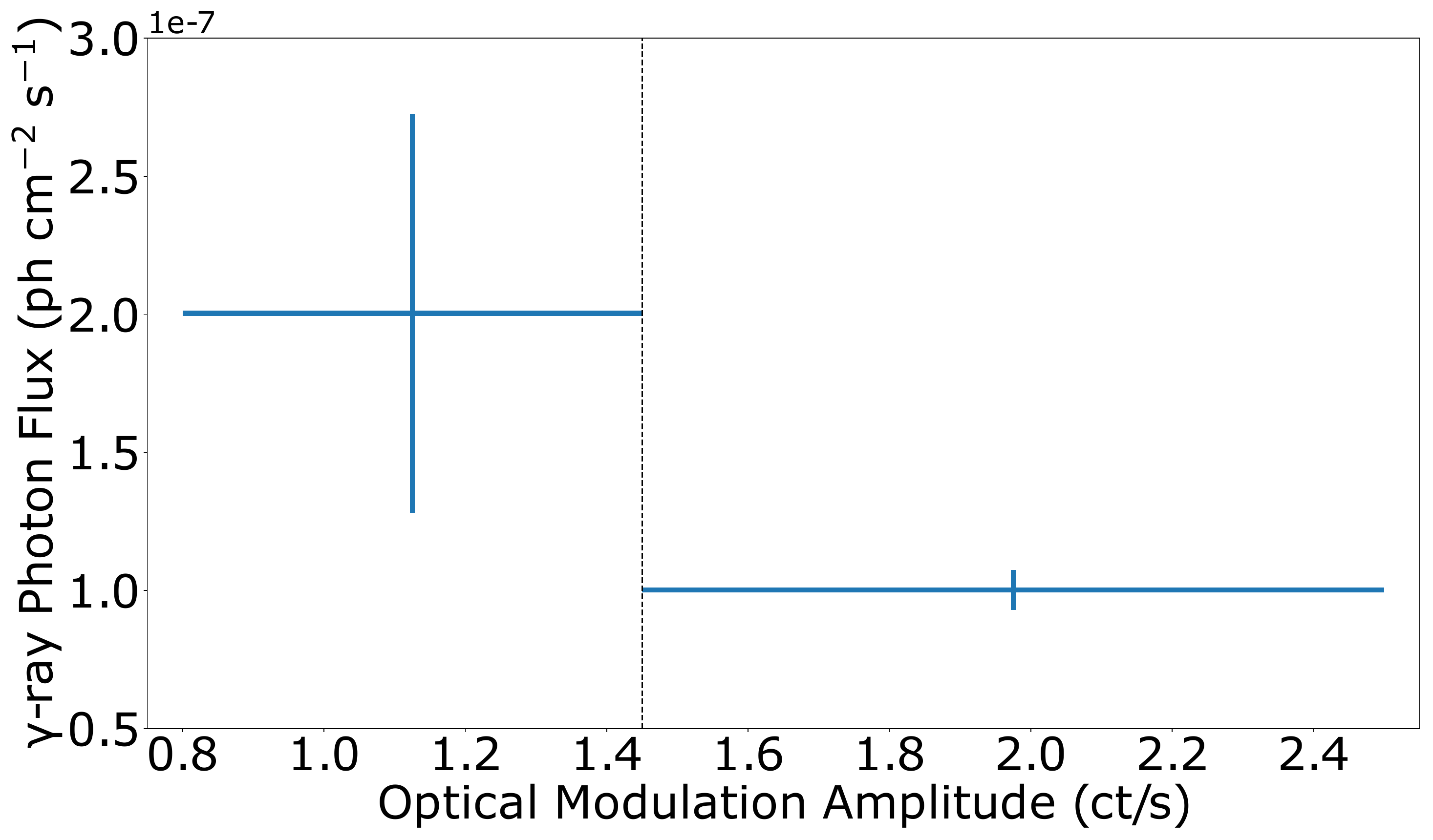}
\end{subfigure}
	\begin{subfigure}{.5\textwidth}
    \centering
   		 \includegraphics[width=.95\linewidth]{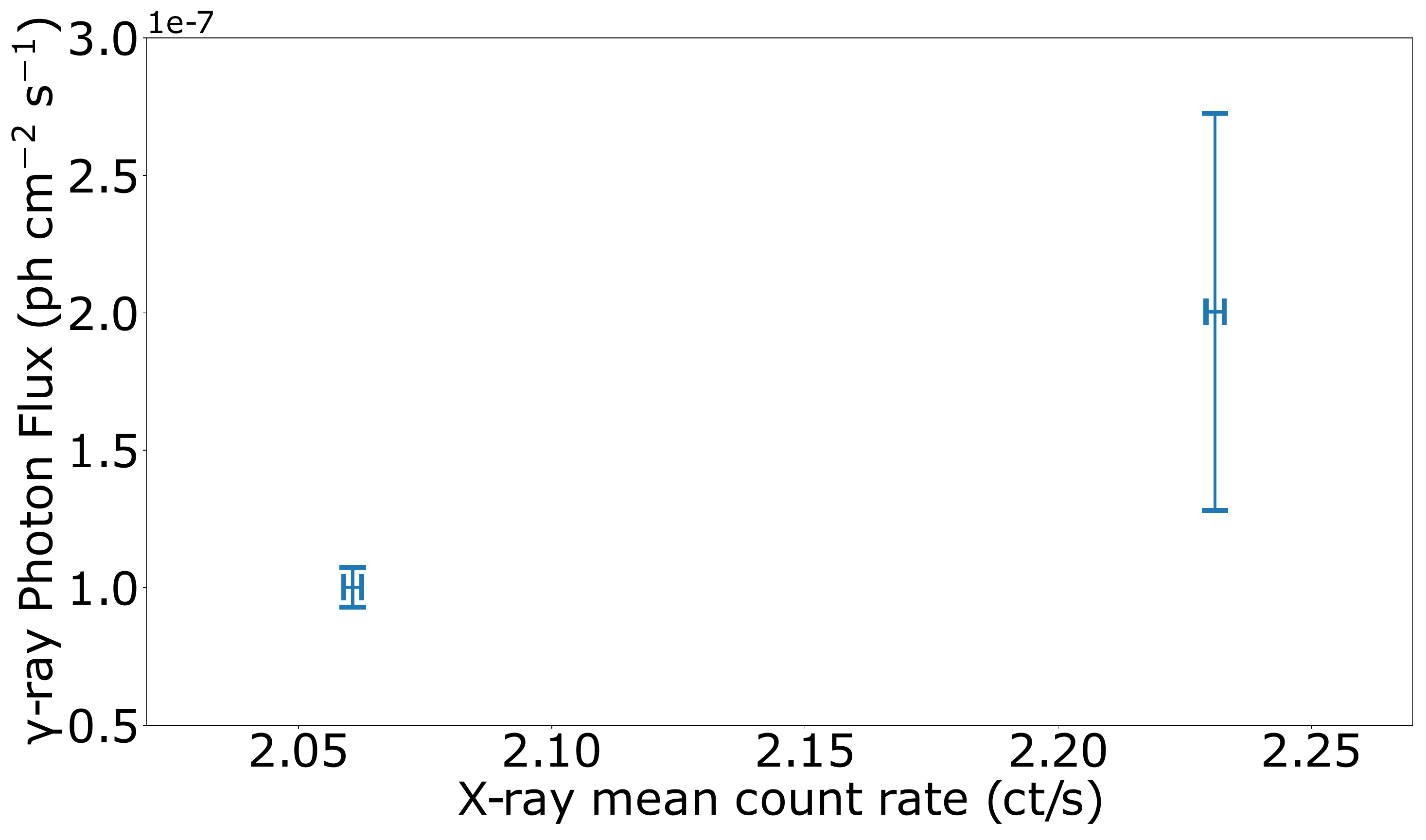}
\end{subfigure}       
\caption{Top: The $\gamma$-ray vs. optical MA plot for the two GTI data groups. The dashed line refers to the separation line of MA = 1.45 cts/s that splits the XMM-Newton observations. Bottom: The $\gamma$-ray vs. X-ray mean count rate for the two GTI data groups.
\label{g_x}}
\end{figure}

With the limitation of the \textit{XMM-Newton} observing windows, which are typically around $\sim10\,$ks per observation, J1023 cannot be detected by \textit{Fermi}-LAT in a single \textit{XMM-Newton} observing window. Therefore, we stacked the windows into two big good time interval (GTI) groups according to the corresponding MA of the GTI. After a few test runs, we decided to split the observations at MA = 1.45 cts/s (Figure \ref{g_x}), mainly considering the balance of the LAT data quantities of the two groups. The first group includes 7 observations (A, B, H, J, K, L, and R; $<$ 1.45 cts/s) with a total exposure time of 413 ks, and the second group includes the rest with a total 304 ks exposure time. We analysed both groups separately using the standard analysis process for the LAT data. In the LAT data analysis, the SOURCE class events (FRONT and BACK) with a zenith angle smaller than 90$^{\circ}$ is chosen. The center of the $14^{\circ} \times 14^{\circ}$ region of interest (ROI) is at ($\alpha$, $\delta$) = ($155^{\circ}.949, 0^{\circ}.645$), which is the position of J1023. A $\gamma$-ray emission model was constructed to describe the $\gamma$-ray photons detected in the observations. The emission model includes the latest Galactic interstellar (gll$\_$iem$\_$v07.fits) and isotropic (iso$\_$P8R3$\_$SOURCE$\_$V3$\_$v1.txt) diffuse components.  A LogParabola model is used for J1023. Besides, the source model contains all the 4FGL sources within 10 degrees from J1023.  We allow the background diffuse components and the sources inside a 5$^{\circ}$ radius circle from J1023 to vary to get 28 free parameters in the emission model. 

We performed a binned likelihood analysis with 37 logarithmically uniform energy bins for both groups. The 0.1--300 GeV $\gamma$-ray flux for the first bin is $F_{0.1-300GeV} = (2.00\pm0.72)\times10^{-7}$ \flux\ with a test statistic (TS) value of 12.9 ($\sim3.6\sigma$). The result for the second bin is $F_{0.1-300GeV} = (1.00\pm0.07)\times10^{-7}$ \flux\ with TS value of 7.9 ($\sim2.8\sigma$). Although the detections are not significant, J1023 is already a well-known LAT source and the detection is likely genuine. Figure \ref{g_x} shows the $\gamma$-ray flux against the MA for the two data groups. The distinction of $\gamma$-ray flux significance is low ($\sim1.4\sigma$) owing to the limited number as well as the high uncertainties of the data bins. We also computed the X-ray mean count rate for this two data group by $\sum_{i} \frac{CR_i\times exp_i}{exp_i}$, where $CR_i$ is the X-ray mean count rate and $exp_i$ is exposure time for each observation, and compared them with the corresponding $\gamma$-ray fluxes. We do not see any significant correlation between them.


\subsection{A Low-Mode Anti-Correlation Between the EPIC and OM Data}\label{sec:lowmodeanti}
\cite{obs10101} reported that J1023 is brighter by about 0.1 mag on average in the B band during the X-ray low mode in November 2013 (corresponding to Obs. A), but the phenomenon disappeared in June 2014 (Obs. B). In this work, we expand the analysis to 18 observations (from November 2013 to June 2021) to investigate the long-term behaviour of the low-mode anti-correlation.

\subsubsection{Quantization of the Anti-Correlation}

In every low-mode episode of each observation (i.e., time periods with consecutive low-mode bins), we took the median value of the EPIC count rates as the representative X-ray flux for the epoch (i.e. $F_{\mathrm{low},x,i}$ for the $i$-th episode observed). Similarly, the median value of the 30--100-second data before/after the episode was taken as the non-low-mode representative (i.e. $F_{\mathrm{non},x,i}$) for the same episode. The 30-second gaps are introduced for avoiding the ambiguous data points between the emission modes, and only non-low-mode data were included in the median computation. If there is no suitable data for the non-low-mode median calculation, no further investigation will be done for the particular low-mode episode. In addition, a flux ratio, $R_{x,i}=F_{\mathrm{low},x,i}/F_{\mathrm{non},x,i}$, will be computed.
The same process was done on the detrended OM B-band light curves, in which the long-term trends (e.g., orbital modulation) and optical flares were removed, resulting in three new variables, $F_{\mathrm{non},b,i}$, $F_{\mathrm{low},b,i}$, and $R_{b,i}$.
Table \ref{tab:low_mode_summary} shows the mean values (without the subscript $i$) of the above six parameters and the number of useful low-mode episodes ($n_\mathrm{low}$), which is also counted in Section \ref{sec:xlhm}, in each \textit{XMM-Newton} observation. 

\begin{table*}
\centering
\caption{Summary of the low-mode properties for the 18 \textit{XMM-Newton} observations}
\begin{tabular}{lcccccccccc}

\hline
 & $n_\mathrm{low}$ & $F_{\mathrm{low},x}$ & $F_{\mathrm{non},x}$ & $R_x$ & $F_{\mathrm{low},b}$ & $F_{\mathrm{non},b}$ & $R_b$ & $P$ & FAP & Visibility$^{a}$ \\
 & & (cts/s) & (cts/s) & & (cts/s) & (cts/s) & & (\%) & (\%) & \\
\hline
A & 139 & 0.70 & 2.53 & 0.28 & 8.43 & 7.13 & 1.19 & 91 & $\sim10^{-23}$ & o \\
B & 133 & 0.56 & 2.33 & 0.25 & 6.78 & 6.68 & 1.02 & 53 & 24 & x \\
C & 27 & 0.42  & 2.34 & 0.18 & 8.57 & 7.30 & 1.18 & 96 & $\sim10^{-5}$ & o \\
D & 56 & 0.53  & 2.23 & 0.24 & 8.80 & 7.22 & 1.23 & 96 & $\sim10^{-12}$ & o \\
E & 19 & 0.46  & 2.33 & 0.20 & 13.30 & 11.50 & 1.16 & 95 & 0.004 & o \\
F & 40 & 0.51  & 2.40 & 0.21 & 7.88 & 6.94 & 1.14 & 90 & $\sim10^{-5}$ & o \\
G & 35 & 0.51  & 2.22 & 0.24 & 7.00 & 6.31 & 1.12 & 86 & 0.001 & x \\
H & 4 & 0.46   & 3.24 & 0.17 & 6.66 & 6.34 & 1.07 & 50 & 69 & x \\
I & 59 & 0.55  & 2.34 & 0.24 & 9.26 & 8.49 & 1.09 & 78 & 0.001 & x \\
J & 16 & 0.52  & 2.44 & 0.22 & 6.20 & 5.85 & 1.06 & 56 & 40 & x \\
K & 20 & 0.46  & 2.39 & 0.19 & 6.88 & 6.70 & 1.03 & 65 & 13 & x \\
L & 6 & 0.48   & 2.32 & 0.21 & 7.29 & 6.64 & 1.10 & 100 & 2 & x \\
M & 28 & 0.49  & 2.20 & 0.23 & 7.60 & 7.06 & 1.08 & 82 & 0.05 & x \\
N & 20 & 0.49  & 2.28 & 0.22 & 7.09 & 6.67 & 1.07 & 70 & 6 & x \\
O & 27 & 0.51  & 2.10 & 0.25 & 7.96 & 6.80 & 1.17 & 100 & $\sim10^{-6}$ & o \\
P & 36 & 0.56  & 2.25 & 0.25 & 9.08 & 7.82 & 1.17 & 92 & $\sim10^{-5}$ & o \\
Q & 39 & 0.57  & 2.29 & 0.25 & 11.26 & 10.49 & 1.07 & 85 & 0.001 & x \\
R & 50 & 0.50  & 2.22 & 0.23 & 7.93 & 6.89 & 1.15 & 98 & $\sim10^{-11}$ & o \\
\hline
\multicolumn{11}{c}{$^{a}$Whether the low-mode anti-correlation can be found by visual inspection (o: visible; or x: invisible)}
\end{tabular}

\label{tab:low_mode_summary}
\end{table*}

In Table \ref{tab:low_mode_summary}, it is clear that $R_b>1$ for all the 18 observations, indicating that, when the pulsar system is in the X-ray low mode, the B-band emission is generally higher. To find out whether this low-mode anti-correlation holds for all the individual low-mode episodes, we checked the fraction of the episodes in an observation that fulfils $R_{b,i}>1$, namely $P$. While not all the episodes show the anti-correlation, $P$ is always larger than or equal to 50\% in all the observations and some are as high as $>90\%$ (Table \ref{tab:low_mode_summary}). Assuming a binomial distribution\footnote{The B-band emission and the X-ray low-mode phenomenon are assumed to be independent, and therefore the chance of $R_{b,i}>1$ is 50\%.}, we calculated the false alarm probability (FAP) that the fraction of $R_{b,i}>1$ is higher than or equal to $P$ in a sample size of $n_\mathrm{low}$. Using a threshold of 0.001\% (over 4$\sigma$ significance), we conclude that the low-mode anti-correlation is significant in seven of the observations (Table \ref{tab:low_mode_summary}). 

The anti-correlation can also be identified by visual inspection (see the two examples in Figure \ref{fig:low_mode_example}). Besides the seven observations that are statistically valid, we find one additional samples (i.e., Obs. IDs 0748390601) that possibly shows the anti-correlation. In 0748390601, the small number of low-mode episodes (i.e., 19) makes the FAP too high to be significant, so that the anti-correlation is hard to be seen statistically. Given the above example, we cannot rule out the possibility that the anti-correlation always exits; perhaps the anti-correlation is just insufficiently strong to be determined statistically or even visually in some observations. Nevertheless, there is no doubt that the anti-correlation varies over time. 

\section{Discussion}
\subsection{The MA--X-Ray Anti-Correlation}
In our \textit{XMM-Newton} analysis, we find a possible anti-correlation between the optical MA and X-ray flux. Several methods were applied to estimate the significance. Although the anti-correlation detection is not strong, the result still unambiguously concludes that there is no positive correlation between the X-ray emission and the MA. There are at least three optical emission components of J1023, which are the millisecond optical pulsations from the pulsar \citep{2017NatAs...1..854A,2019ApJ...882..104P,2023A&A...669A..26I}, the accretion disk \citep{takata2014,2023A&A...677A..30B}, and the irradiated companion \citep{2014MNRAS.444.1783C,2018MNRAS.477.1120K}. The optical pulsed emission is thought to be synchrotron emission from a shock-driven mini pulsar nebula located $\sim$100 km from the pulsar \citep{2023A&A...669A..26I}. This component was faint with a pulse fraction of around 1\% or less \citep{2017NatAs...1..854A}, which is too weak compared to the measured MA. In contrast, the accretion disk and the irradiated companion are both strong enough to result in the observed MA variation. However, the accretion disk, although irradiated by the central pulsar, should not be orbitally modulated as we observed in the \textit{XMM-Newton} data because the same fraction of the heated inner edge of the disk is seen at different orbital phases. Occultation would be possible to cause eclipses if the inclination is high, but J1023 is not viewed edge-on with an inclination of 40--48 degrees \citep{Archibald,jay}. Therefore, it is almost certain that the MA is caused by the irradiated companion as the star rotates. Nevertheless, the variations of the other emission components can still affect the MA measurement, and we suspect that this is one of the reasons why some observations are scattered from the main trend in Figures \ref{fig:mode_occur}, \ref{fig:mean}, and \ref{fig:median}. The effect is hard to be eliminated based on the current dataset. Hopefully, the outliers can be distinguished and eliminated statistically when more observations are available in the future. 

Provided that the MA of J1023 is a consequence of pulsar heating, which have also been shown in many redback MSP systems \citep{j1048,1227lc,j1910}, we can rule out that the X-ray emission as the primary source for the heating effect. We also checked the relation between $\gamma$-rays and MA, but the simultaneous \textit{Fermi}-LAT and \textit{XMM-Newton} MA data do not show a significant finding (Figure \ref{g_x}).

Apparently, given the X-ray--MA anti-correlation found in this work, the irradiation and the X-ray emission should also be anti-correlated. In addition, if the X-ray emission is not the primary irradiation source, the pulsar likely heats up the companion through $\gamma$-rays and/or pulsar winds \citep{pwheating2,gheating2,gheating,pwheating,xor,2018MNRAS.477.1120K,1227lc}. 

There have been theoretical studies predicting a proportional relationship between X-rays and $\gamma$-rays in J1023. For example: the increases of the X-ray and $\gamma$-ray emission produced by the IBS and inverse Compton scattering by the pulsar wind, respectively, as the accretion disk is growing \citep{takata2014,li2014}; the disk in-flow-magnetosphere shock accelerating electrons to generate X-rays and $\gamma$-rays by the synchrotron and self-synchrotron Compton in a propeller mode \citep{propeller}. The positive correlation is also marginally seen in our analysis. Therefore, if the X-ray emission is not the main heating source, the $\gamma$-ray emission is likely not either.

In contrast, pulsar wind could be the varying heating source in some circumstances, which agrees with the notion in \cite{2018MNRAS.477.1120K}. This also supports some shock models to explain the mode-switching phenomenon in J1023 \citep{2019ApJ...882..104P,2023A&A...677A..30B}. When an IBS is formed between the pulsar winds and the accretion disk, it partially blocks the pulsar winds, leading to a decrease in the irradiation on the companion. Moreover, the shock provides an environment for the enhanced X-ray emission and probably also the $\gamma$-ray emission that explains the MA--X-ray anti-correlation found in this work. If the disk in-flow is weak (i.e, weak or no shock), less pulsar winds would be stopped, and hence, a greater fraction of the winds would contribute to the companion heating.

Figure \ref{fig:mode_occur} has shown a 2.3--3.0$\sigma$ anti-correlation between the non-low-mode occurrence rate and the MA. Therefore, the irradiation is likely stronger in the X-ray low mode. Combining this with the pulsar wind scenario aforementioned, the low/high-mode phenomenon should be related to the shock formation. However, the X-ray mode-switching is on a time-scale of $\sim10$s \citep{xranti}, and whether the disk can be extended/compressed that quickly is still questionable. Alternatively, the changing MA could be caused by the strong radio-emitting outflow/jet observed in the X-ray low mode \citep{xranti}, but the physical origin of the outflow/jet is still largely unknown.

\subsection{The Varying Low-Mode Anti-Correlation}

The X-ray and radio variabilities were observed to be anti-correlated using simultaneous \textit{Chandra} and VLA observations \citep{xranti}. The authors also found an archival \textit{XMM-Newton}/VLA data pair taken in November 2013 (Obs. ID 0720030101) that overlapped with each other for about 50 minutes. In this dataset, the X-ray and radio luminosities were also anti-correlated. Interestingly, the B-band/X-ray anti-correlation is observed in the same \textit{XMM-Newton} observation as well (Table \ref{tab:low_mode_summary}). \cite{xranti} proposed that the enhanced radio emission originates from a radio-emitting outflow, which could be driven by accretion, propeller effect, or pulsar wind. The enhanced B-band light is possibly an emission component of the outflow. If it is true, the launch of the outflow during the X-ray low mode is probably unstable, given the instability of the B-band/X-ray anti-correlation. More simultaneous \textit{XMM-Newton} and VLA observations can test the scenario by checking whether the radio/X-ray and B-band/X-ray anti-correlations vanish at the same times.

Alternatively, an existence of a precessing warped disk can also explain the occasional B-band/X-ray low-mode anti-correlation. Through the MA--X-ray analysis, it is known that the irradiation power of the system is likely stronger during the X-ray low mode. The variable irradiation source could also shine on the accretion disk to create the observed low-mode anti-correlation. Similar re-radiated disk models have been proposed to explain the observed thermal X-ray emission in some pulsar systems, e.g., SMC X-1 and LMC X-4 \citep{2005ApJ...633.1064H,2020ApJ...888..125B,2022MNRAS.512.3422A}. The irradiation power of J1023 in the sub-luminous state is presumably much lower than that of these X-ray pulsars (i.e., $\sim10^{38}$\lum\ for SMC X-1 and LMC X-4; \citealt{2020ApJ...888..125B}), and therefore an illuminated region with lower temperatures that can be observed in the B-band, but not soft X-rays, would be formed. As the disk precesses, we get and lose the best observing angles for the disk ``reflection'' from time to time. This normally explains the disappearance of the low-mode anti-correlation in some of the \textit{XMM-Newton} observations.

\section{Conclusion}

Based on the analyses in the $\gamma$-ray, X-ray, optical bands of the \psr, we summarise the results as below:

\begin{itemize}
\item A possible anti-correlation is shown between the non-low-mode occurrence rate and the optical MA with 2.2--2.9$\sigma$.
\item An anti-correlation is also shown between the X-ray emission and the MA with significances of 1.7--3.1$\sigma$.
\begin{itemize}	
\item Some observations are filtered out to minimize the optical/X-ray flaring effect by considering the following conditions. 
\begin{itemize}
\item (1) The difference between the mean and median count rates (with 2.2--2.7$\sigma$ significance).

\item (2) The difference in the optical light curve slope with/without flare filtering (with 1.7--2.4$\sigma$ significance).
\end{itemize}
\end{itemize}	
\item Only the photon index among the X-ray spectral parameters shows a potential anti-correlation.
\item We do not find any significant correlation in the $\gamma$-ray and the optical MA because of the lack of data.
\item A low-mode anti-correlation is found in some \textit{XMM-Newton} observations between EPIC and OM data.
\end{itemize}		

The optical MA-X-ray flux anti-correlation shows that the X-ray emission cannot be responsible for the variation of the optical modulation amplitude. Moreover, the X-ray/optical low-mode anti-correlation was found in some of the \textit{XMM-Newton} observation. These anti-correlations likely suggest that the irradiation is generally stronger when the X-ray flux is in a fainter state, indicating that there is a more dominant irradiation source than the X-ray emission. More and deeper observations are needed to support our results and investigate the heating and mode-switching phenomenon in J1023.

\newpage

\section*{Acknowledgements}
\addcontentsline{toc}{section}{Acknowledgements}

The \textit{Fermi}-LAT Collaboration acknowledges generous ongoing support from a number of agencies and institutes that have supported both the development and the operation of the LAT as well as scientific data analysis. These include the National Aeronautics and Space Administration and the Department of Energy in the United States, the Commissariat \`{a} l'Energie Atomique and the Centre National de la Recherche Scientifique / Institut National de Physique Nucl\'{e}aire et de Physique des Particules in France, the Agenzia Spaziale Italiana and the Istituto Nazionale di Fisica Nucleare in Italy, the Ministry of Education, Culture, Sports, Science and Technology (MEXT), High Energy Accelerator Research Organization (KEK) and Japan Aerospace Exploration Agency (JAXA) in Japan, and the K. A. Wallenberg Foundation, the Swedish Research Council and the Swedish National Space Board in Sweden. Additional support for science analysis during the operations phase from the following agencies is also gratefully acknowledged: the Istituto Nazionale di Astrofisica in Italy and and the Centre National d'Etudes Spatiales in France. This work performed in part under DOE Contract DE-AC02-76SF00515. Based on observations obtained with \textit{XMM-Newton}, an ESA science mission with instruments and contributions directly funded by ESA Member States and NASA.

K.Y.A., K.L.L., and L.C.C.L. are supported by the National Science and Technology Council of the Republic of China (Taiwan) through grants NSTC 113-2636-M-006-003 and NSTC 114-2112-M-006-035. This research is partially supported by the Yushan Fellow Program by the Ministry of Education (MOE), Taiwan (MOE-114-YSFMS-0005-002-P2). A.K.H.K. are supported by the National Science and Technology Council of the Republic of China (Taiwan) through grants NSTC 113-2112-M-007-001. J.T. is supported by the National Key Research and Development Program of China (grant No. 2020YFC2201400) and the National Natural Science Foundation of China (grant No. 12173014). C.Y.H. is supported by the National Research Foundation of Korea through grants 2016R1A5A1013277 and 2022R1F1A1073952.

\section*{Data Availability}
All the data of \textit{XMM-Newton} and \textit{Fermi}-LAT are available in HEASARC (\url{https://heasarc.gsfc.nasa.gov/cgi-bin/W3Browse/w3browse.pl}) and FSSC (\url{https://fermi.gsfc.nasa.gov/ssc})


\bibliographystyle{mnras}
\bibliography{reference.bib} 

\newpage


\appendix
\section{B-band optical light curves of the 18 \textit{XMM-Newton} observations}\label{appA}

\begin{figure*}
\centering

	\begin{minipage}[c]{0.5\textwidth}
		\centering
		\includegraphics[width=1\textwidth]{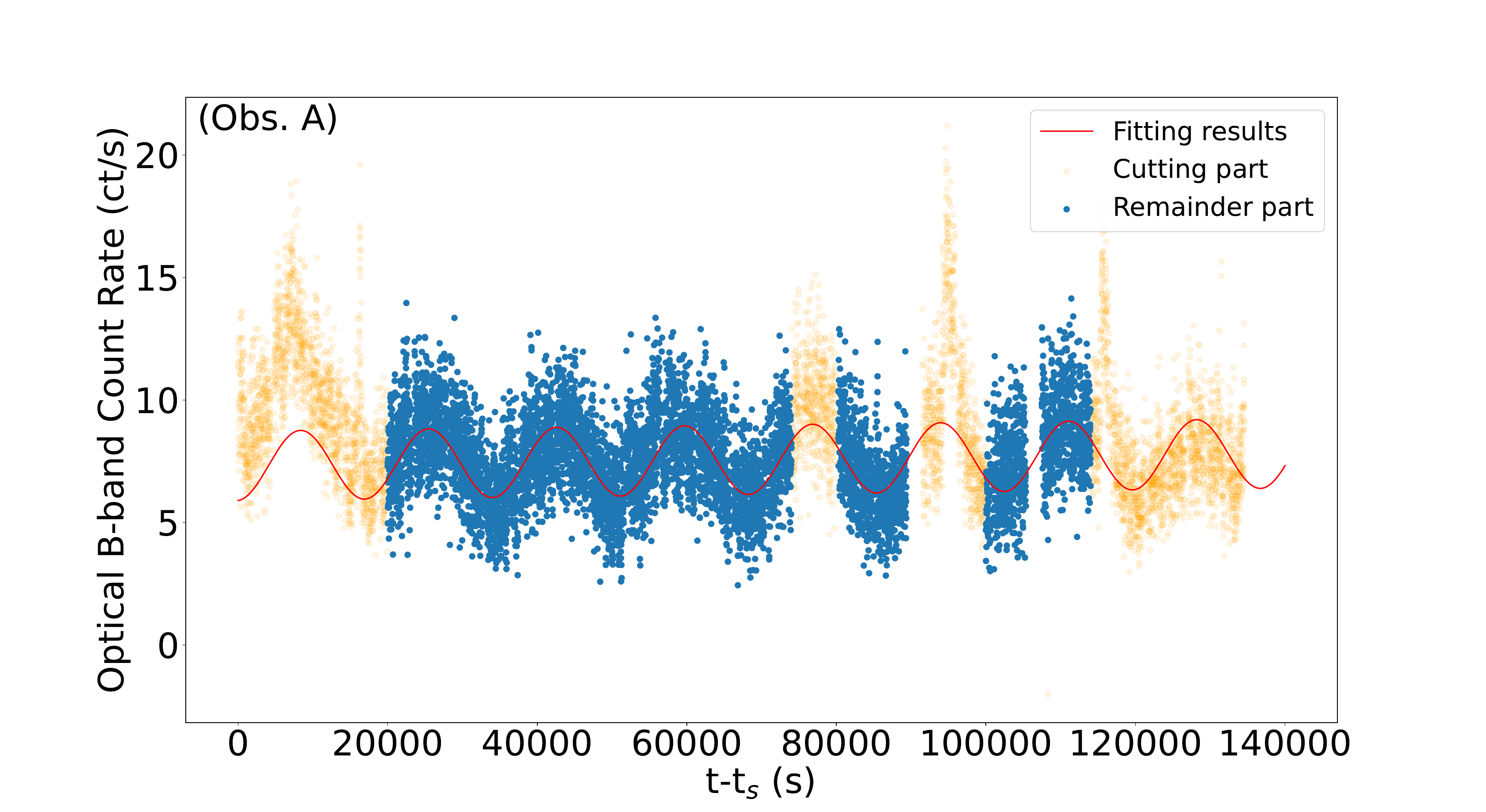}
	\end{minipage}
	\hspace{-0.5cm}
	\begin{minipage}[c]{0.5\textwidth}
		\centering
		\includegraphics[width=1\textwidth]{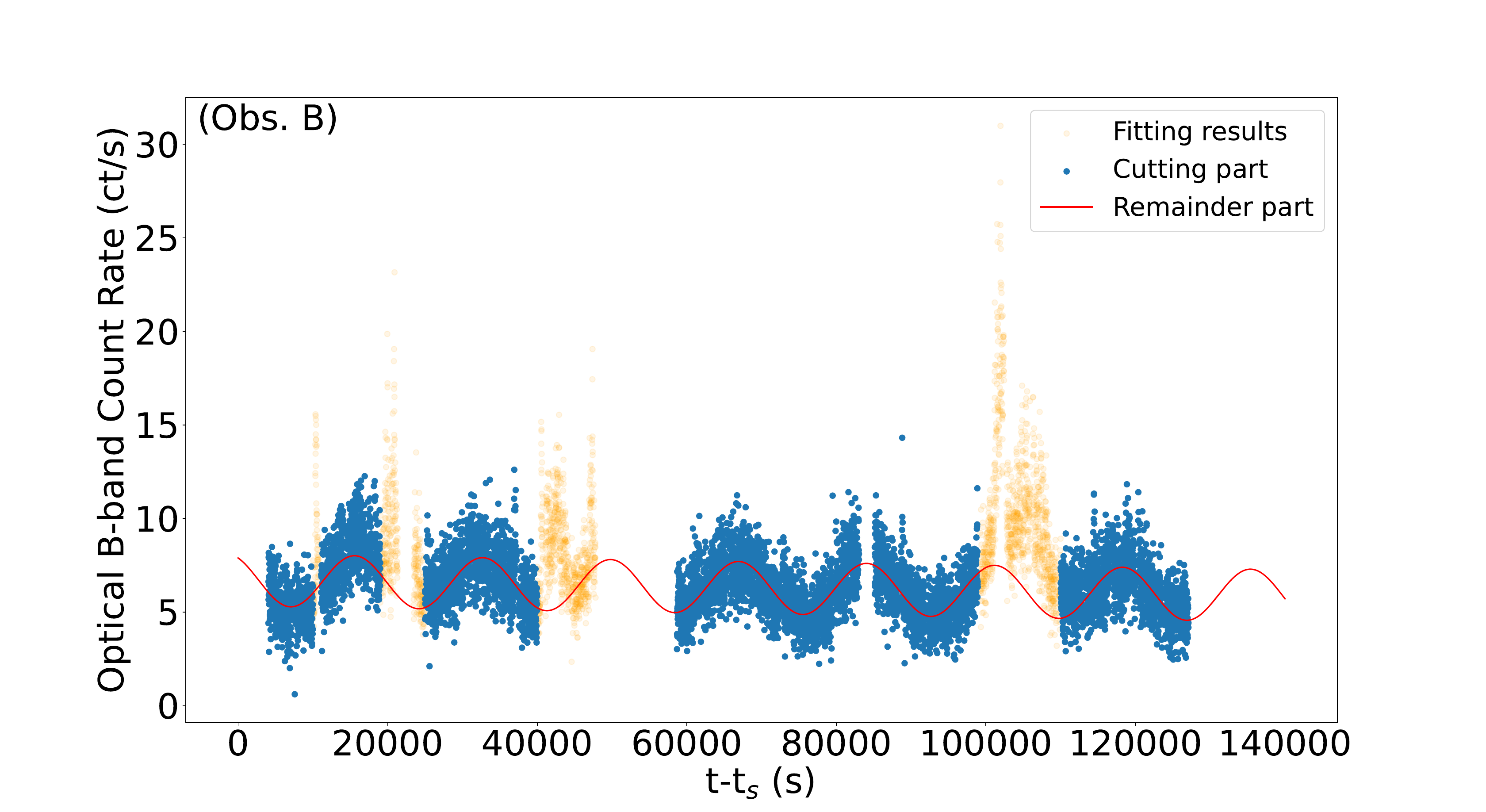}
	\end{minipage}
	\begin{minipage}[c]{0.5\textwidth}
		\centering
		\includegraphics[width=1\textwidth]{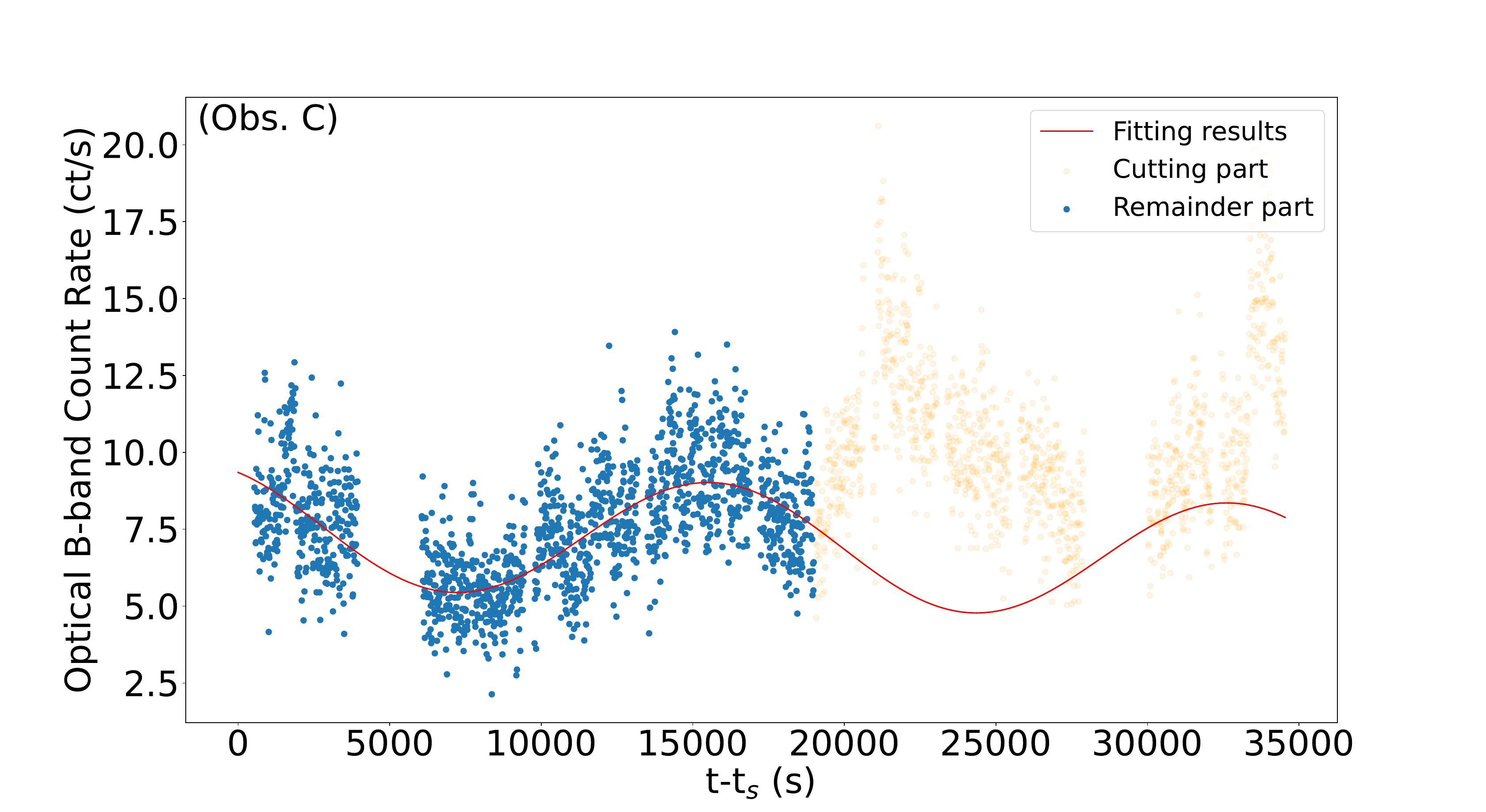}
	\end{minipage}
	\hspace{-0.5cm}
	\begin{minipage}[c]{0.5\textwidth}
		\centering
		\includegraphics[width=1\textwidth]{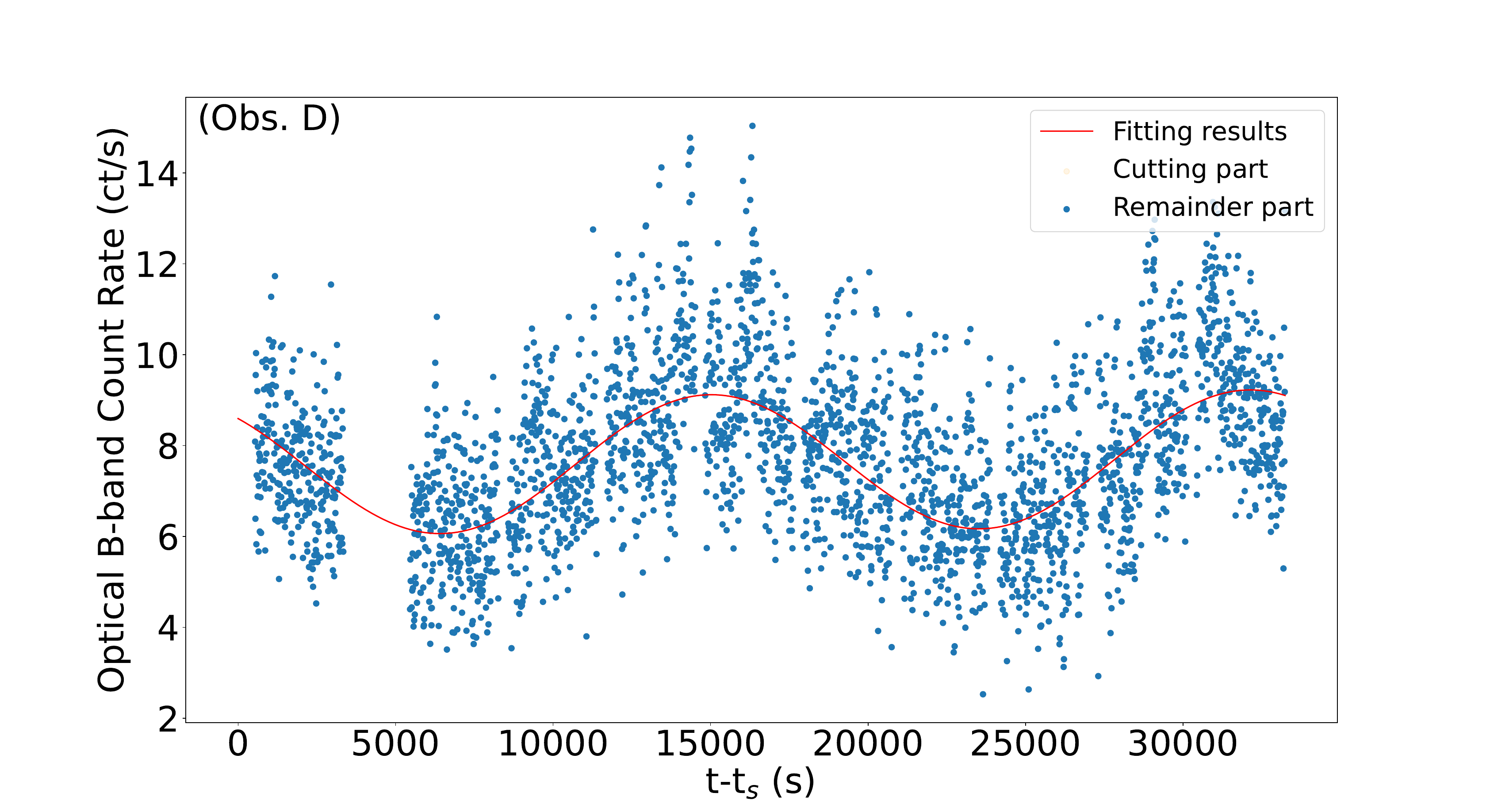}
	\end{minipage}
	\begin{minipage}[c]{0.5\textwidth}
		\centering
		\includegraphics[width=1\textwidth]{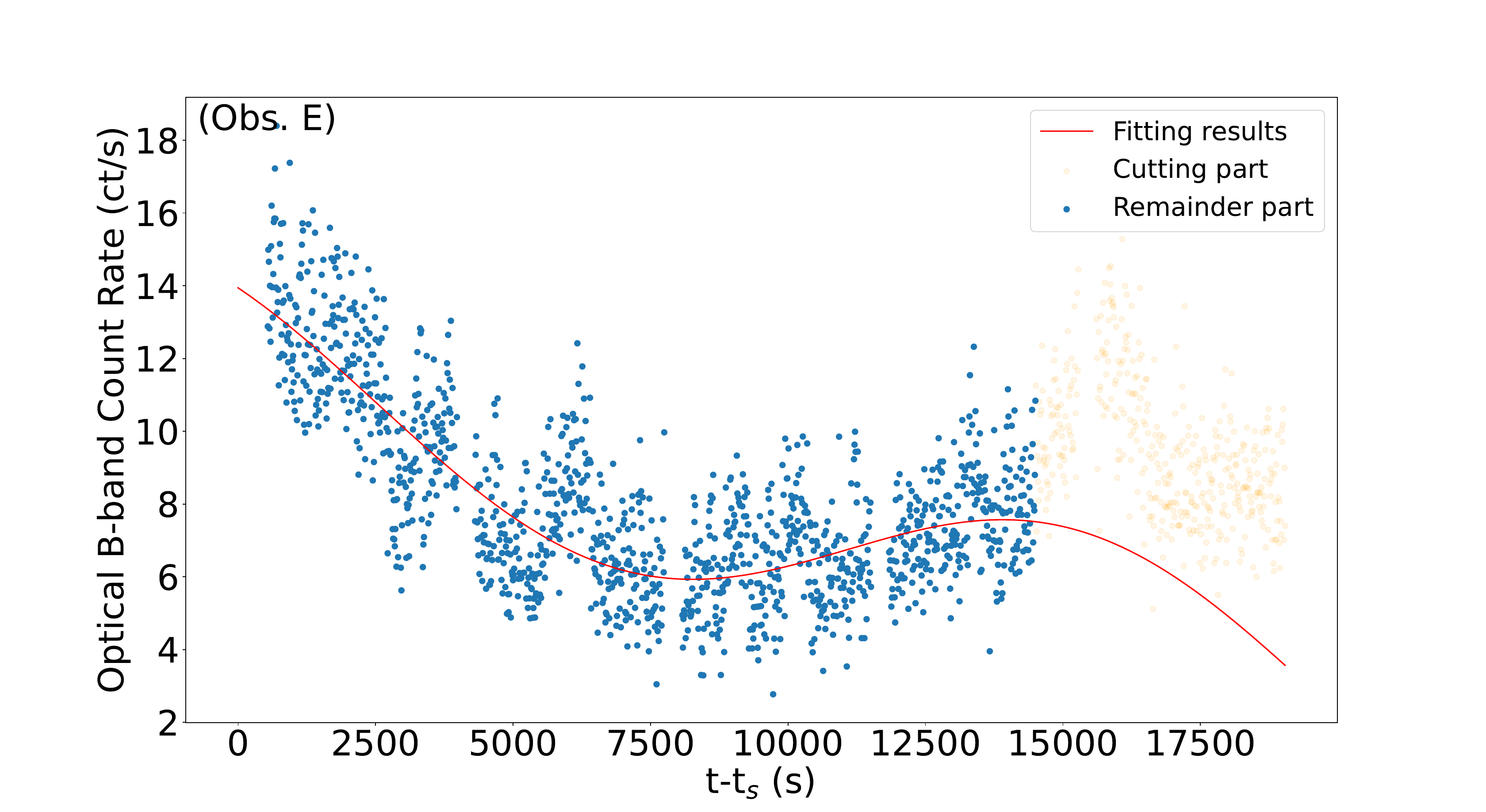}
	\end{minipage}
	\hspace{-0.5cm}
	\begin{minipage}[c]{0.5\textwidth}
		\centering
		\includegraphics[width=1\textwidth]{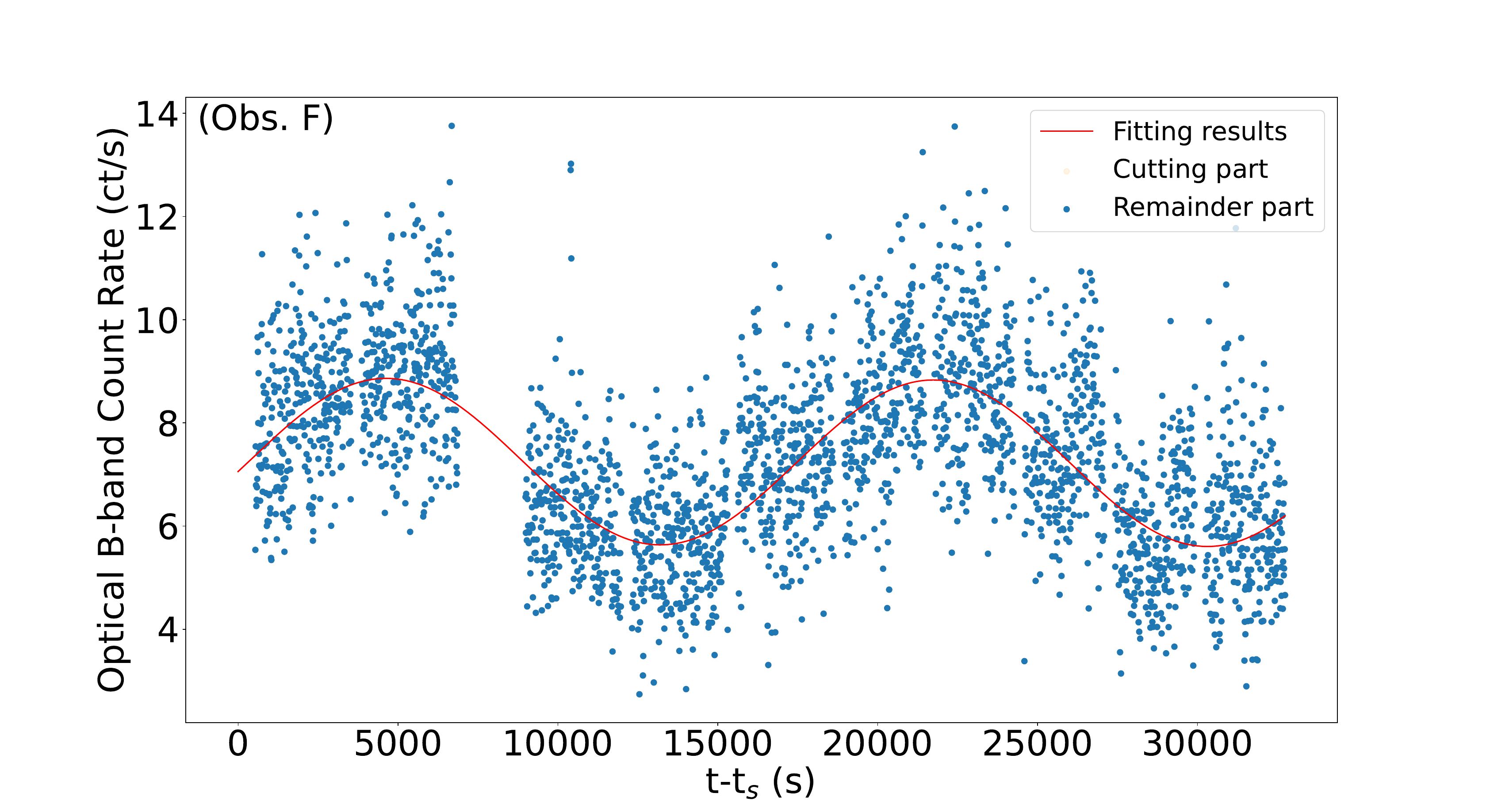}
	\end{minipage}
	\begin{minipage}[c]{0.5\textwidth}
		\centering
		\includegraphics[width=1\textwidth]{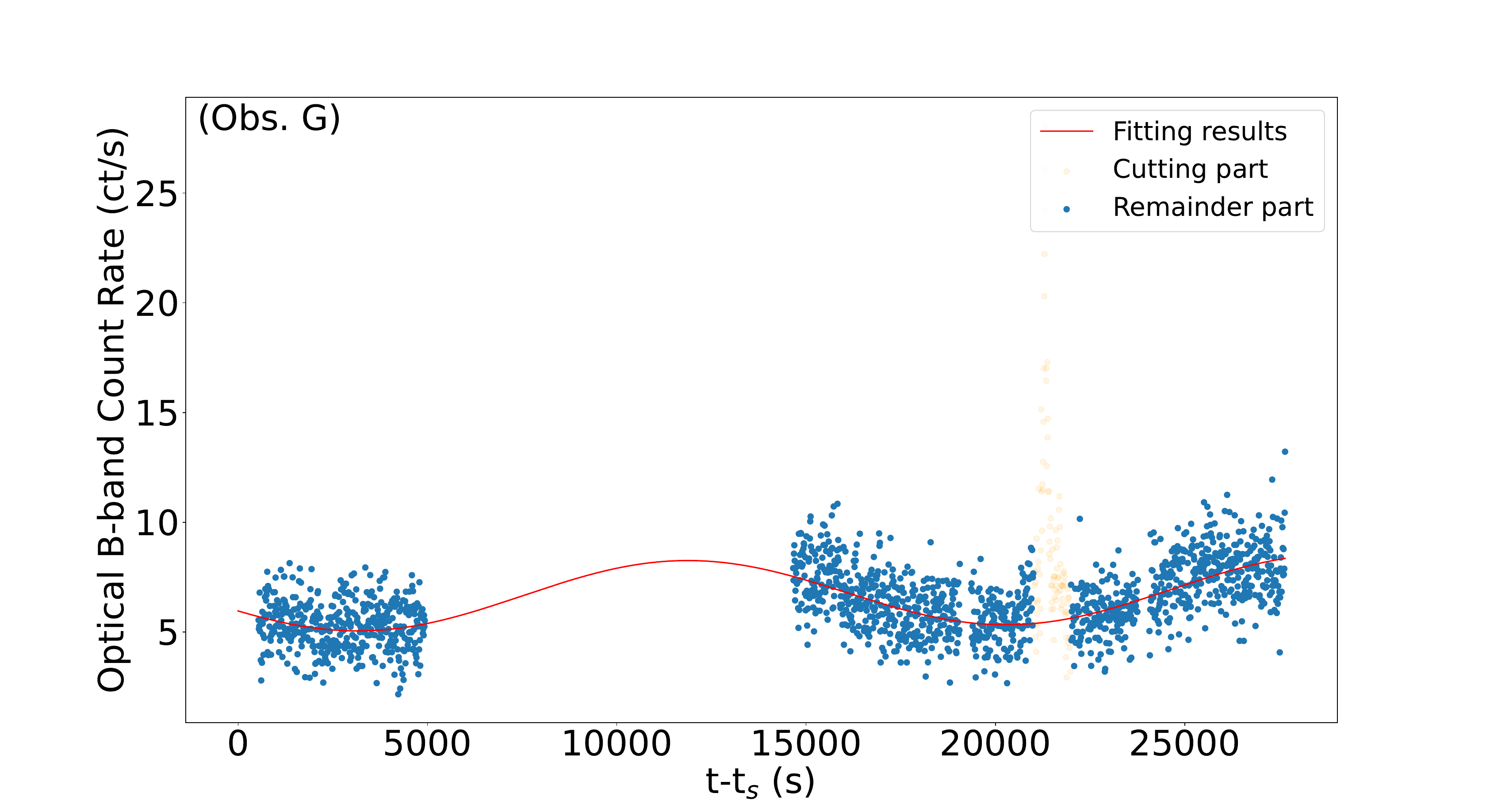}
	\end{minipage}
	\hspace{-0.5cm}
	\begin{minipage}[c]{0.5\textwidth}
		\centering
		\includegraphics[width=1\textwidth]{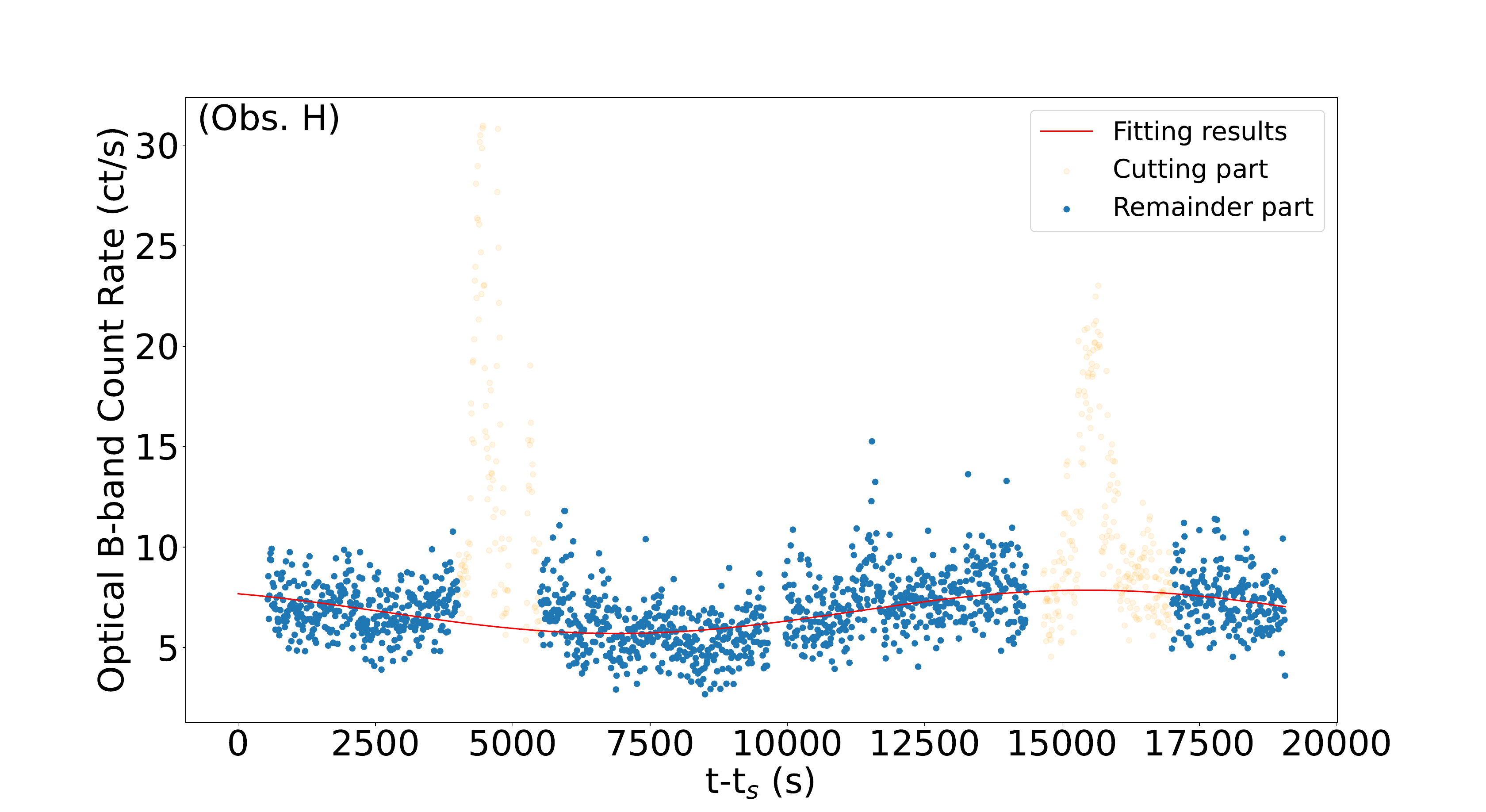}
	\end{minipage}
	\begin{minipage}[c]{0.5\textwidth}
		\centering
		\includegraphics[width=1\textwidth]{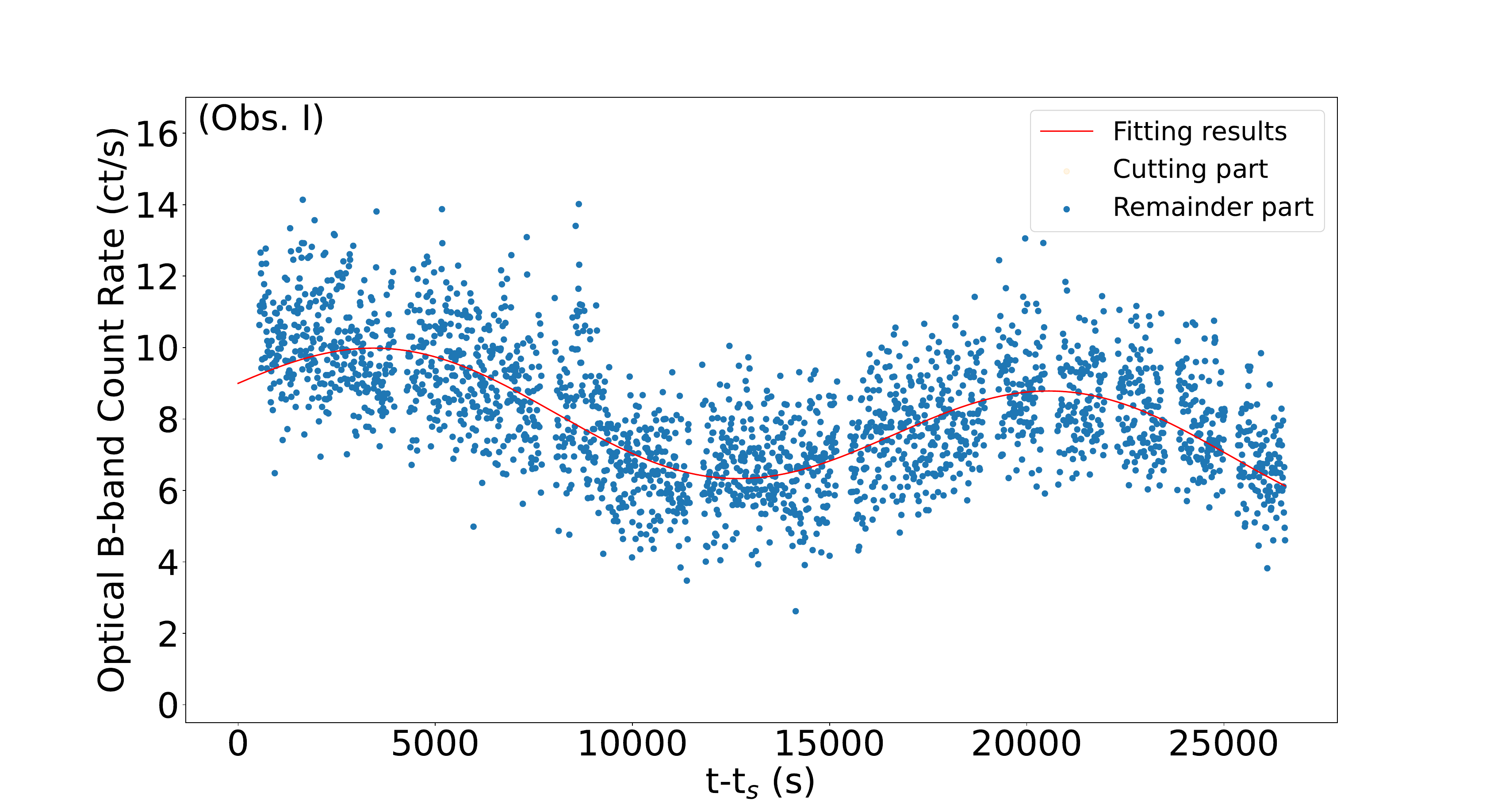}
	\end{minipage}
	\hspace{-0.5cm}
	\begin{minipage}[c]{0.5\textwidth}
		\centering
		\includegraphics[width=1\textwidth]{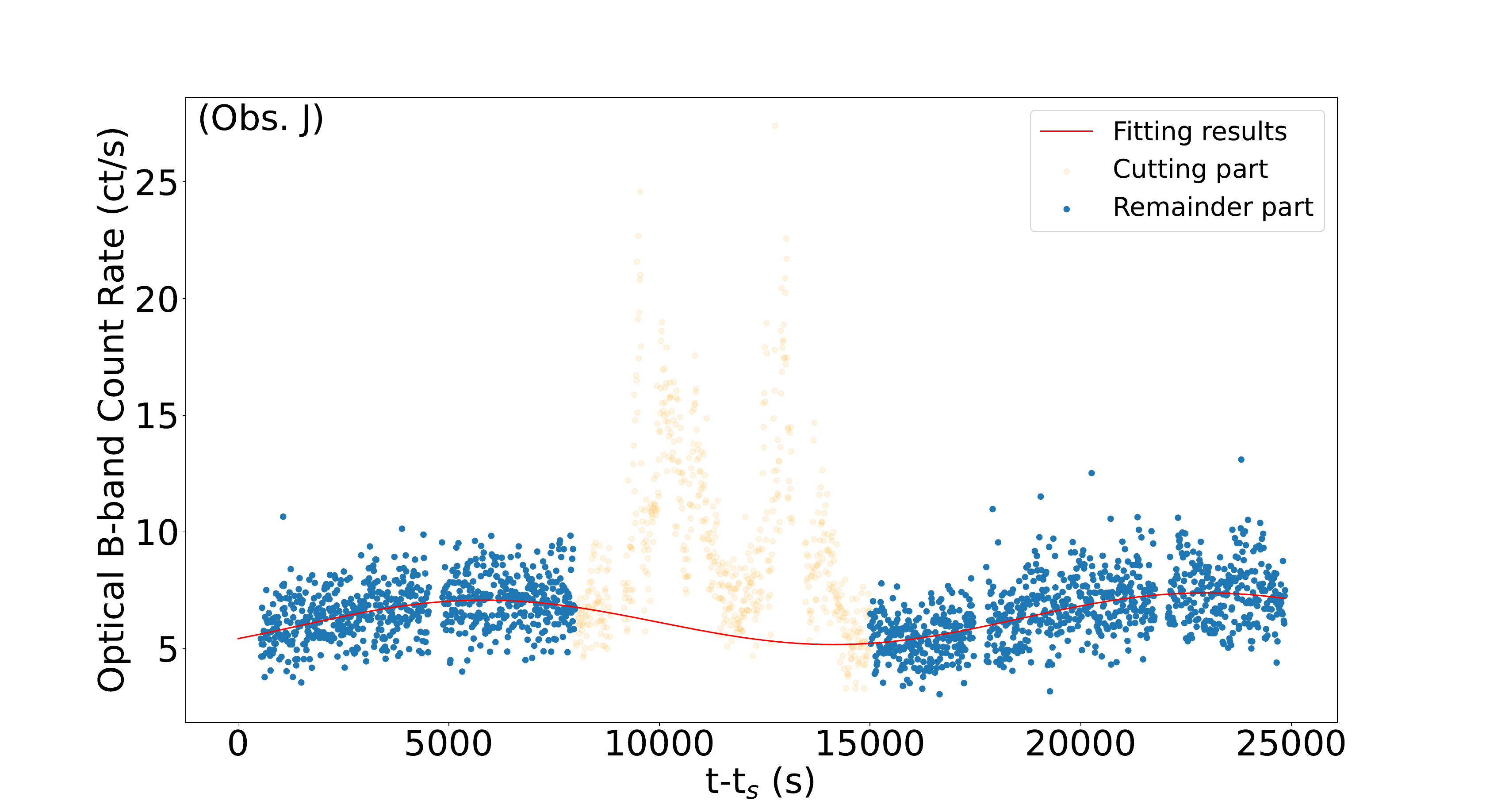}
	\end{minipage}
\end{figure*}

\begin{figure*}
\vspace{-0.25cm}
	
	\begin{minipage}[c]{0.5\textwidth}
		\centering
		\includegraphics[width=1\textwidth]{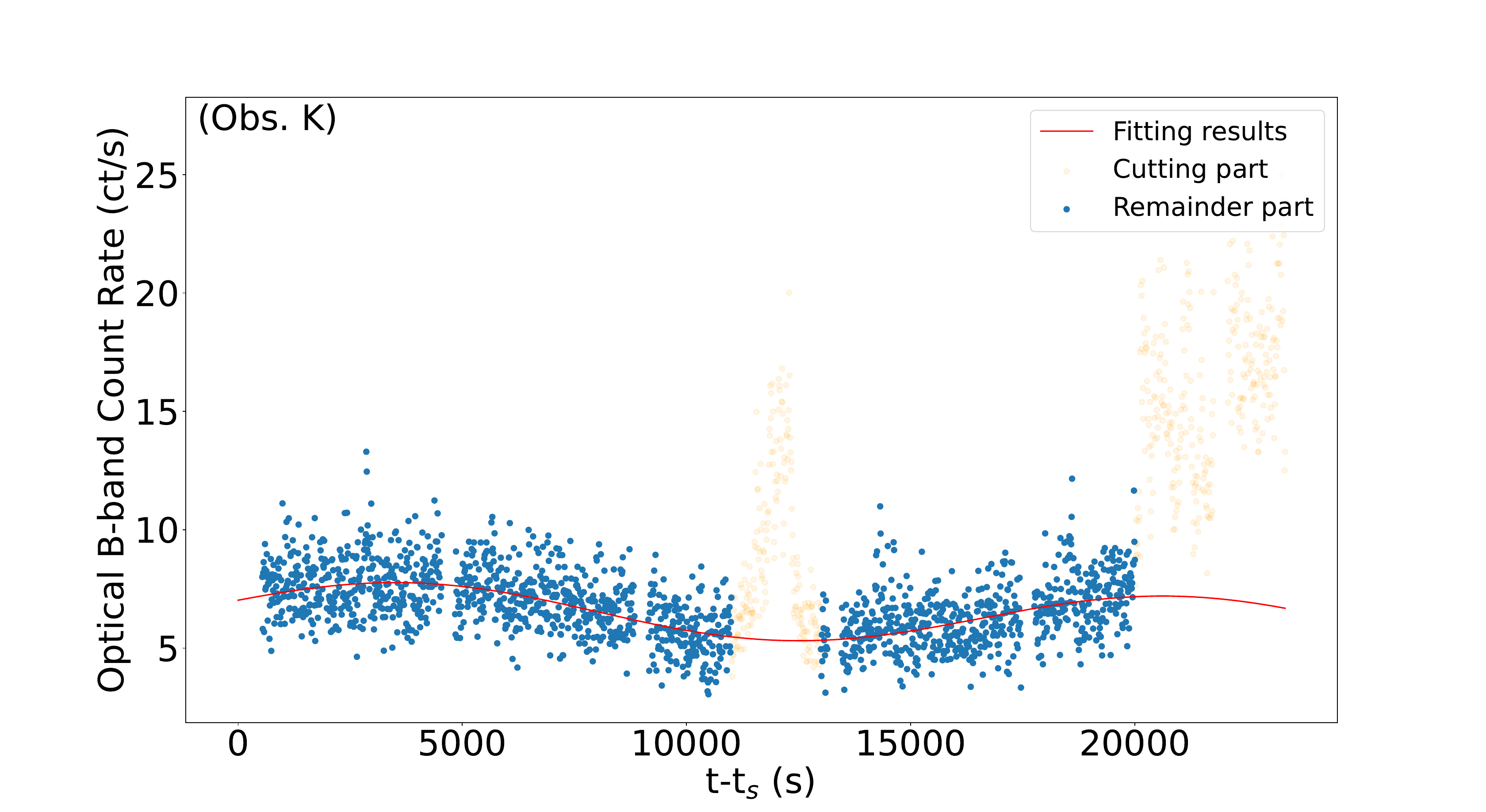}
	\end{minipage}
	\hspace{-0.5cm}
	\begin{minipage}[c]{0.5\textwidth}
		\centering
		\includegraphics[width=1\textwidth]{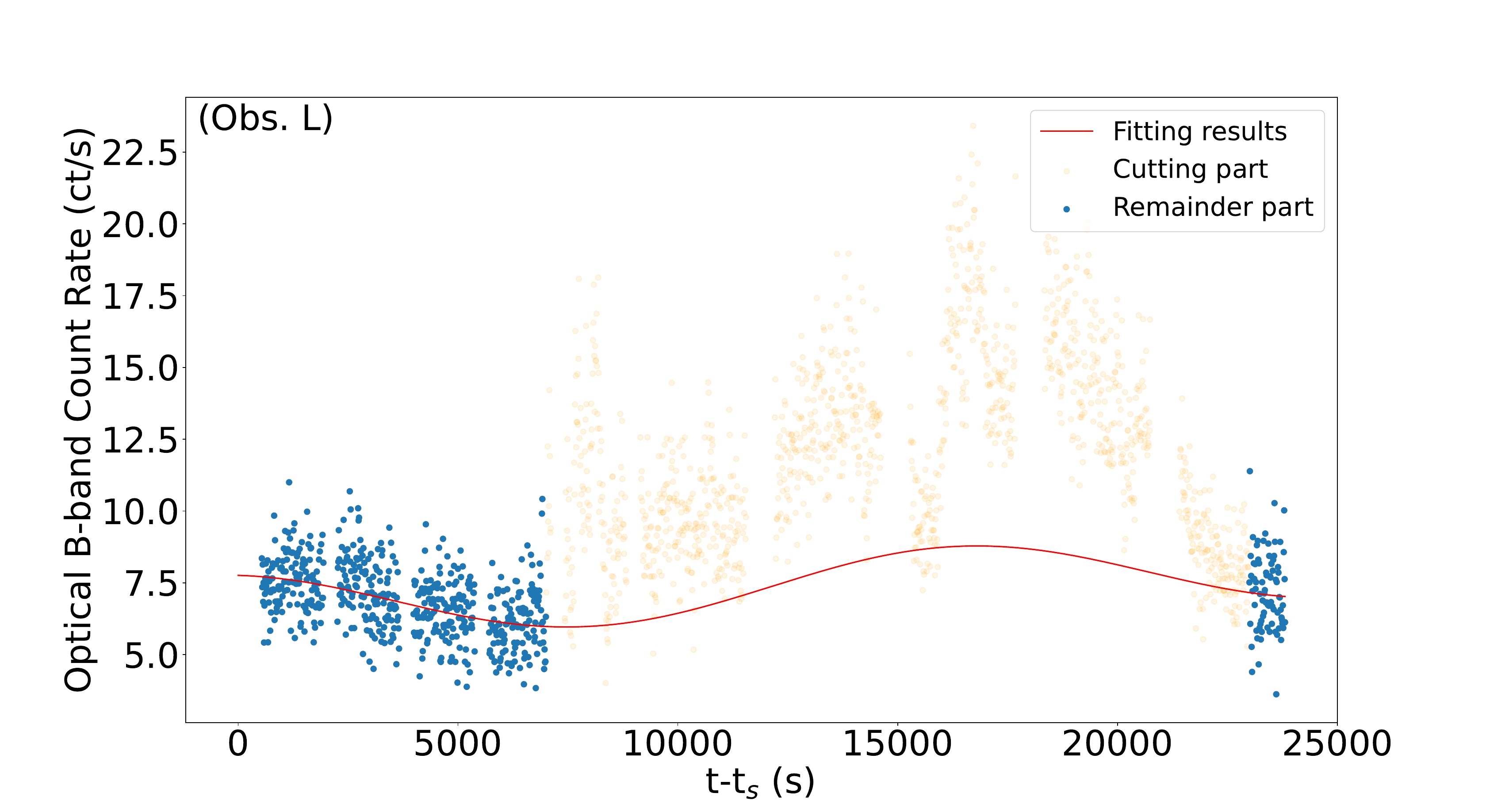}
	\end{minipage}\\
	\begin{minipage}[c]{0.5\textwidth}
		\centering
		\includegraphics[width=1\textwidth]{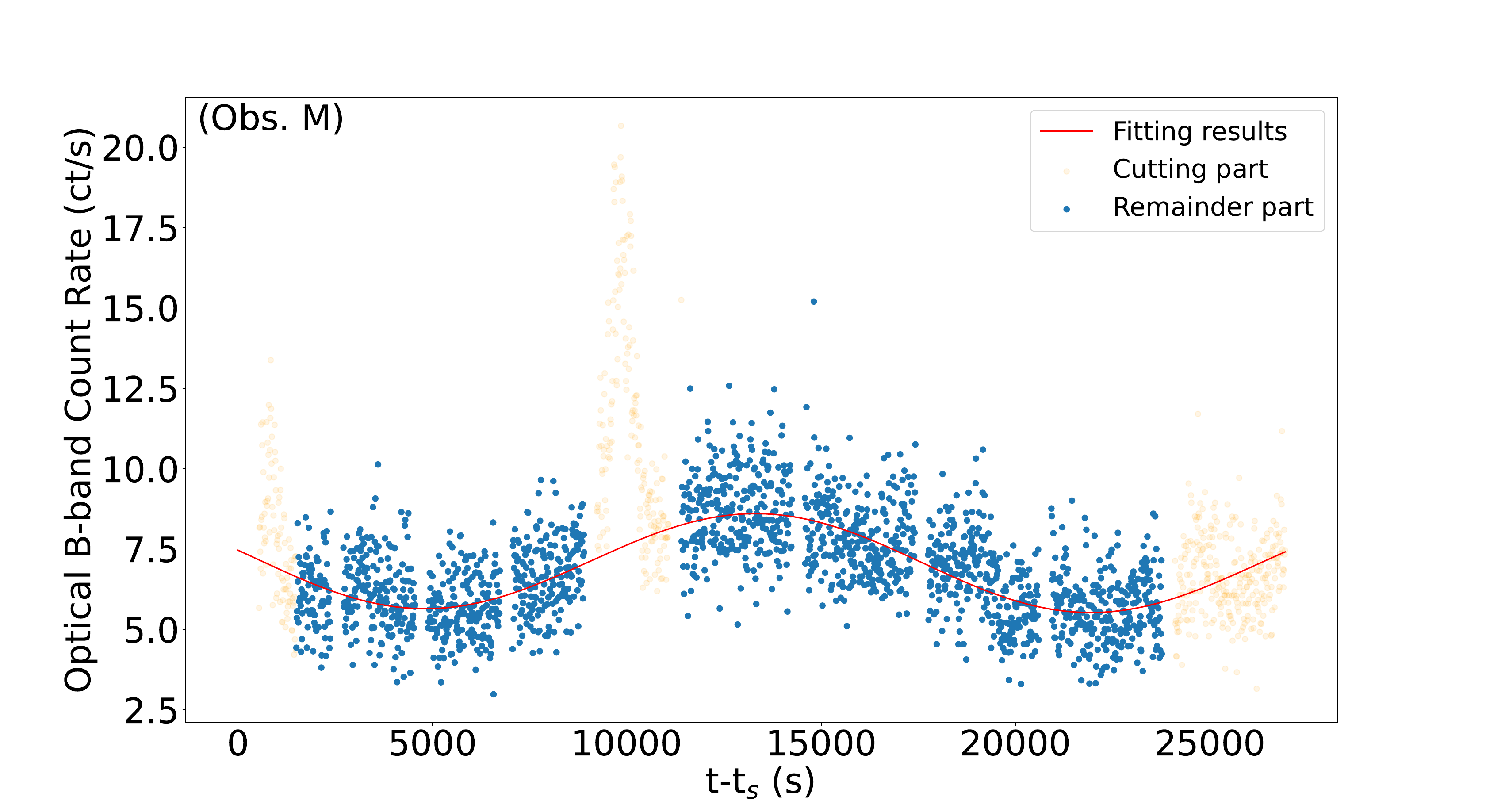}
	\end{minipage}
	\hspace{-0.5cm}
	\begin{minipage}[c]{0.5\textwidth}
		\centering
		\includegraphics[width=1\textwidth]{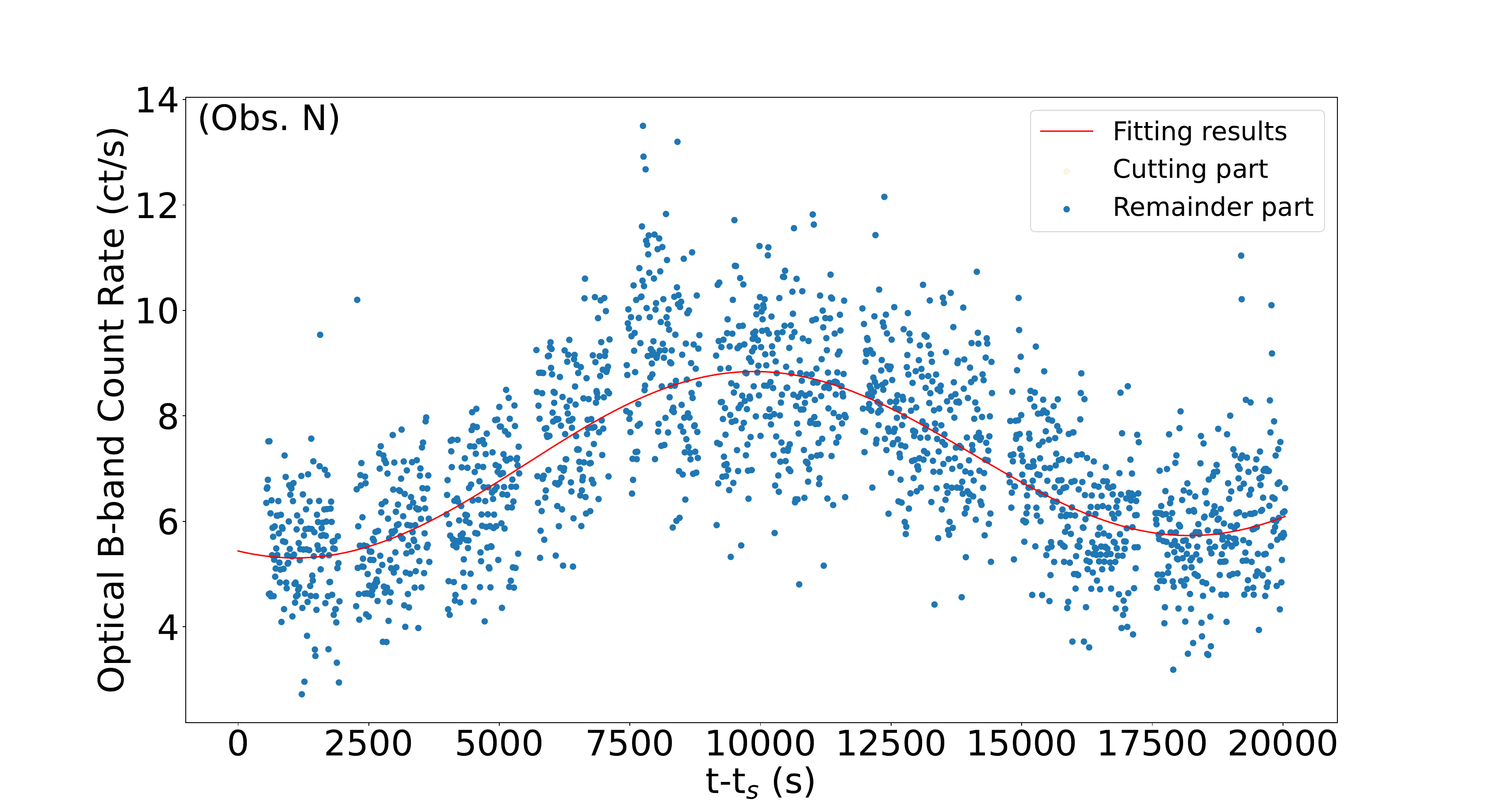}
	\end{minipage}\\
	\begin{minipage}[c]{0.5\textwidth}
		\centering
		\includegraphics[width=1\textwidth]{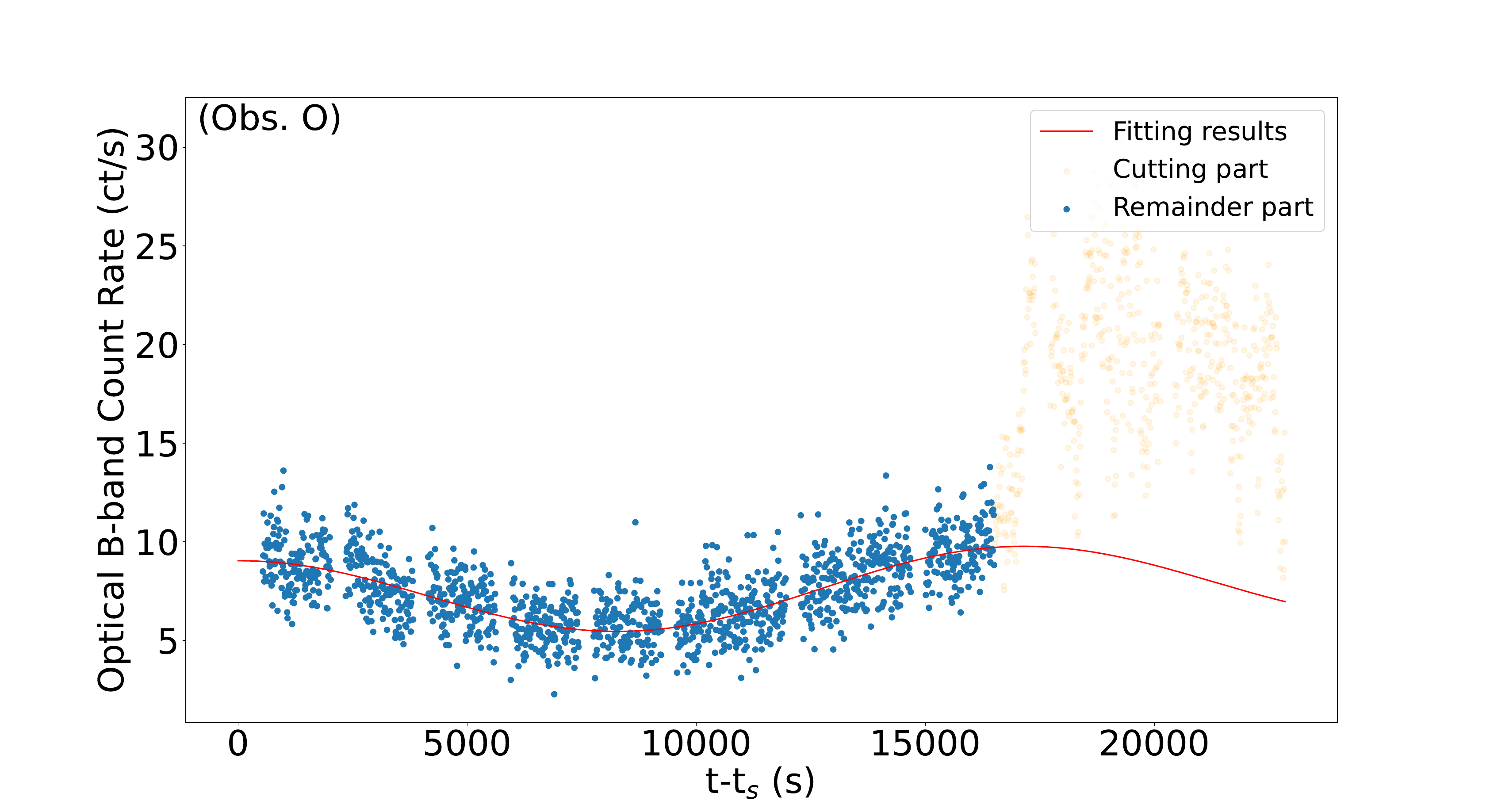}
	\end{minipage}
	\hspace{-0.5cm}
	\begin{minipage}[c]{0.5\textwidth}
		\centering
		\includegraphics[width=1\textwidth]{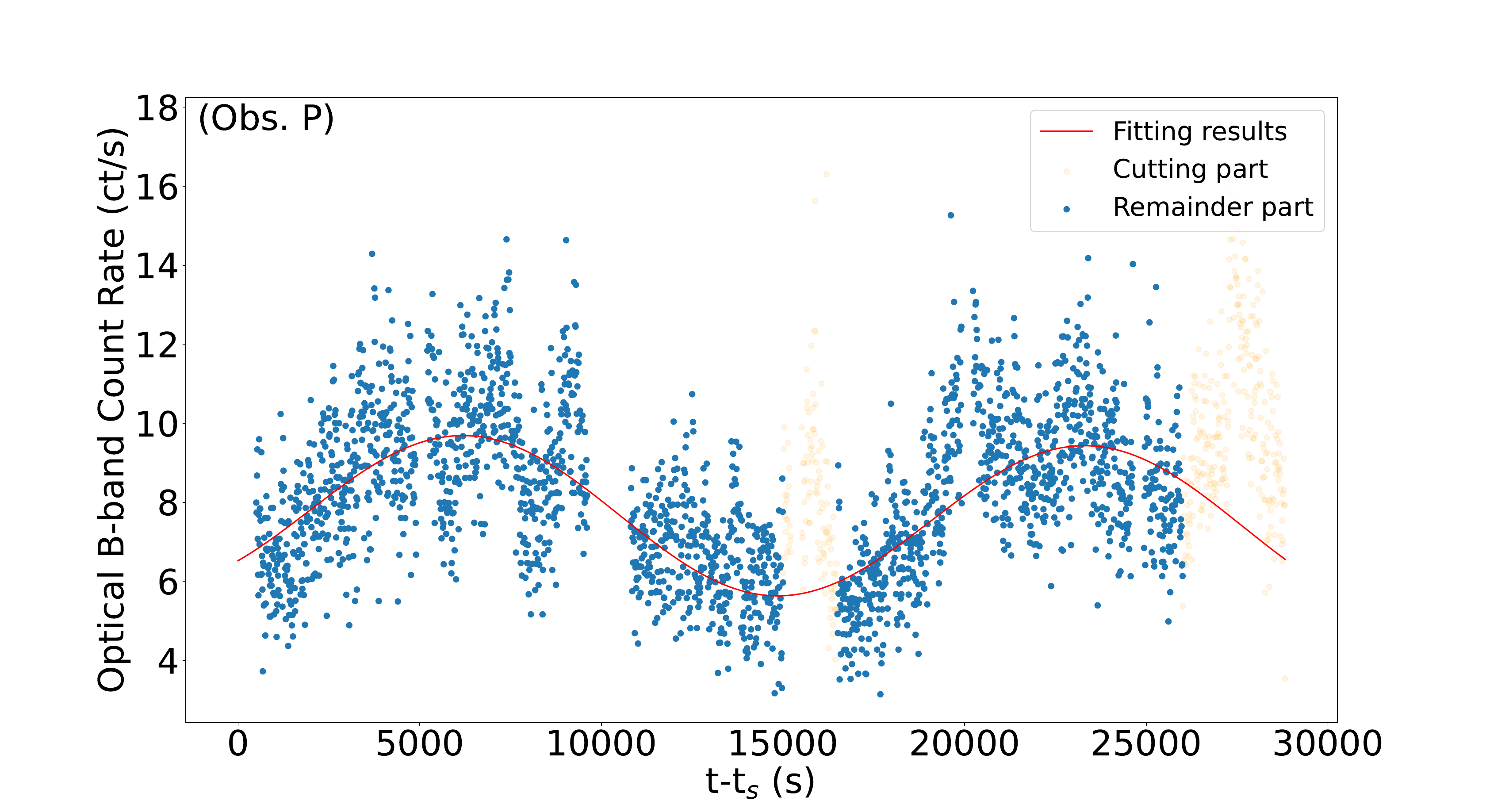}
	\end{minipage}\\
	\begin{minipage}[c]{0.5\textwidth}
		\centering
		\includegraphics[width=1\textwidth]{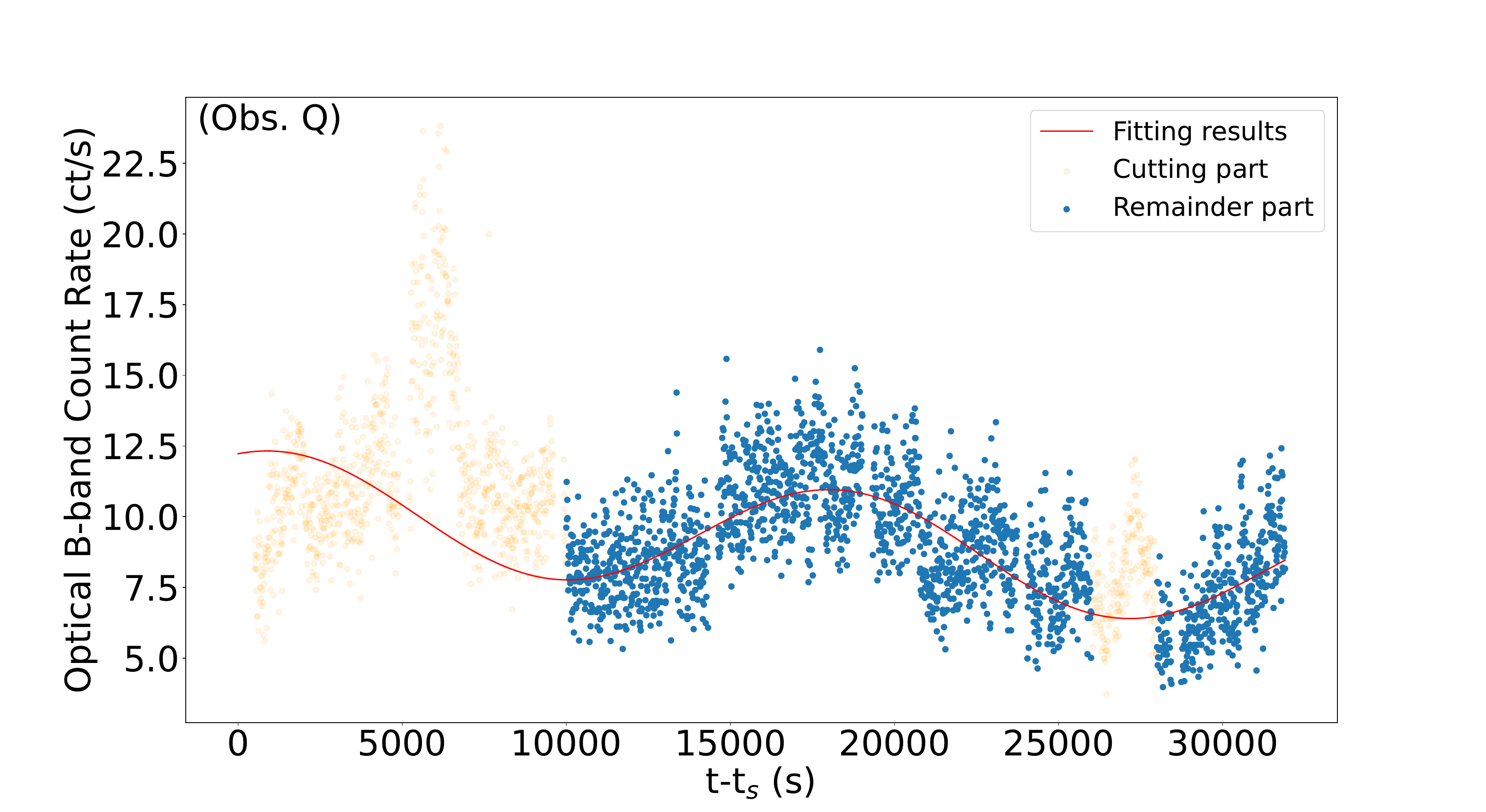}
	\end{minipage}
	\hspace{-0.5cm}
	\begin{minipage}[c]{0.5\textwidth}
		\centering
		\includegraphics[width=1\textwidth]{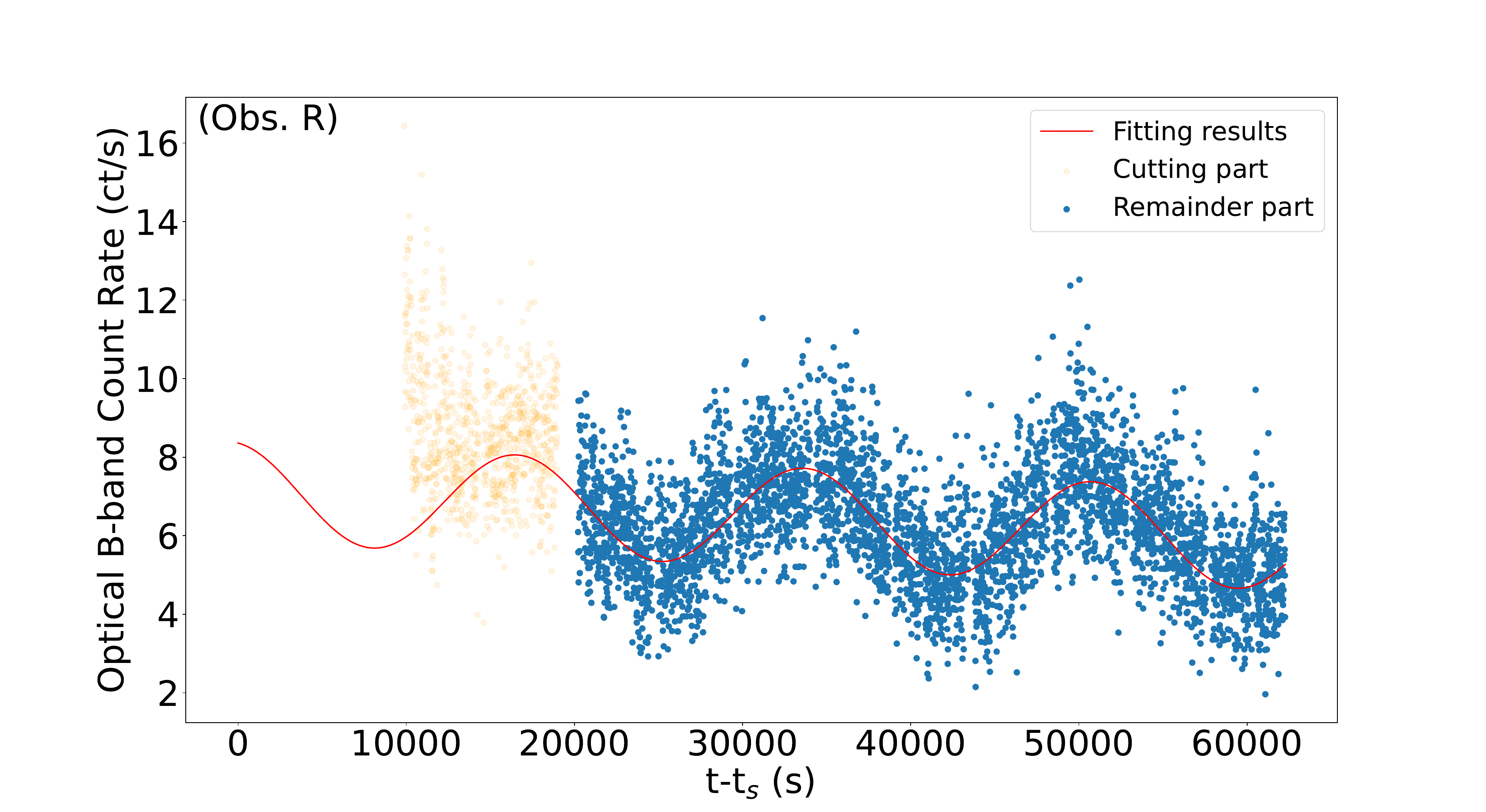}
	\end{minipage}\\

\caption{The B-band optical light curves of the 18 \textit{XMM-Newton} Observations. The translucent orange points are the flare/variable region cut by visual inspection. The red curves are the fitting results mentioned in the Section \ref{sec:o}.
\label{app_a}}
\end{figure*}

\clearpage 
\section{The X-ray light curves of the 18 \textit{XMM-Newton} observations}\label{appB}

\begin{figure*}
\centering
	\begin{minipage}[c]{1\textwidth}
		\centering
		\includegraphics[width=1\textwidth]{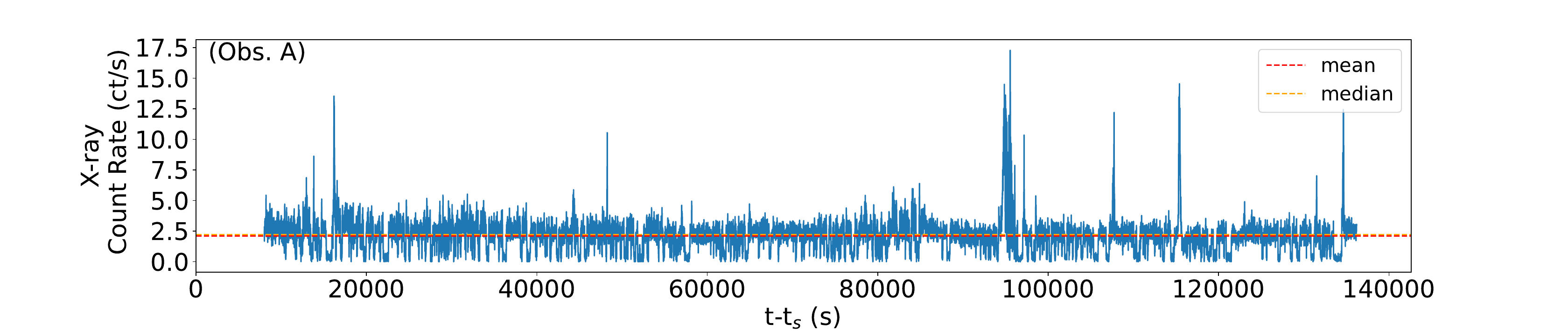}
	\end{minipage}
	\hspace{-0.5cm}
	\begin{minipage}[c]{1\textwidth}
		\centering
		\includegraphics[width=1\textwidth]{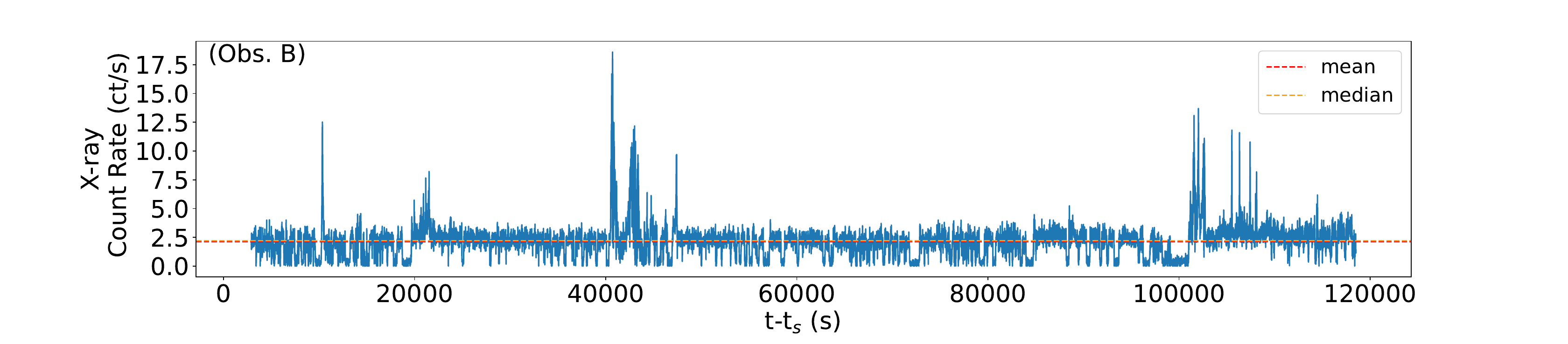}
	\end{minipage}
	\begin{minipage}[c]{0.5\textwidth}
		\centering
		\includegraphics[width=1\textwidth]{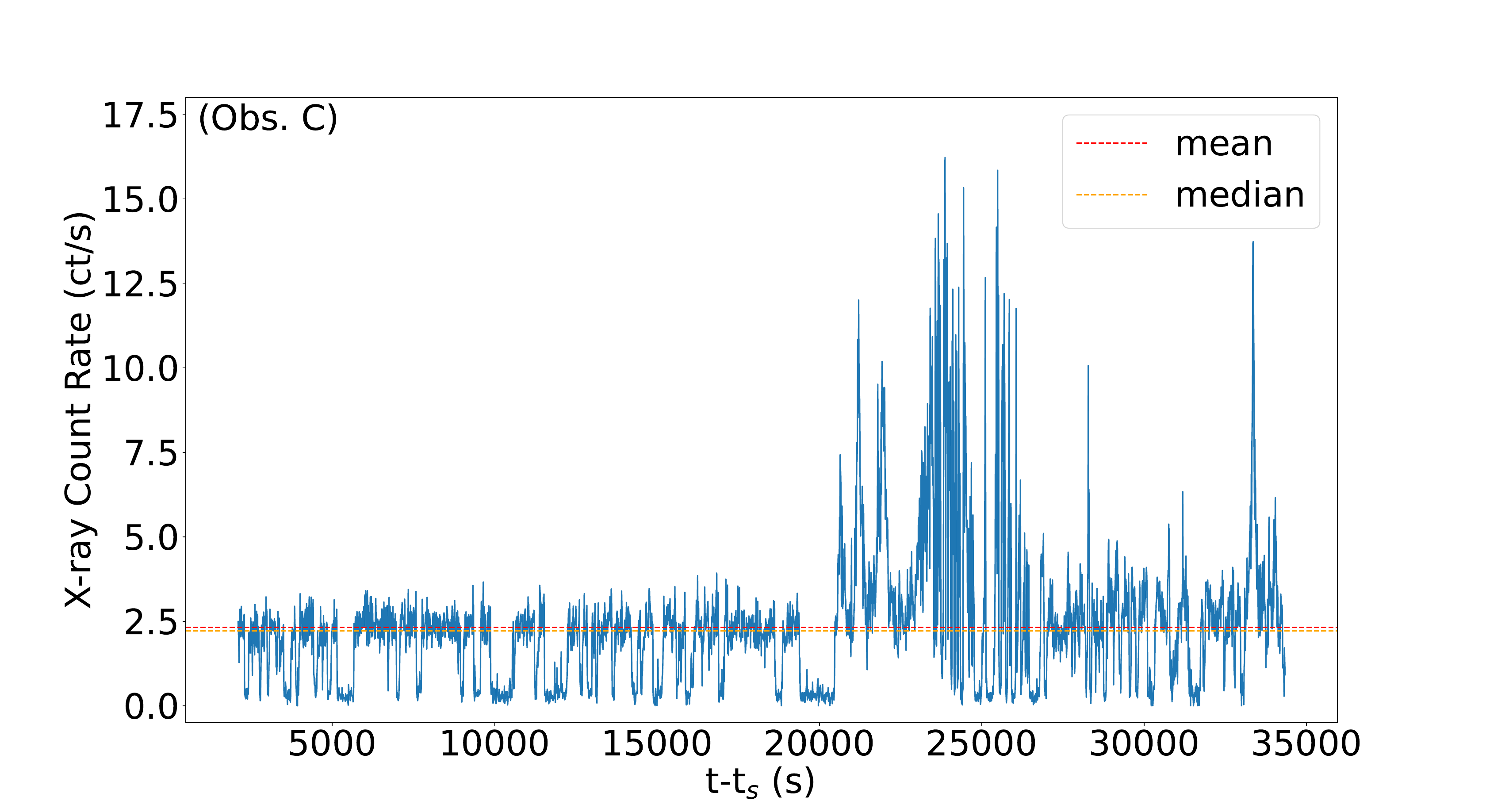}
	\end{minipage}
	\hspace{-0.5cm}
	\begin{minipage}[c]{0.5\textwidth}
		\centering
		\includegraphics[width=1\textwidth]{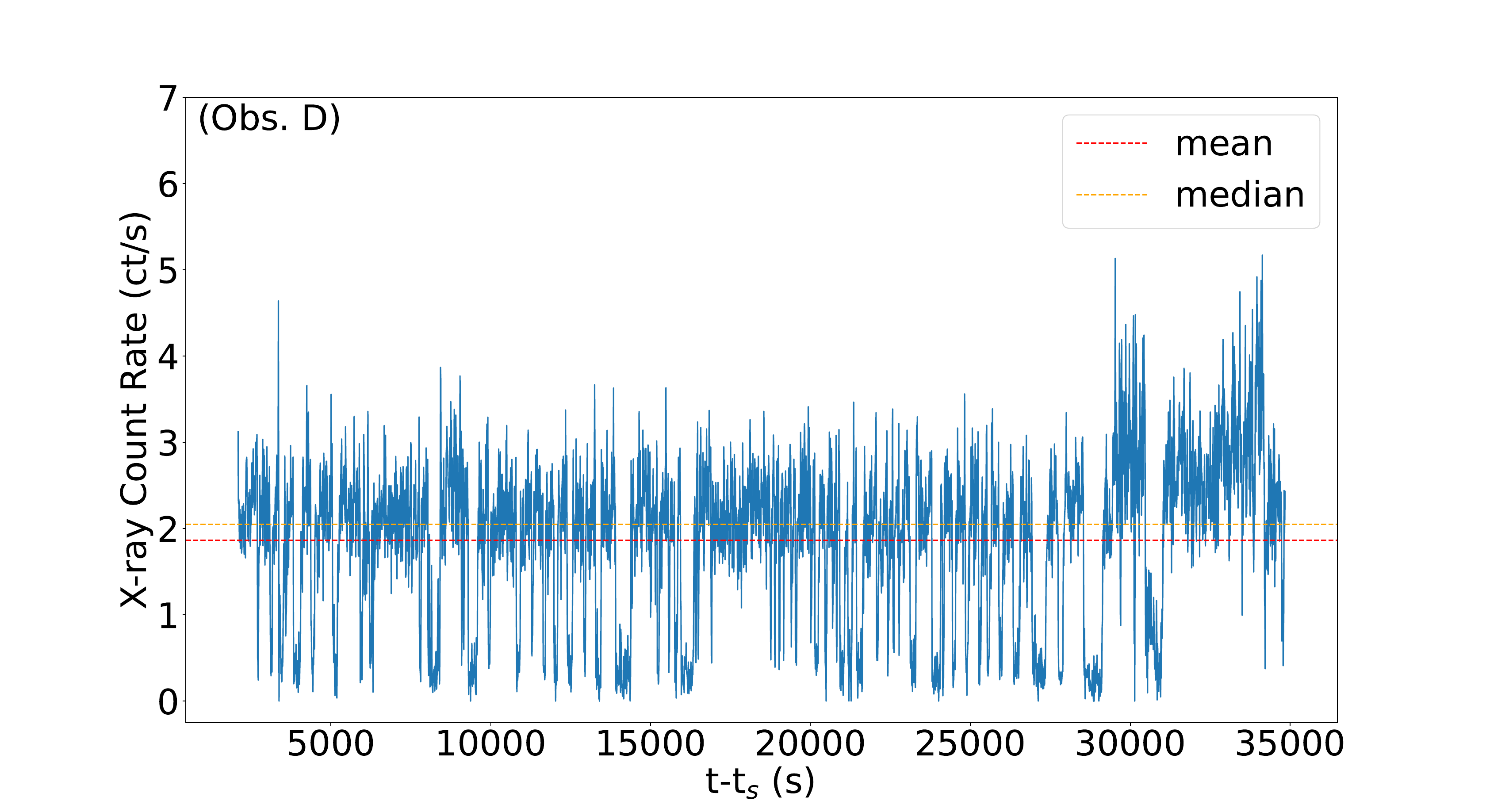}
	\end{minipage}
	\begin{minipage}[c]{0.5\textwidth}
		\centering
		\includegraphics[width=1\textwidth]{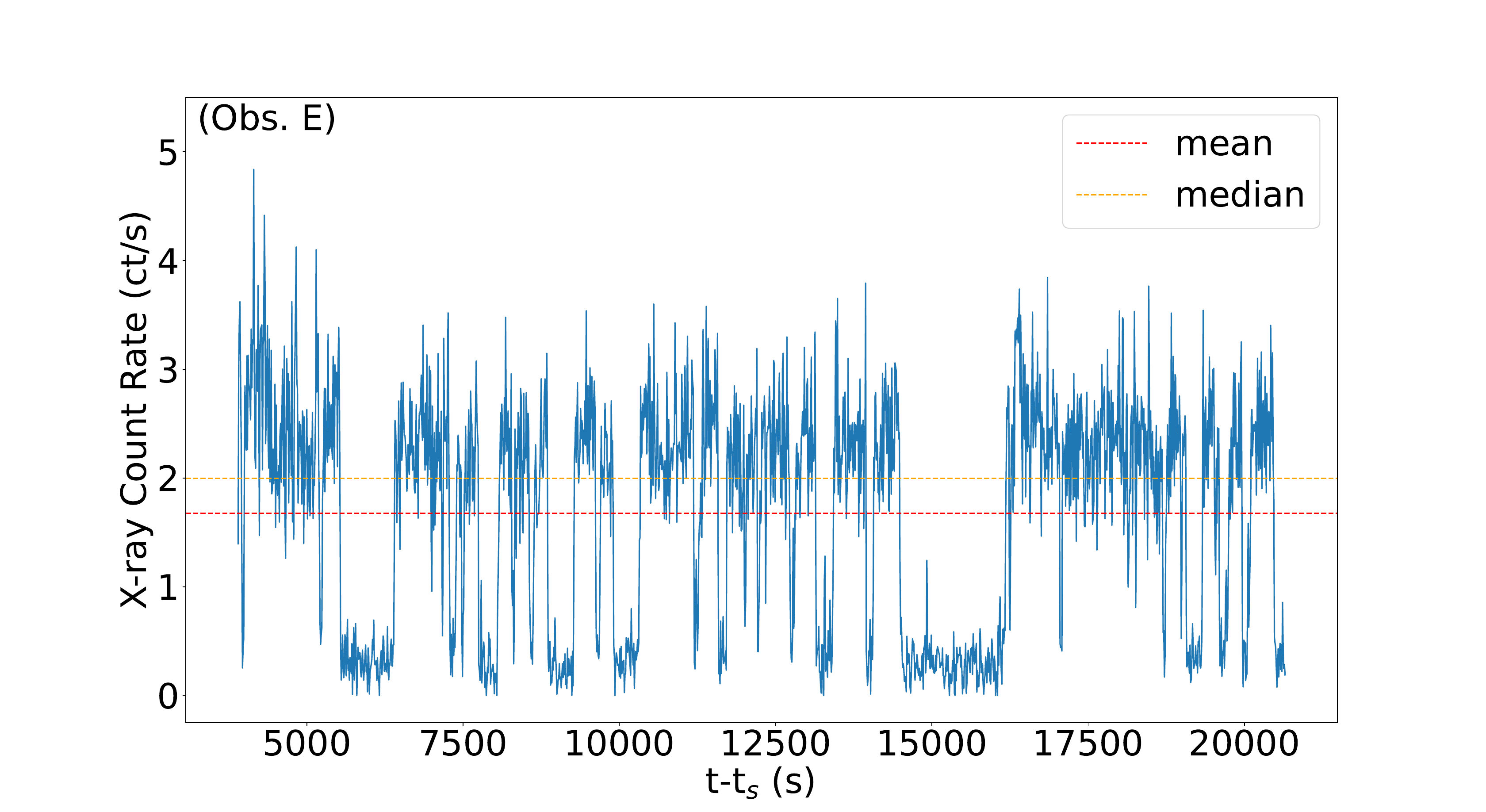}
	\end{minipage}
	\hspace{-0.5cm}
	\begin{minipage}[c]{0.5\textwidth}
		\centering
		\includegraphics[width=1\textwidth]{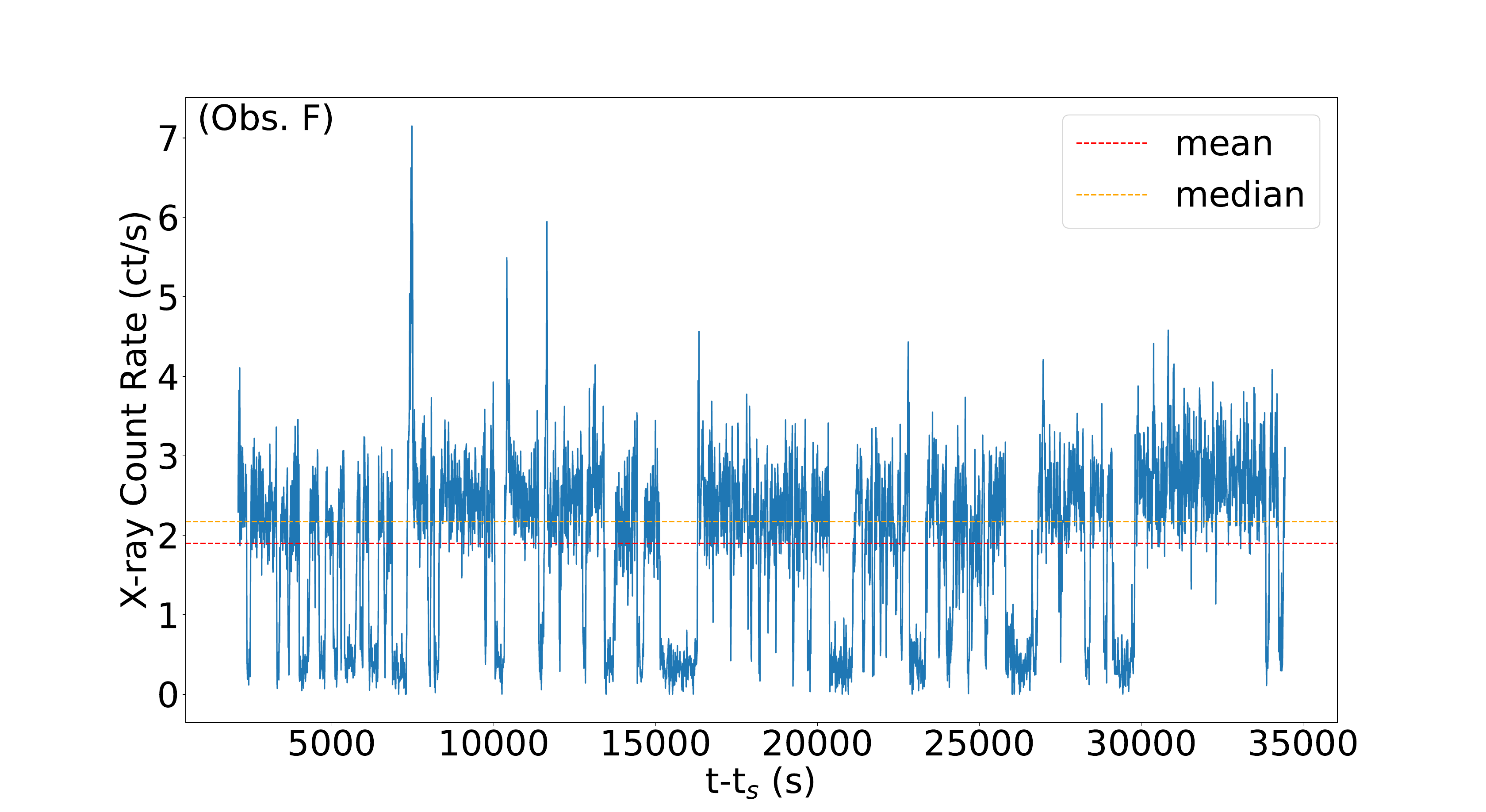}
	\end{minipage}
	\begin{minipage}[c]{0.5\textwidth}
		\centering
		\includegraphics[width=1\textwidth]{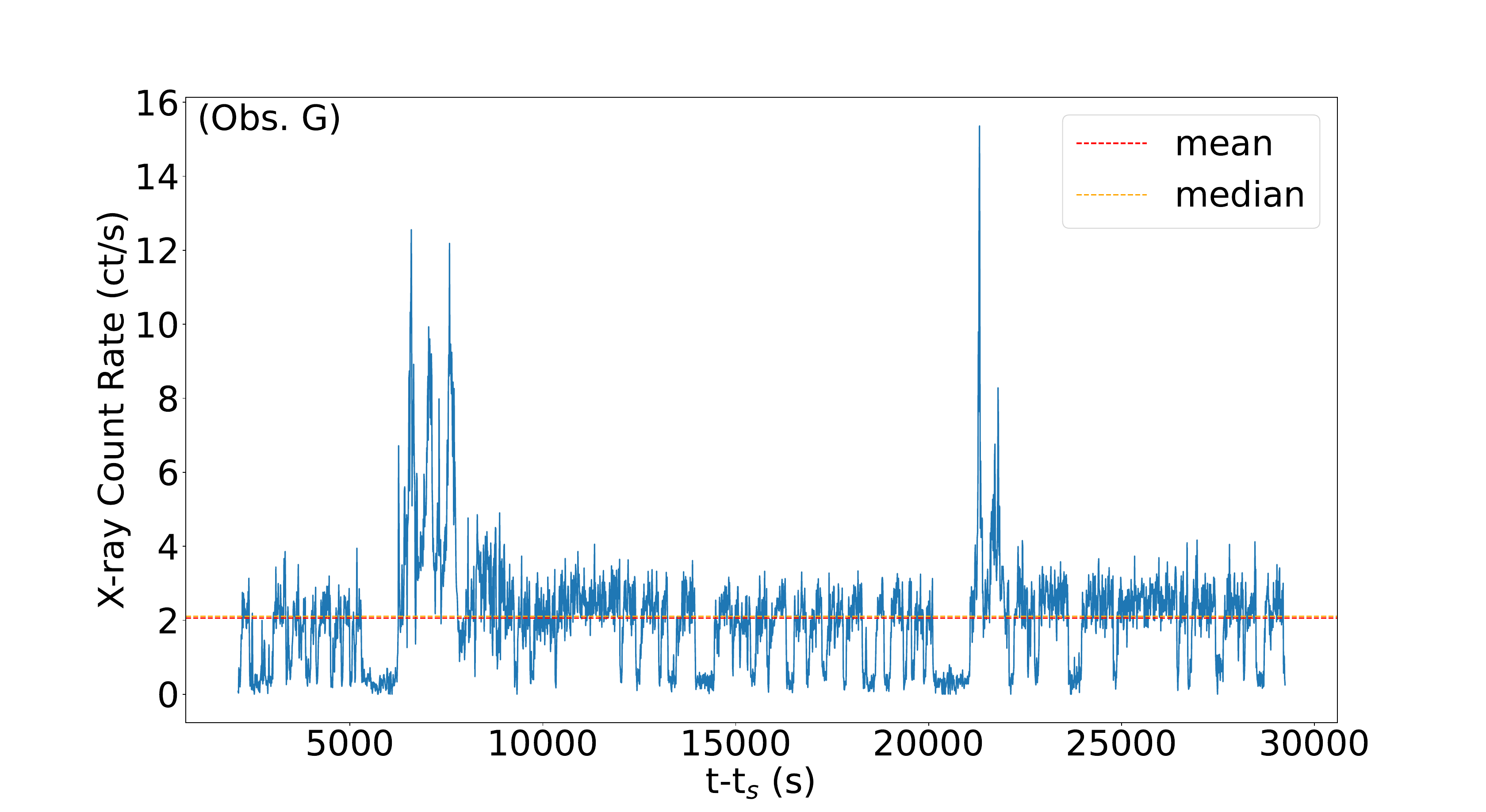}
	\end{minipage}
	\hspace{-0.5cm}
	\begin{minipage}[c]{0.5\textwidth}
		\centering
		\includegraphics[width=1\textwidth]{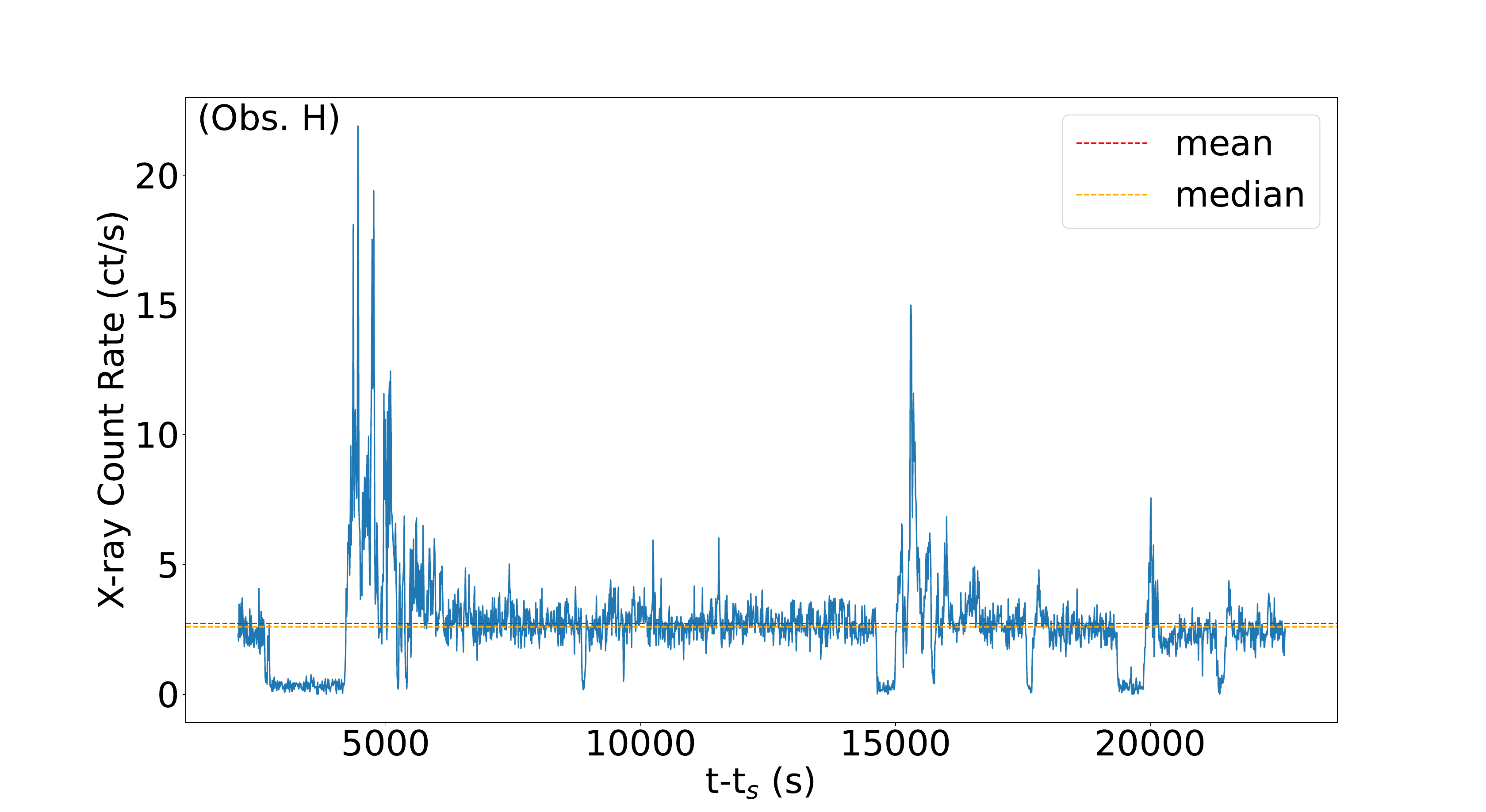}
	\end{minipage}
	
\end{figure*}

\begin{figure*}
\vspace{-0.75cm}
	\begin{minipage}[c]{0.5\textwidth}
		\centering
		\includegraphics[width=1\textwidth]{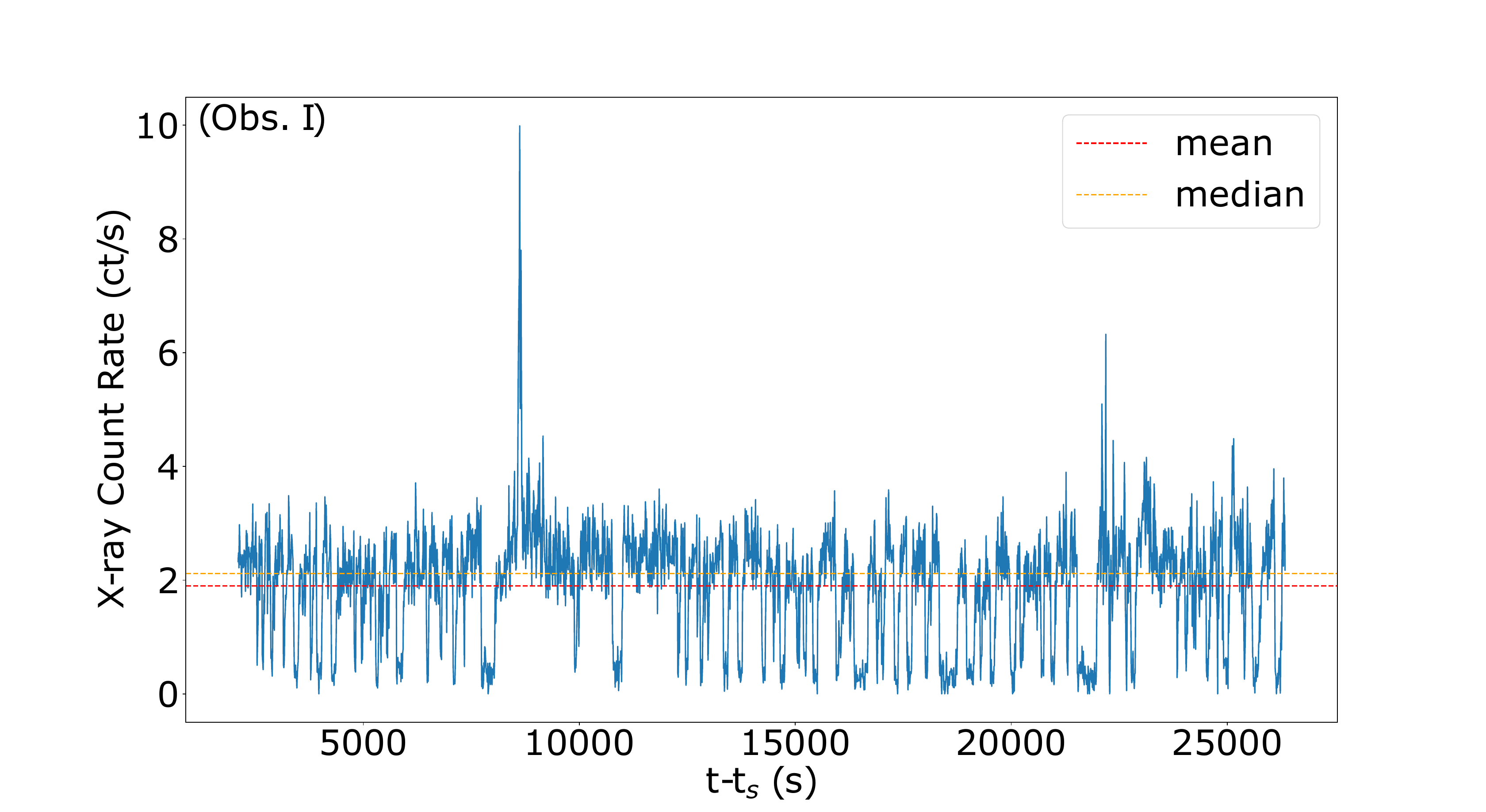}
	\end{minipage}
	\hspace{-1cm}
	\begin{minipage}[c]{0.5\textwidth}
		\centering
		\includegraphics[width=1\textwidth]{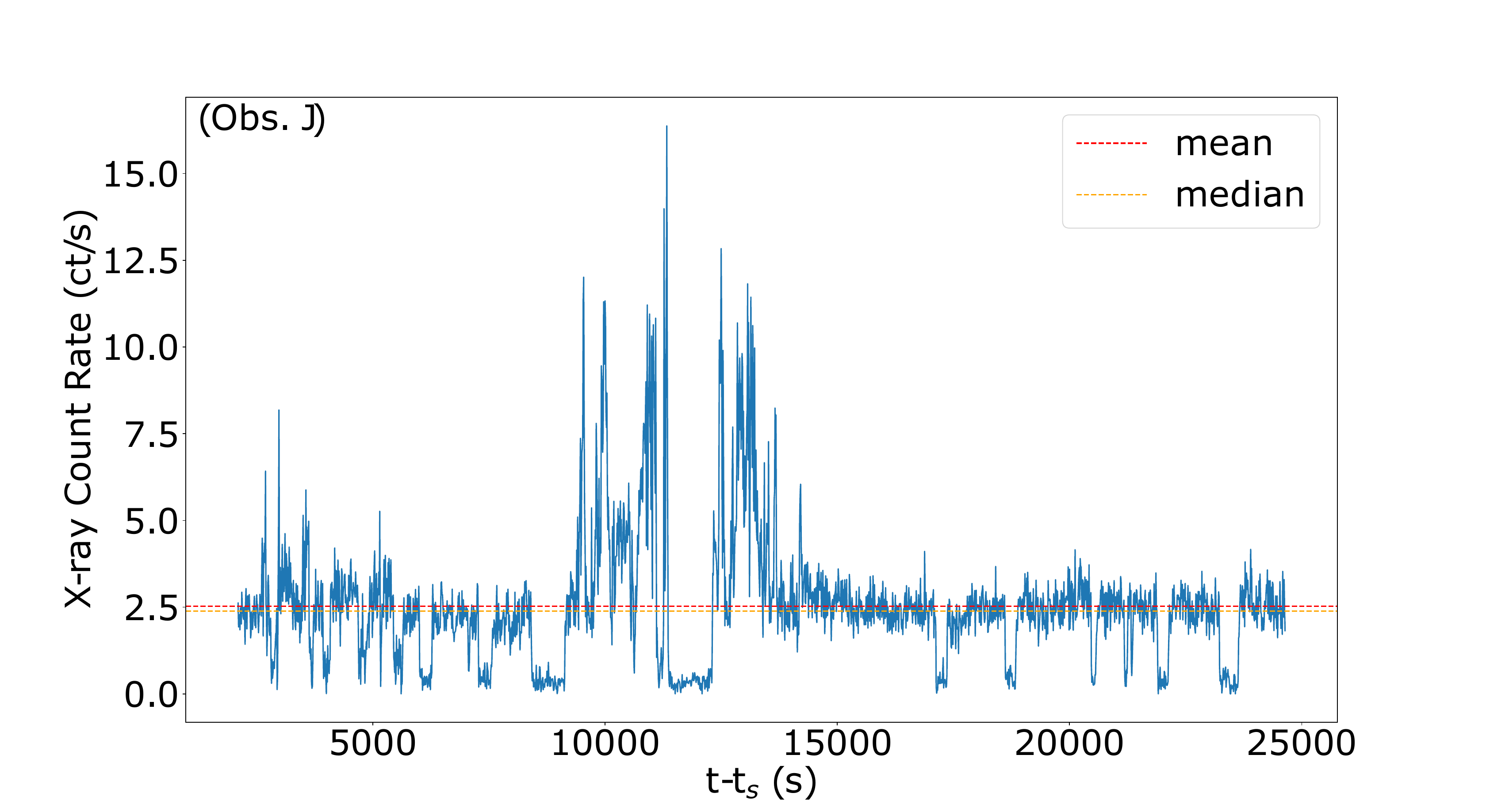}
	\end{minipage}
	\begin{minipage}[c]{0.5\textwidth}
		\centering
		\includegraphics[width=1\textwidth]{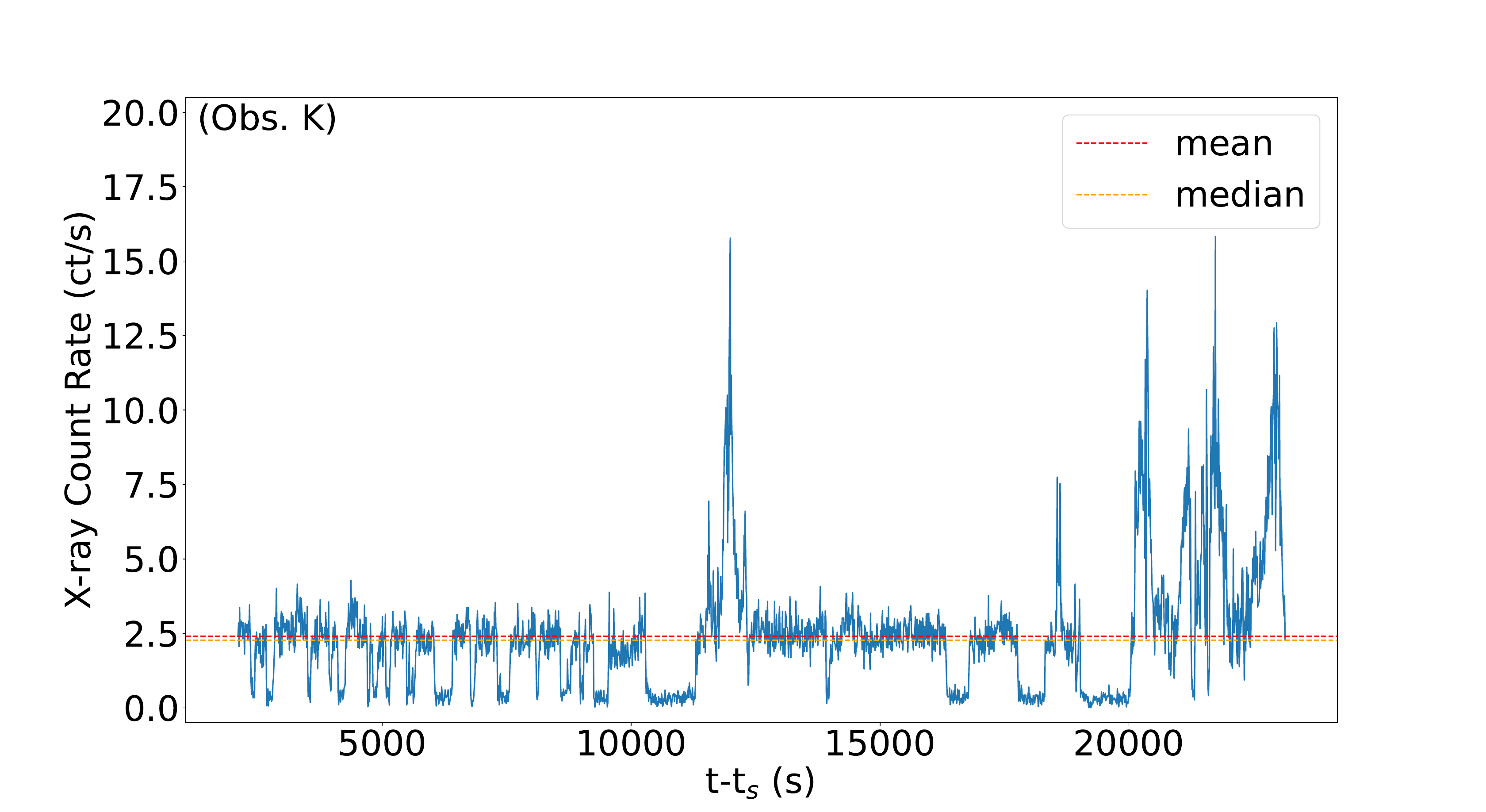}
	\end{minipage}
	\hspace{-1cm}
	\begin{minipage}[c]{0.5\textwidth}
		\centering
		\includegraphics[width=1\textwidth]{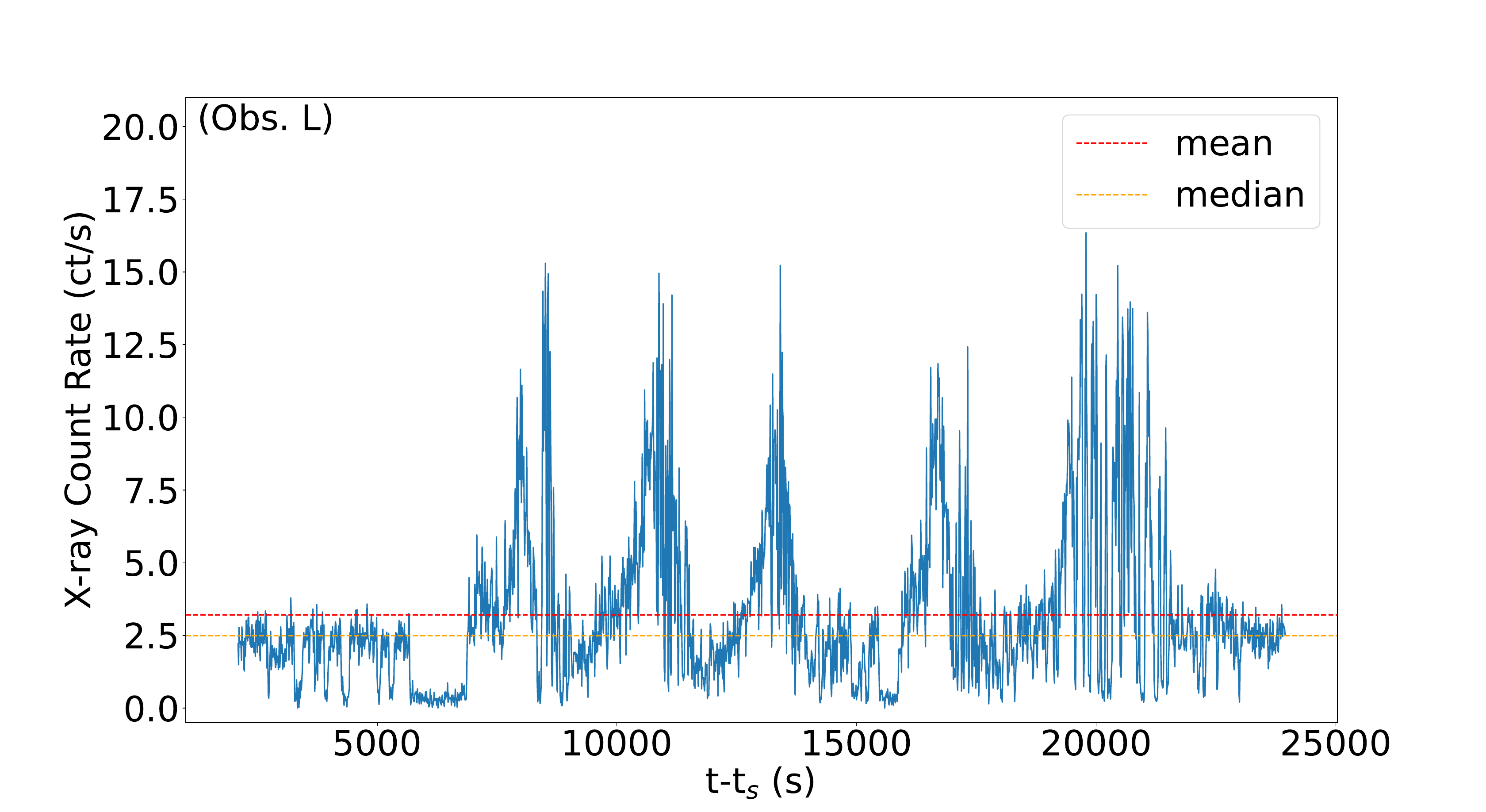}
	\end{minipage}\\
	\begin{minipage}[c]{0.5\textwidth}
		\centering
		\includegraphics[width=1\textwidth]{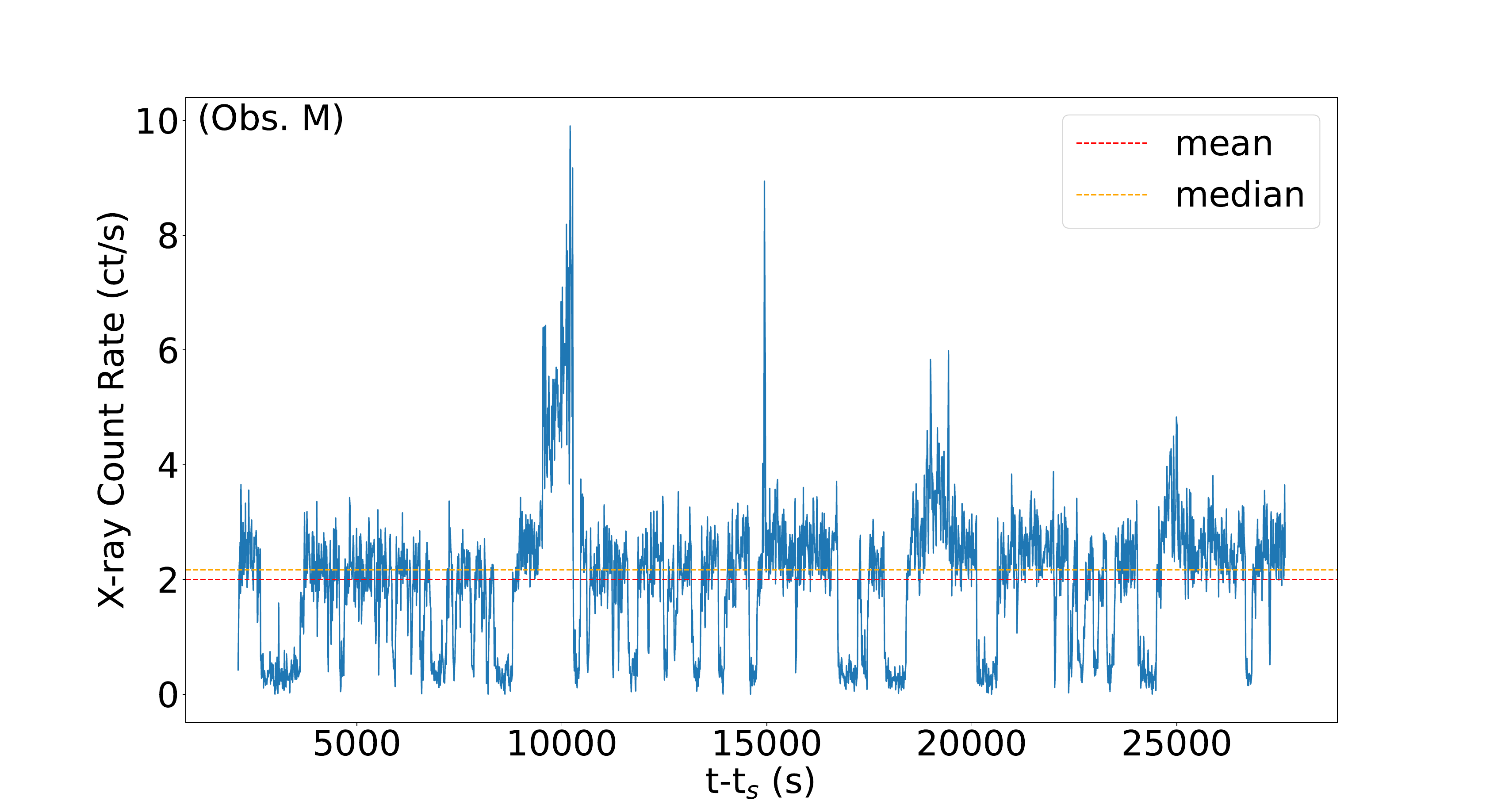}
	\end{minipage}
	\hspace{-1cm}
	\begin{minipage}[c]{0.5\textwidth}
		\centering
		\includegraphics[width=1\textwidth]{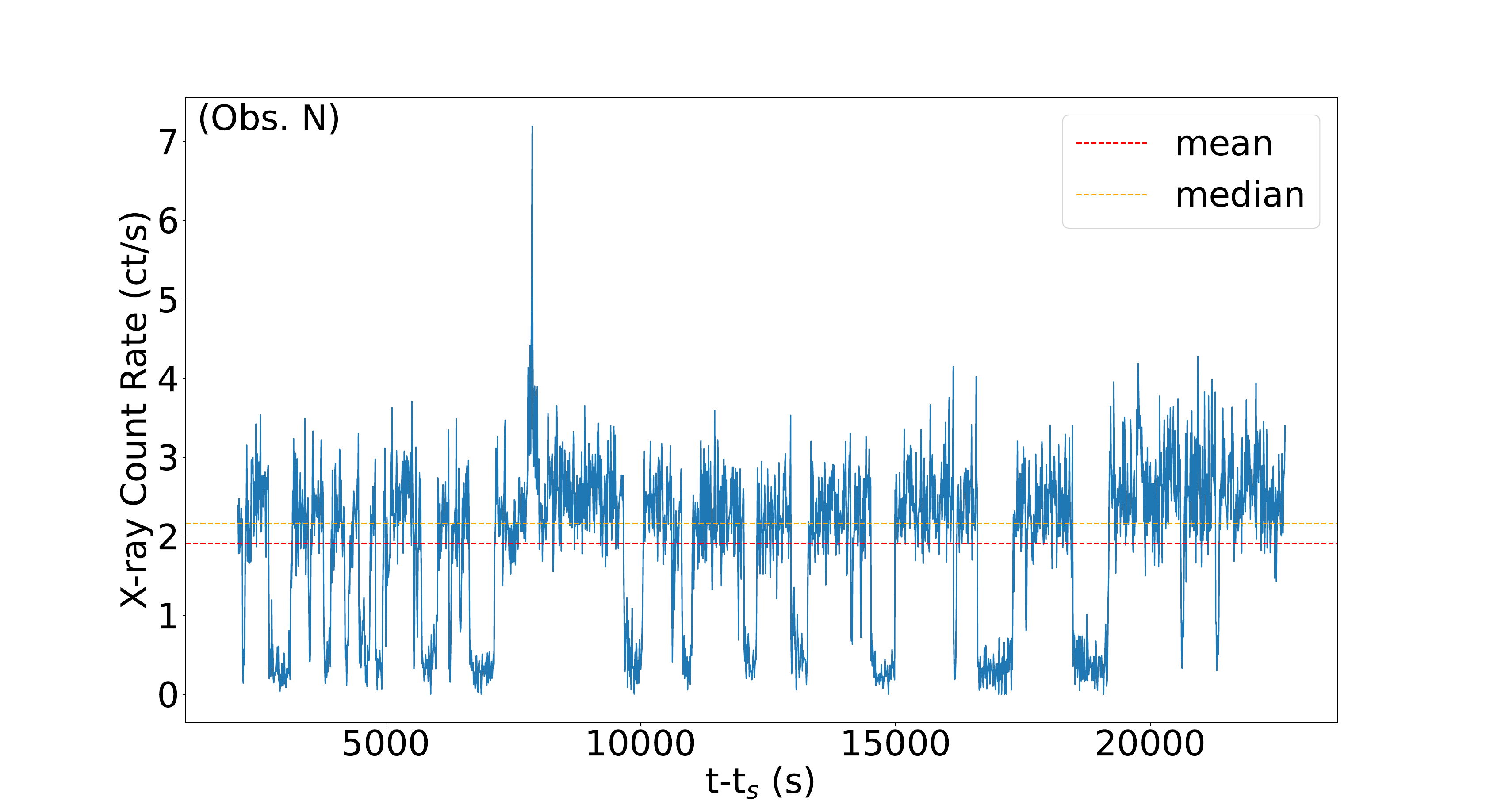}
	\end{minipage}\\
	\begin{minipage}[c]{0.5\textwidth}
		\centering
		\includegraphics[width=1\textwidth]{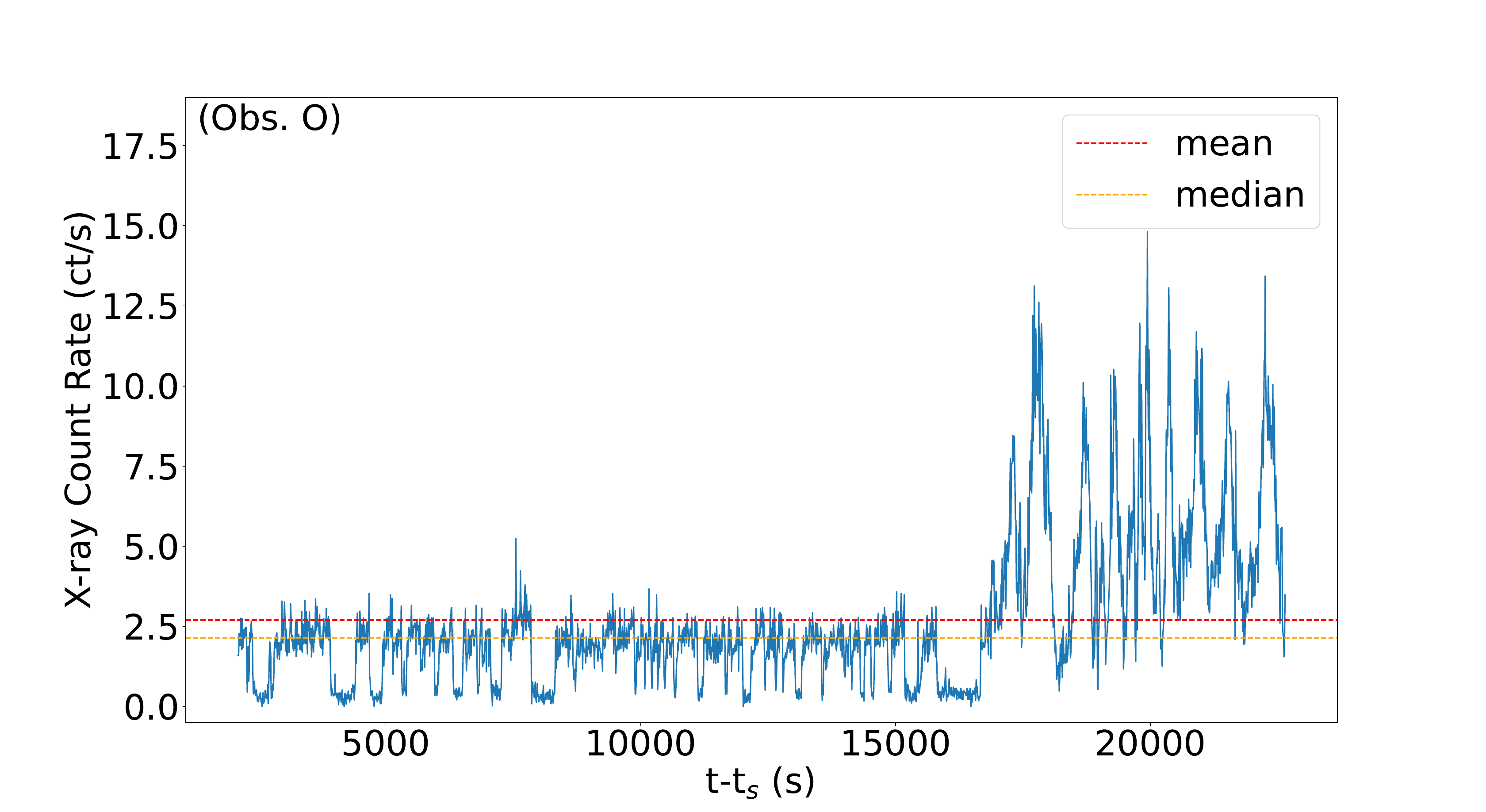}
	\end{minipage}
	\hspace{-1cm}
	\begin{minipage}[c]{0.5\textwidth}
		\centering
		\includegraphics[width=1\textwidth]{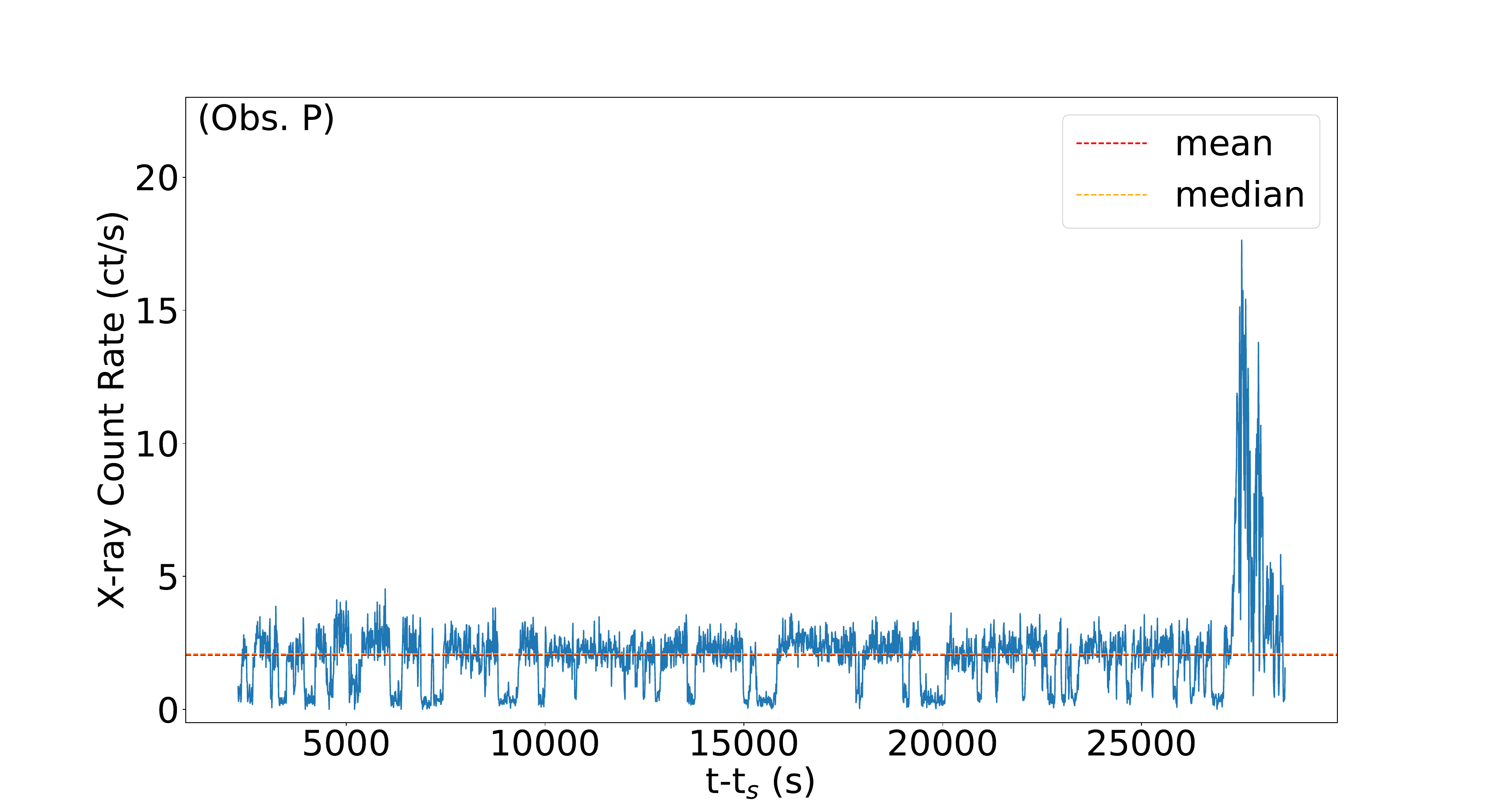}
	\end{minipage}\\
	\begin{minipage}[c]{0.5\textwidth}
		\centering
		\includegraphics[width=1\textwidth]{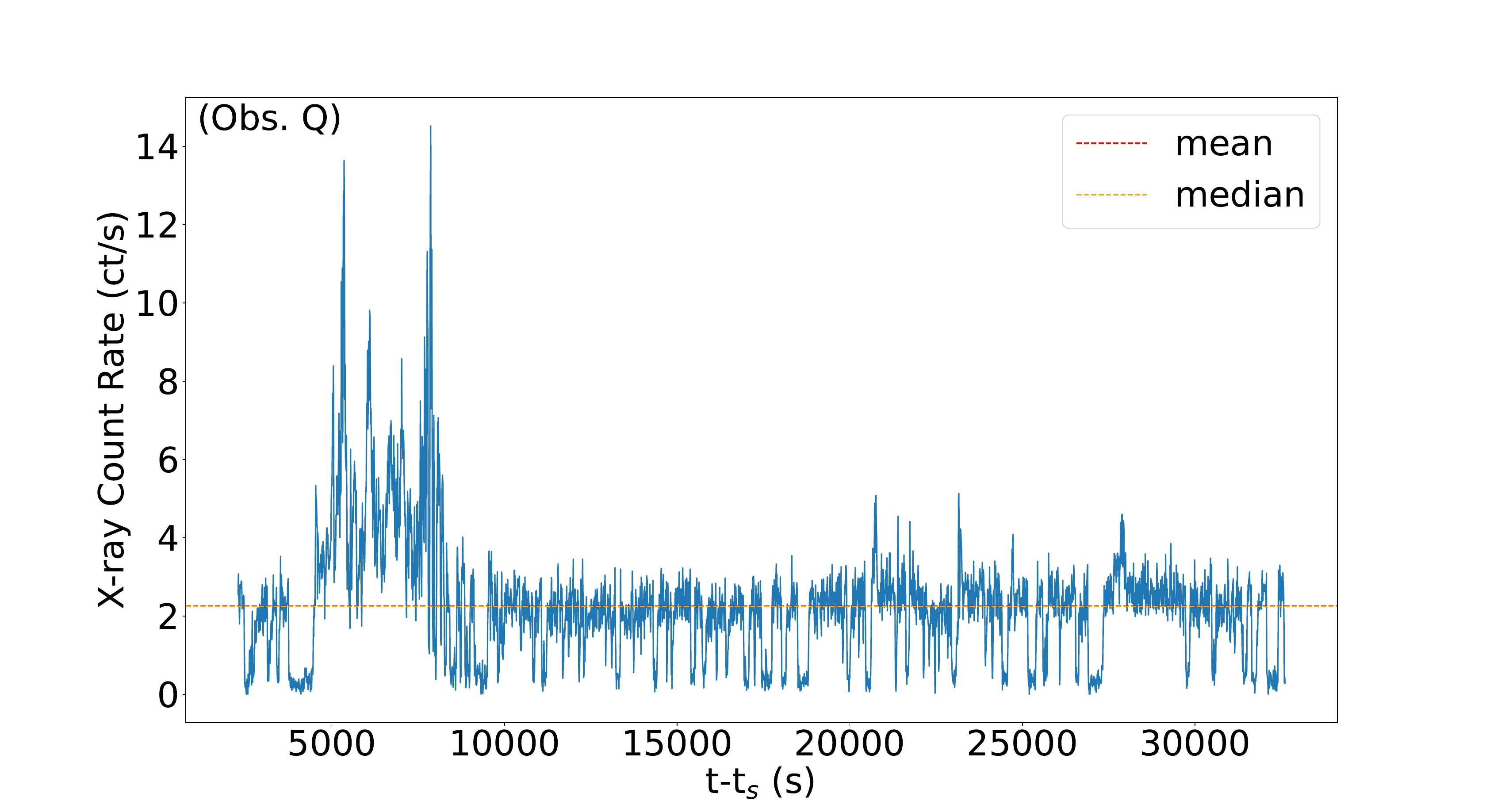}
	\end{minipage}
	\hspace{-1cm}
	\begin{minipage}[c]{0.5\textwidth}
		\centering
		\includegraphics[width=1\textwidth]{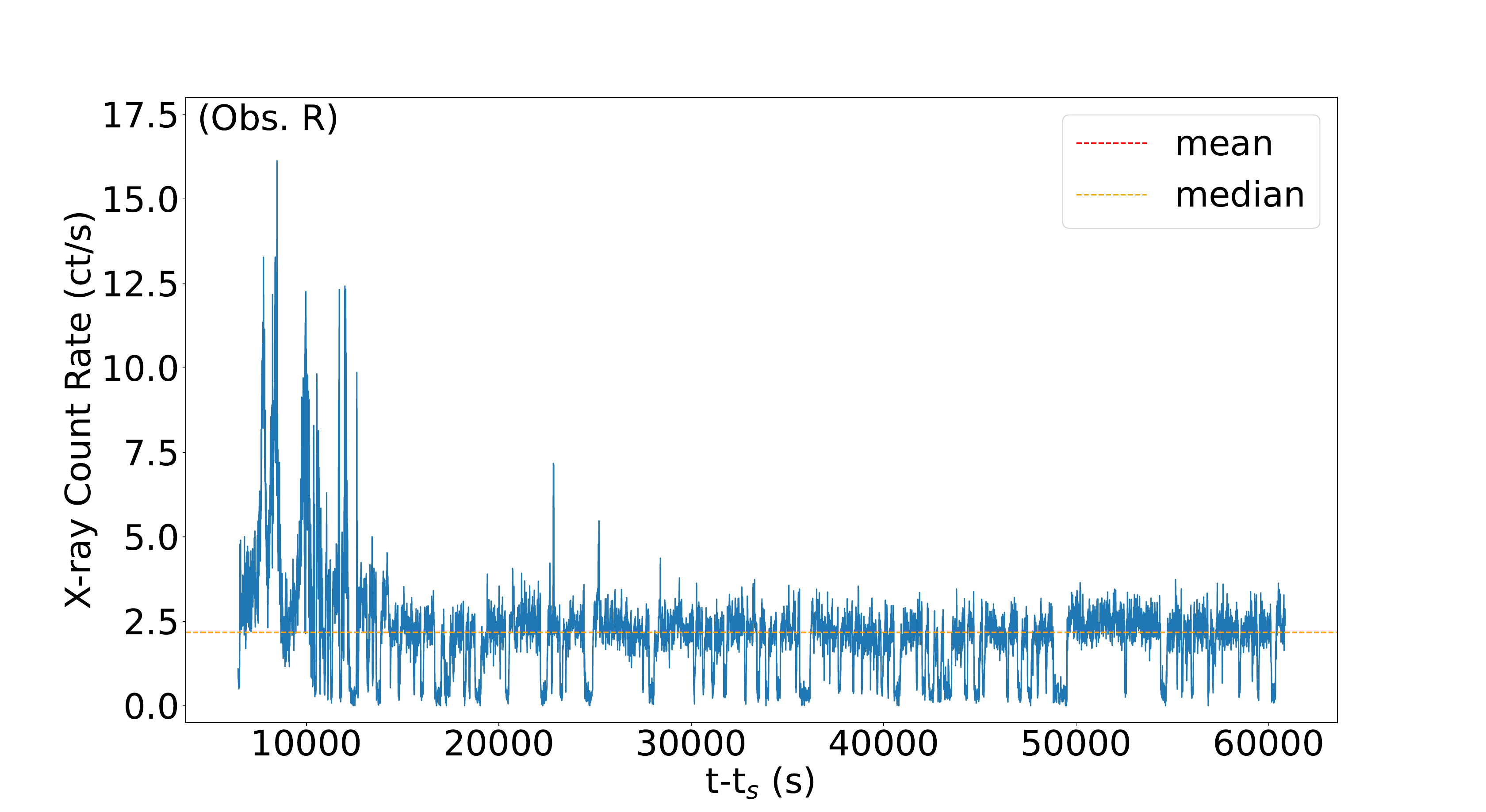}
	\end{minipage}\\

\caption{The X-ray light curves of the 18 \textit{XMM-Newton} observations. The red/orange dashed lines are the mean/median count rates, respectively.
\label{app_b}}
\end{figure*}

\clearpage 
\section{The judgment in the parameter we used in X-ray mode identification}\label{appC}
\begin{figure*}
\includegraphics[width=0.9\textwidth]{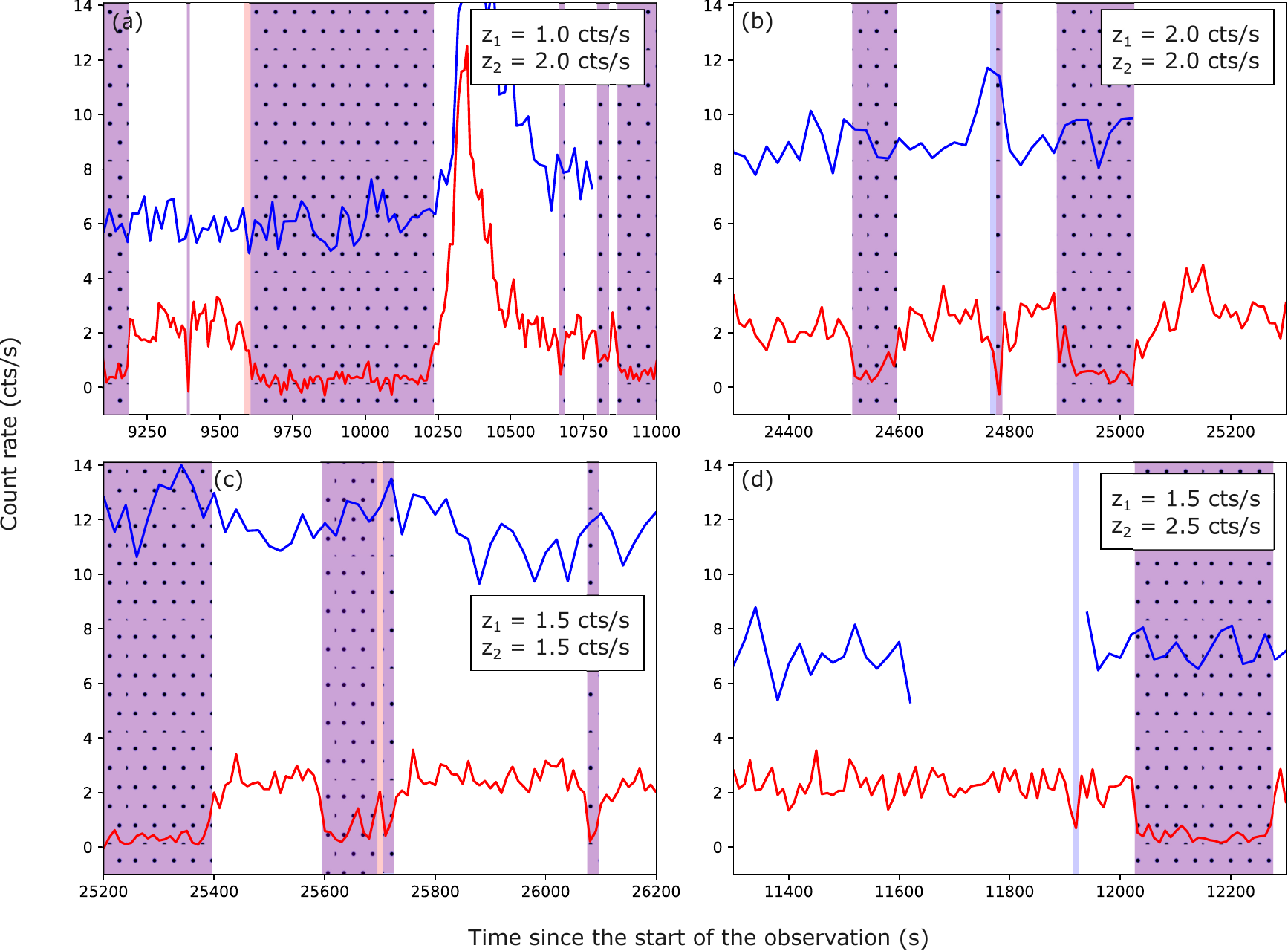}
\caption{Examples of the low-mode classification results with different sets of threshold values $z_1$ and $z_2$ (a: Obs. B; b: Obs. I; c: Obs. Q; d: Obs. N). The dotted purple shadows are low-mode regions agreed by both the parameters sets. The pink shadows indicate the low-mode intervals classified only by the adopted values (i.e., $z_1=1.5$ counts/s and $z_2=2.0$ counts/s), and the blue ones are those classified only by the parameters shown at the top right corners. As shown in the examples, the differences are raised by ambiguous cases. In general, the pink regions looks slightly more convincing than the blue ones, and therefore $z_1=1.5$ counts/s and $z_2=2.0$ counts/s were adopted. It is important noting that no discrepancy was found in many observations among the tested parameters sets, and those with discrepancy are still closely consistent with each other. 
\label{fig:lowmode_classification}}
\end{figure*}

\bsp	
\label{lastpage}
\end{document}